\documentclass[authoryear,5p,twocolumn]{elsarticle}

\usepackage{graphicx}
\usepackage{natbib}
\usepackage{color}
\usepackage{psfrag}
\usepackage{amssymb}
\usepackage{float}
\usepackage{longtable}
\usepackage{deluxetable}
\usepackage[authoryear]{natbib}

\def\astrobj#1{#1}

\def\jh{\mbox{$(J-H)$}}

\def\jk{\mbox{$(J-K_s)$}}

\journal{New Astronomy}

\begin{document}

\begin{frontmatter}

\title{New Galactic embedded clusters and candidates from a WISE Survey}

\author{D. Camargo, E. Bica, C. Bonatto}

\address{Departamento de Astronomia, Universidade Federal do Rio Grande do Sul, 
Av. Bento Gon\c{c}alves 9500\\
Porto Alegre 91501-970, RS, Brazil\\}

\begin{abstract}
We carried out  a search  for  new infrared star clusters, stellar groups and candidates using  WISE images, which  are  very sensitive to dust emission nebulae. We report the  discovery of  437  embedded clusters and stellar groups that show a variety of  structures, both in the stellar and  nebular components. Pairs or  small groupings of clusters are observed, suggesting multiple generations at the early
formation stages. The resulting catalogue  provides Galactic and equatorial coordinates, together with  angular sizes for all objects. The nature of  a representative test sub-sample of 14  clusters is investigated in detail by means of 2MASS photometry. The colour magnitude diagrams and radial density distributions characterize  them as stellar clusters.   The    437 new objects were found in the  ranges $145^\circ\,\leq\,\ell\,\leq 290^\circ$   and   $-25^\circ\,\leq\,b\,\leq 20^\circ$, and  they appear to be a major object source for future studies of star cluster formation and their early evolution.  WISE  is   a powerful tool to  further probe  for  very young clusters throughout the disk. 
\end{abstract}

\end{frontmatter}

\section{Introduction}
\label{Intro}

Dust enshrouded star clusters have been systematically studied with relatively large samples in the near infrared since \citet{Hodapp94}, as detectors became efficient enough. In the early 2000's several lists of new dusty clusters were provided \citep{Ivanov02, Bica03a, Dutra03}. At that time efforts were made to compile catalogues  of this object class \citep{Lada03, Bica03b}.  Some  young clusters have been   studied in  X-rays with Chandra \citep[e.g.][]{Ybarra13}. Recently, \citet{Majaess13} used WISE together with  other data to discover  229 new dust related stellar clusters. The latter findings involving WISE near and mid infrared bands indicate  the importance to carry out  new searches for stellar clusters in formation and early  evolutionary stages.

The Galaxy appears to contain   $\sim10^5$ open clusters \citep{Bonatto06a, Piskunov06} but  only $\sim2000$ have astrophysical  parameters determined (DAML02\footnote{http://www.astro.iag.usp.br/~wilton/ \citep{Dias02}.}, WEBDA\footnote{http://obswww.unige.ch/webda \citep{Mermelliod96}.}). 
The completeness effects in the OC system are due mainly to the higher extinction and stellar crowding in the  disc \citep{Lada91, Hodapp94, Bica03b}.

Stellar clusters are fundamental tools to improve our understanding of the Galactic structure, star formation history, and evolution \citep{Friel95, Bonatto06a, Piskunov06, Bobylev07, Camargo13, Bobylev14}. Most star formation occurs in crowded environments, such as in embedded clusters (ECs) and embedded stellar groups. ECs are formed  in giant molecular clouds (GMCs), in dense clumps with  size of $\sim1$ pc. Since ECs can be heavily obscured , most of  them are  detected in the infrared. The development of infrared array detectors in recent decades has boosted our knowledge of these objects. As a result sky surveys provided new objects  \citep{Dutra03, Bica03a, Bica03b, Bica05, Kharchenko05a, Kharchenko05b, Mercer05, Kronberger06, Froebrich07, Koposov08, Glushkova10, Borissova11}.  Recently,  we discovered 16 new ECs (CBB 1 - CBB 16), some of them within groups of clusters \citep{Camargo11, Camargo12, Camargo13}. A fundamental contribution of such detailed studies of individual clusters is  homogenously derived astrophysical parameters.

Most (if not all) stars form in star clusters \citep{Lada91, Hodapp94, Lada03}. In this sense, they  are frequently considered as  the building blocks in galaxy construction.  \citet{Lada03} define a cluster as a group of physically related stars that survive longer  than 100 Myr as a bound system. This requires at least  35 stars and stellar-mass density larger than $1M_{\odot}/pc^{-3}$. \citet{Adams01} indicate three ways to form stars: isolated singles and binaries, stellar groups (10 to 100 members), and clusters ($> 100$ stars). They argue that stellar groups are the most common way to form stars. \citet{Hodapp94} considers cluster the stellar groups with five or more stars.   

\begin{figure*}[!ht]
\begin{minipage}[b]{0.328\linewidth}
\psfragscanon
\includegraphics[width=\textwidth]{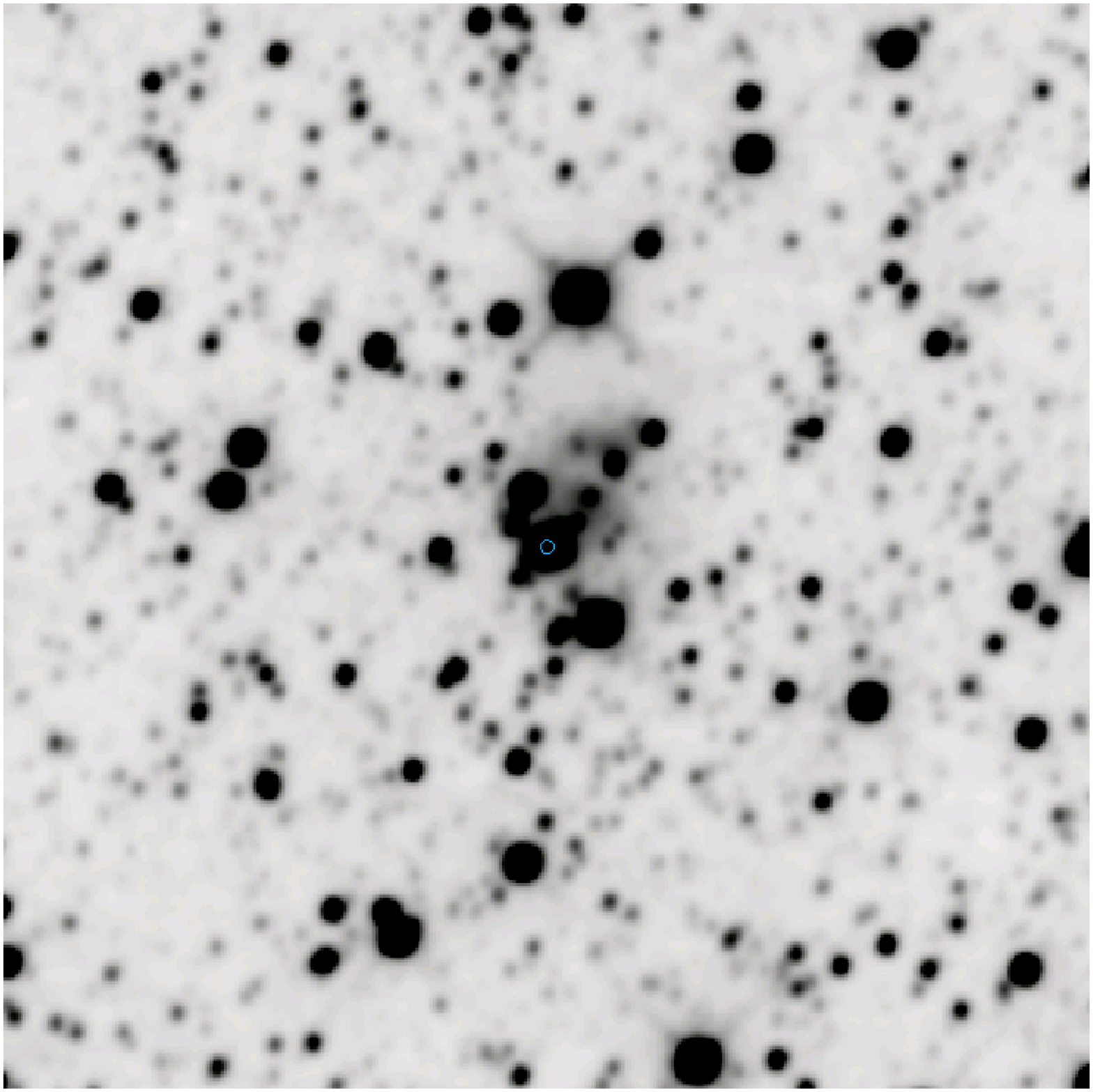}
\put(-140.0,155.0){\makebox(0.0,0.0)[5]{\fontsize{16}{16}\selectfont {\bf C 62}}}
\end{minipage}\hfill
\hspace{0.03cm}
\begin{minipage}[b]{0.328\linewidth}
\includegraphics[width=\textwidth]{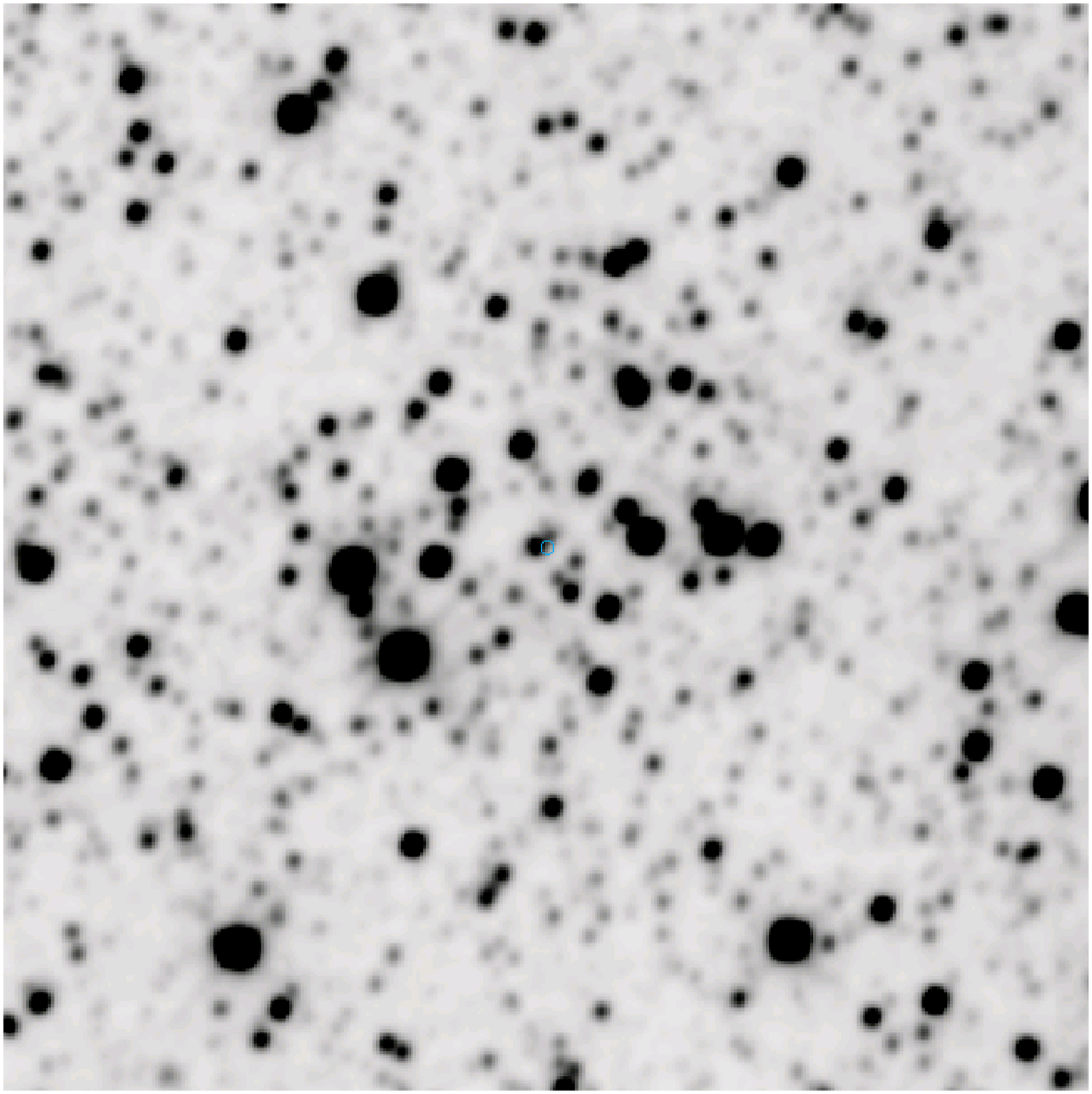}
\put(-140.0,155.0){\makebox(0.0,0.0)[5]{\fontsize{16}{16}\selectfont {\bf C 72}}}
\end{minipage}\hfill
\hspace{0.03cm}
\begin{minipage}[b]{0.328\linewidth}
\includegraphics[width=\textwidth]{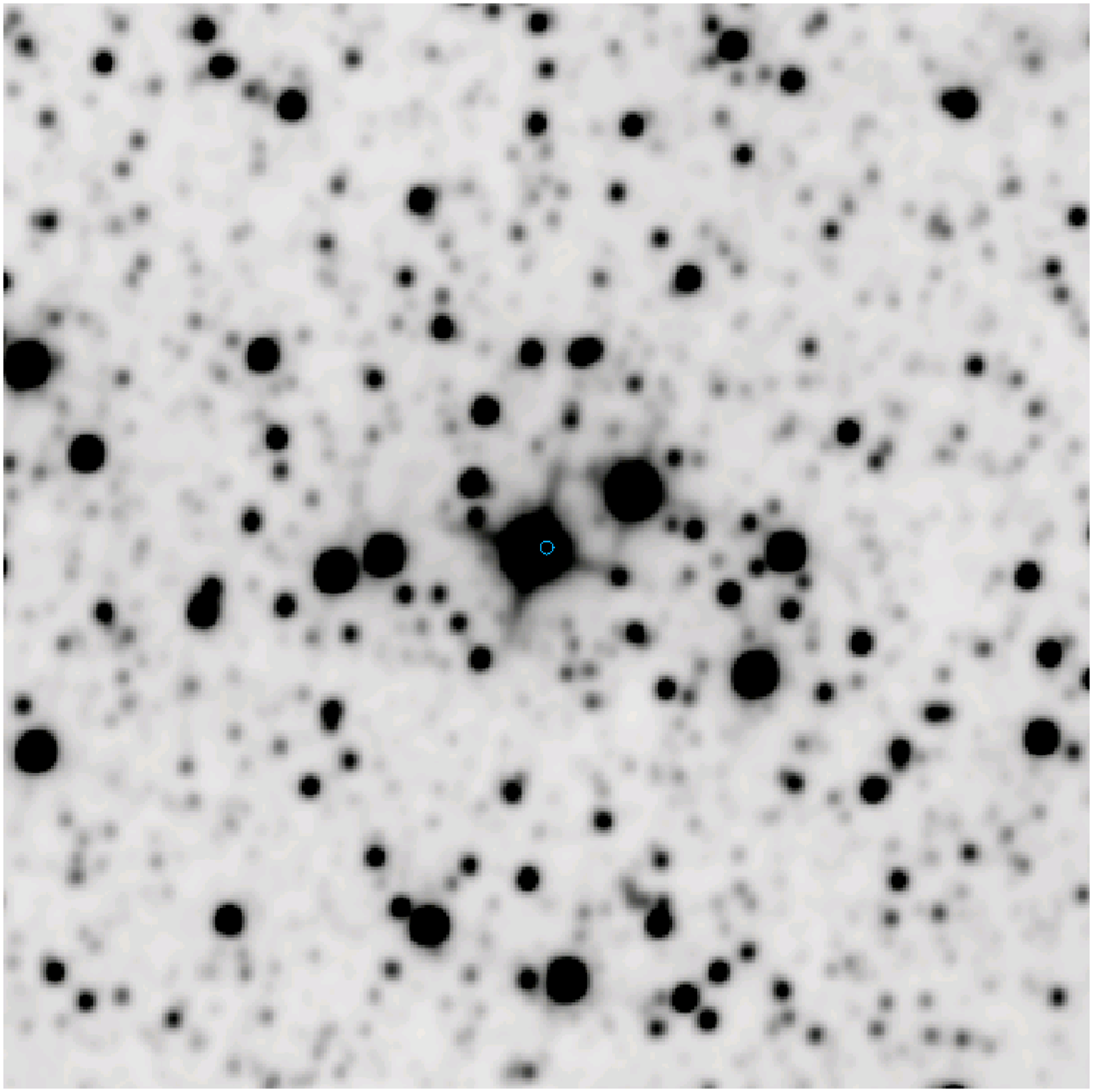}
\put(-140.0,155.0){\makebox(0.0,0.0)[5]{\fontsize{16}{16}\selectfont {\bf C 258}}}
\end{minipage}\hfill
\vspace{0.03cm}
\begin{minipage}[b]{0.328\linewidth}
\includegraphics[width=\textwidth]{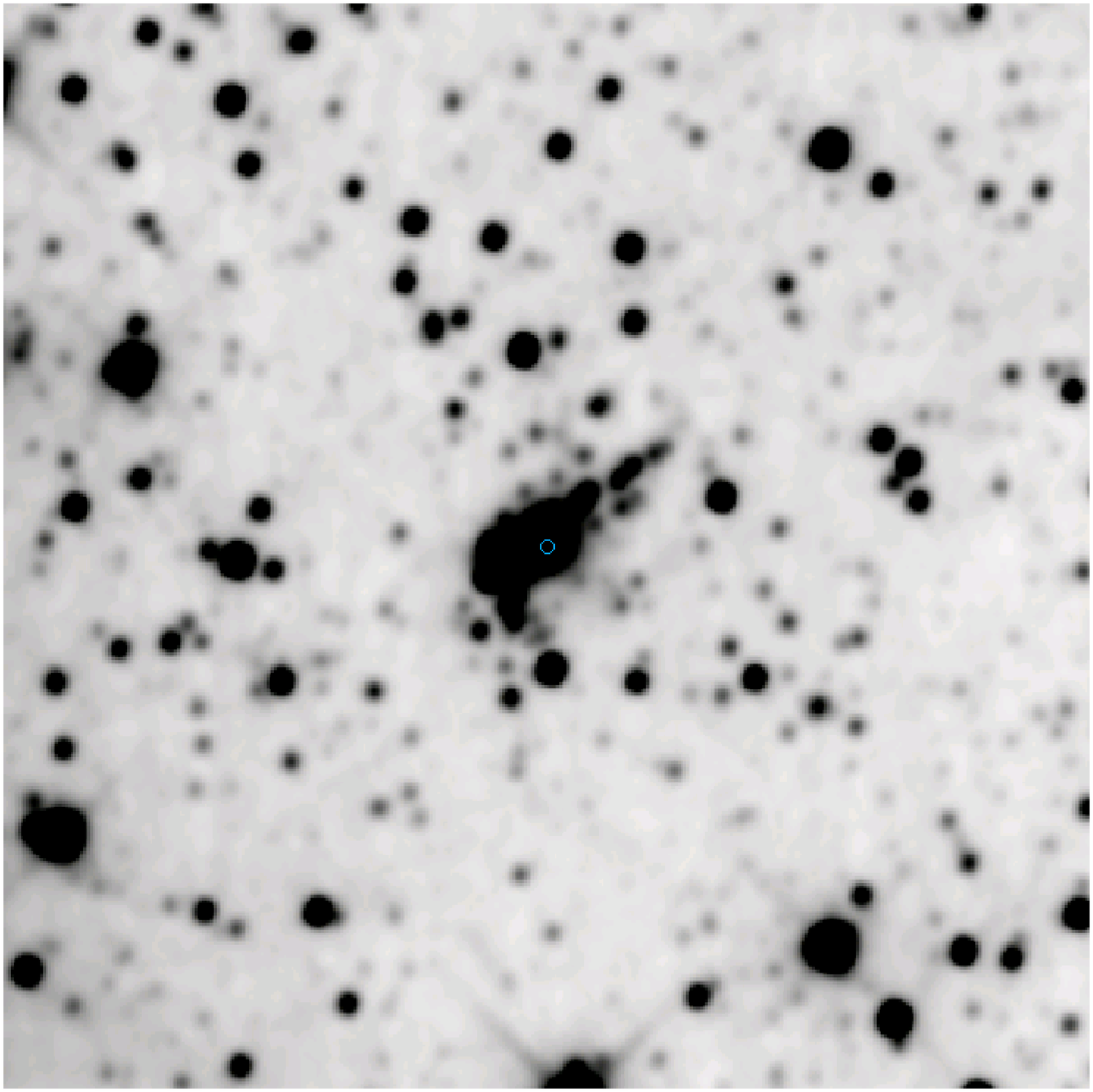}
\put(-140.0,155.0){\makebox(0.0,0.0)[5]{\fontsize{16}{16}\selectfont {\bf C 44}}}
\end{minipage}\hfill
\hspace{0.03cm}
\begin{minipage}[b]{0.328\linewidth}
\includegraphics[width=\textwidth]{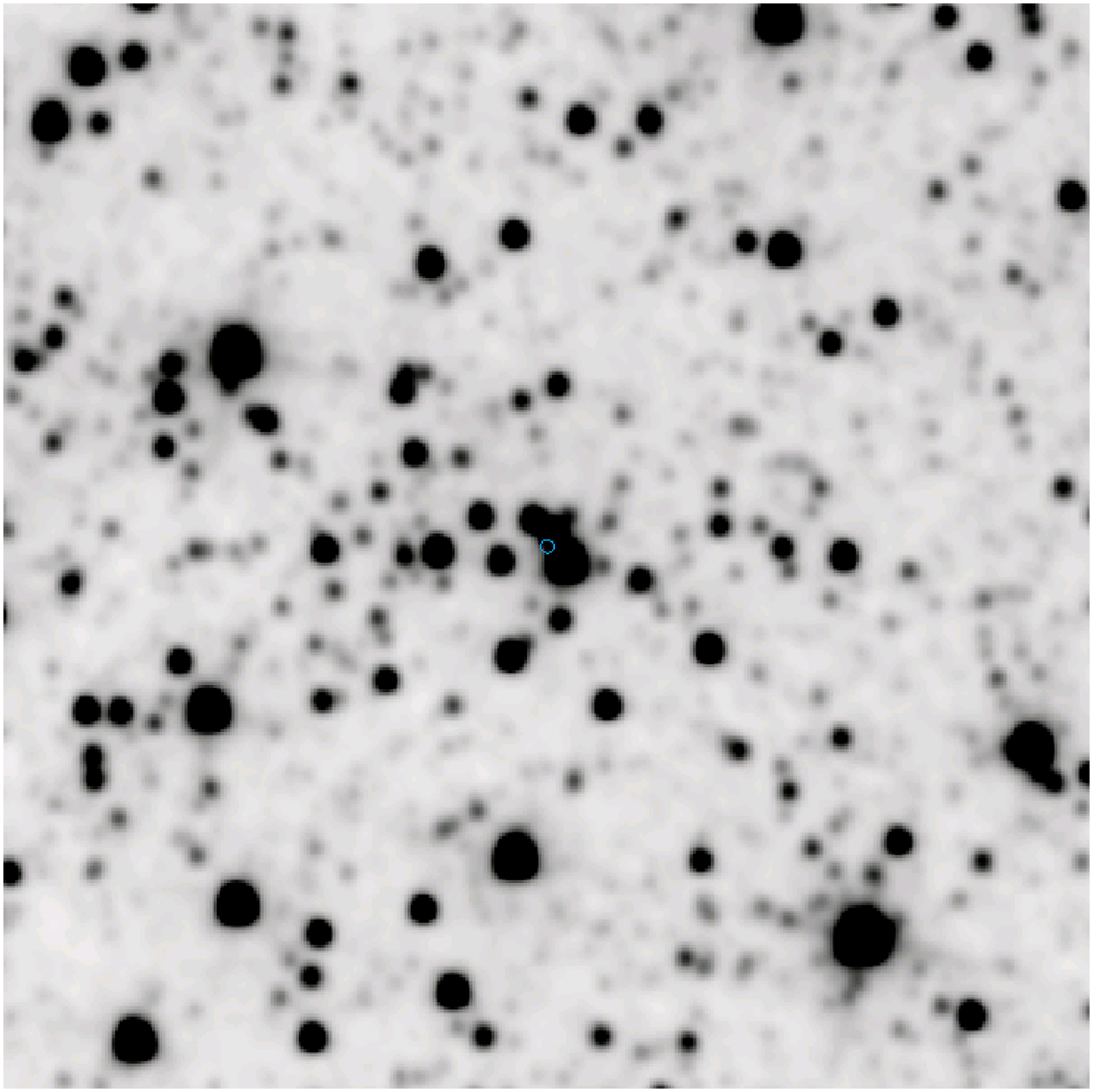}
\put(-140.0,155.0){\makebox(0.0,0.0)[5]{\fontsize{16}{16}\selectfont {\bf C 284}}}
\end{minipage}\hfill
\hspace{0.03cm}
\begin{minipage}[b]{0.328\linewidth}
\includegraphics[width=\textwidth]{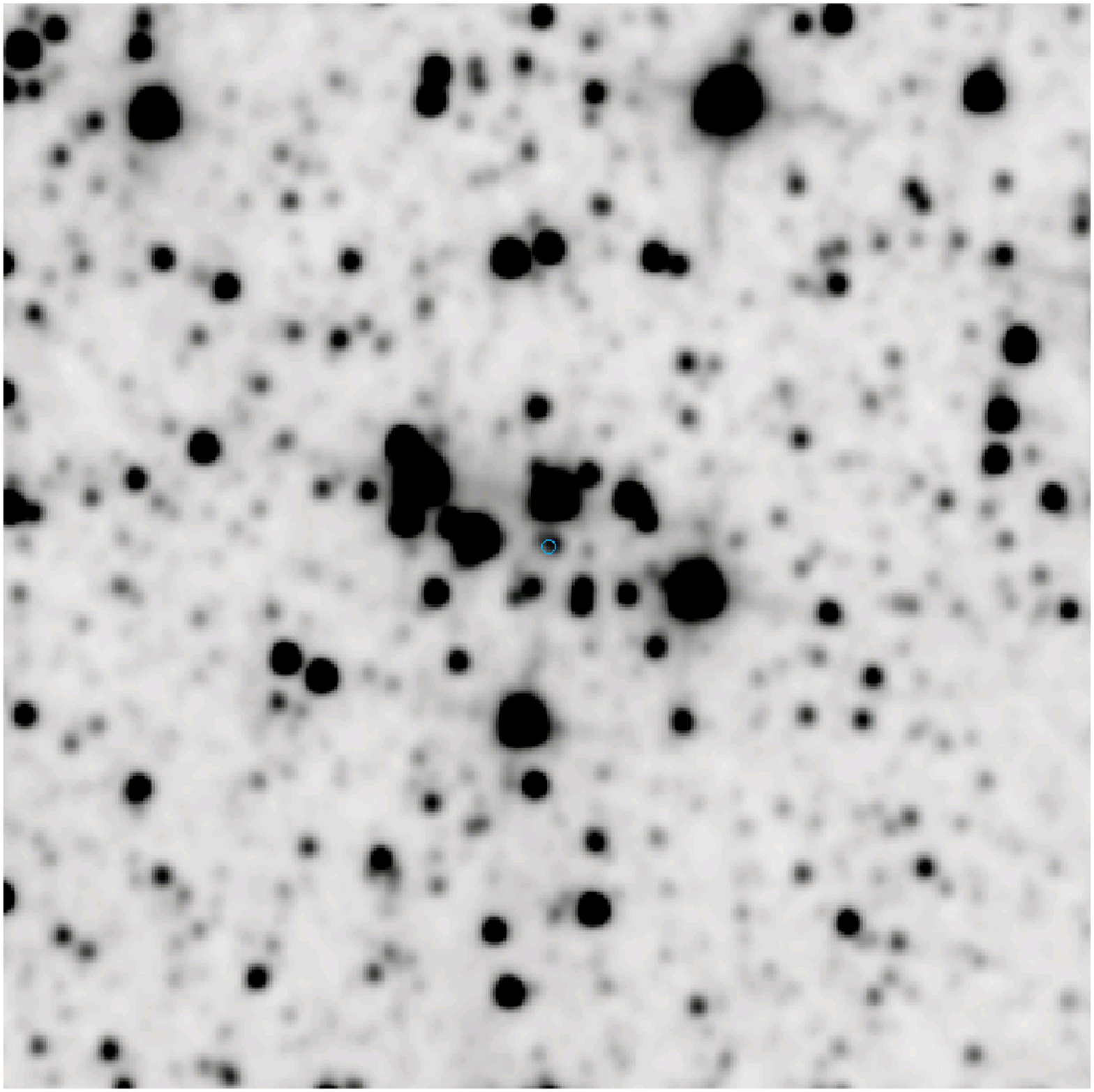}
\put(-130.0,155.0){\makebox(0.0,0.0)[10]{\fontsize{16}{16}\selectfont {\bf C 288}}}
\end{minipage}\hfill
\caption[]{Examples of sample clusters. Top panels: W1 WISE images ($10'\times10'$)  centred on C 62, C 72, and C 258. Bottom: W2 images centred on C 44, C 284, and C 288. These objects are classified as ECs.}
\label{f2}
\end{figure*}

In this  paper  we carry out a search for new embedded clusters using WISE band images. We were encouraged by the impressive large number of recent discoveries by \citet{Majaess13}. However, we employ  a complementary different method.

This paper is organized as follows. In Sect. \ref{sec:2} we present the new clusters found in this study. In Sect. \ref{sec:3} we discuss the stellar density  and dust distributions. In  Sect. \ref{sec:3} we  highlight star formation stages that are  possibly related  to the  dust distribution, and in turn  infer  forming processes themselves.  In Sect. \ref{sec:4} we employ  2MASS photometry for a selection of 14 new clusters to test their reality. In Sect. \ref{sec:5} we analyse the corresponding CMDs and RDPs. Finally, in Sect. \ref{sec:6} we give the concluding remarks.

\begin{figure*}[!ht]
\begin{minipage}[b]{0.328\linewidth}
\psfragscanon
\includegraphics[width=\textwidth]{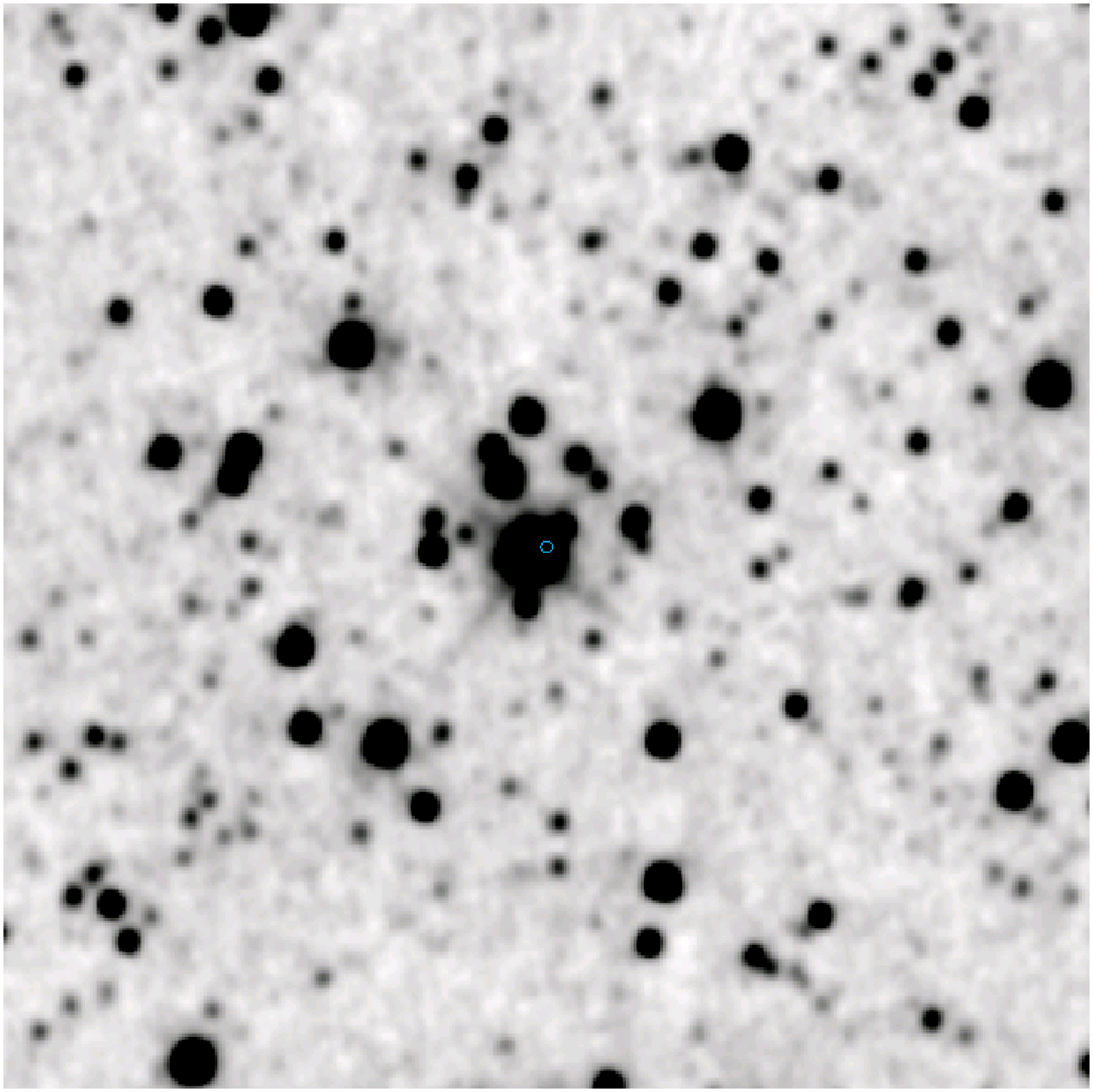}
\put(-120.0,155.0){\makebox(0.0,0.0)[5]{\fontsize{16}{16}\selectfont {\bf C 79  (OC)}}}
\end{minipage}\hfill
\hspace{0.03cm}
\begin{minipage}[b]{0.328\linewidth}
\includegraphics[width=\textwidth]{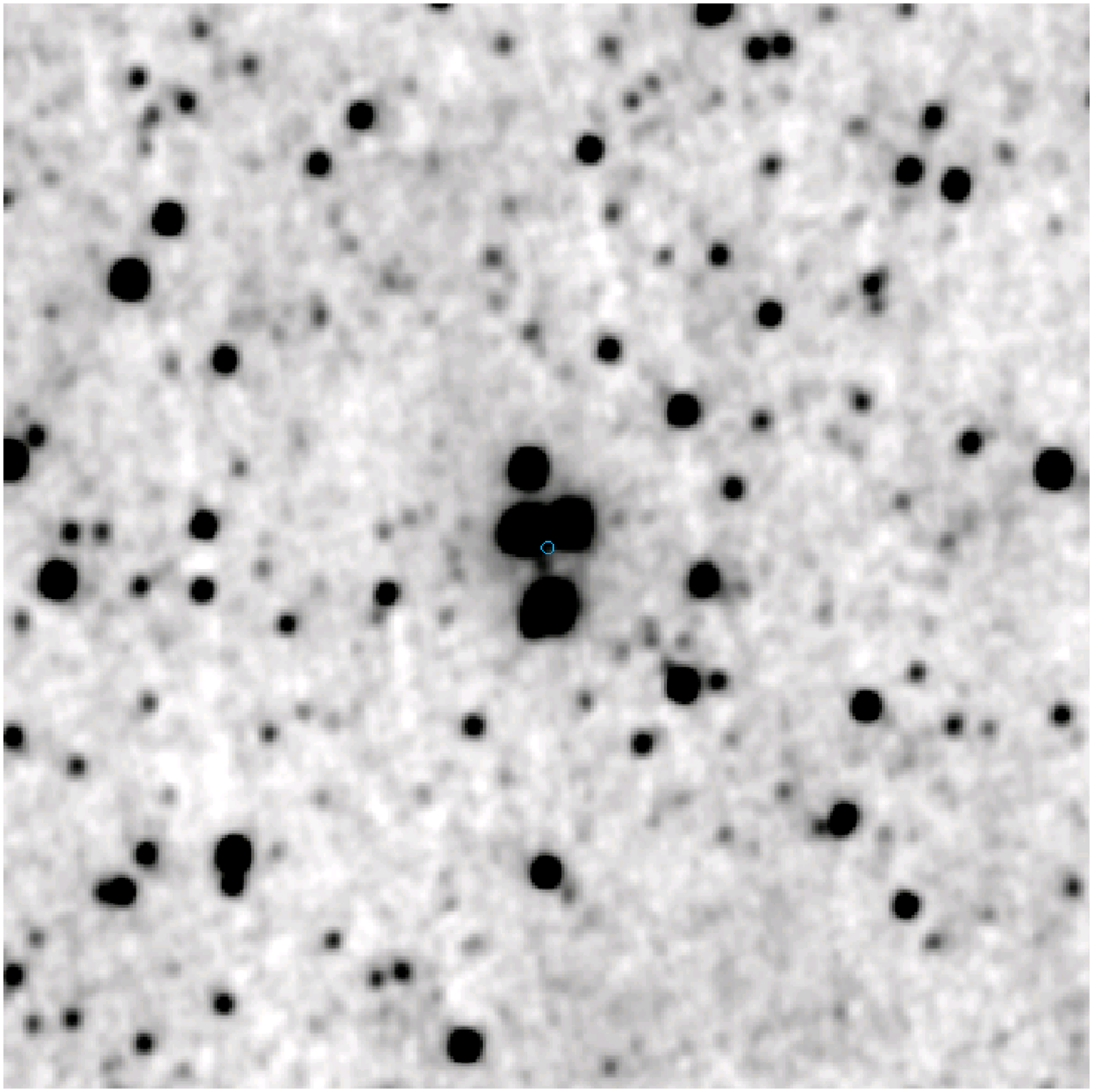}
\put(-110.0,155.0){\makebox(0.0,0.0)[5]{\fontsize{16}{16}\selectfont {\bf C 70 (OCC)}}}
\end{minipage}\hfill
\hspace{0.03cm}
\begin{minipage}[b]{0.328\linewidth}
\includegraphics[width=\textwidth]{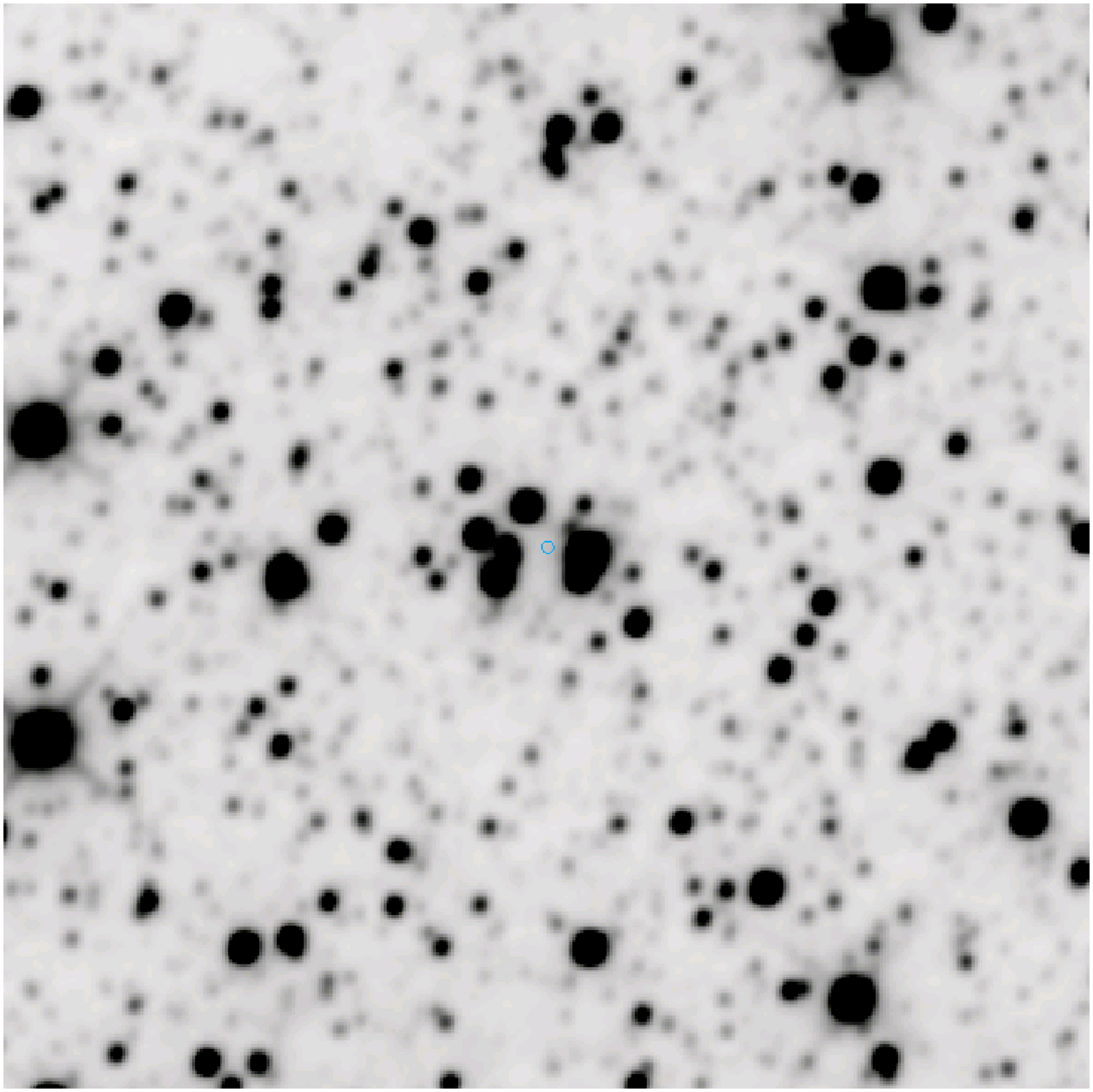}
\put(-105.0,155.0){\makebox(0.0,0.0)[5]{\fontsize{16}{16}\selectfont {\bf C 193 (OCC)}}}
\end{minipage}\hfill
\vspace{0.03cm}
\begin{minipage}[b]{0.328\linewidth}
\includegraphics[width=\textwidth]{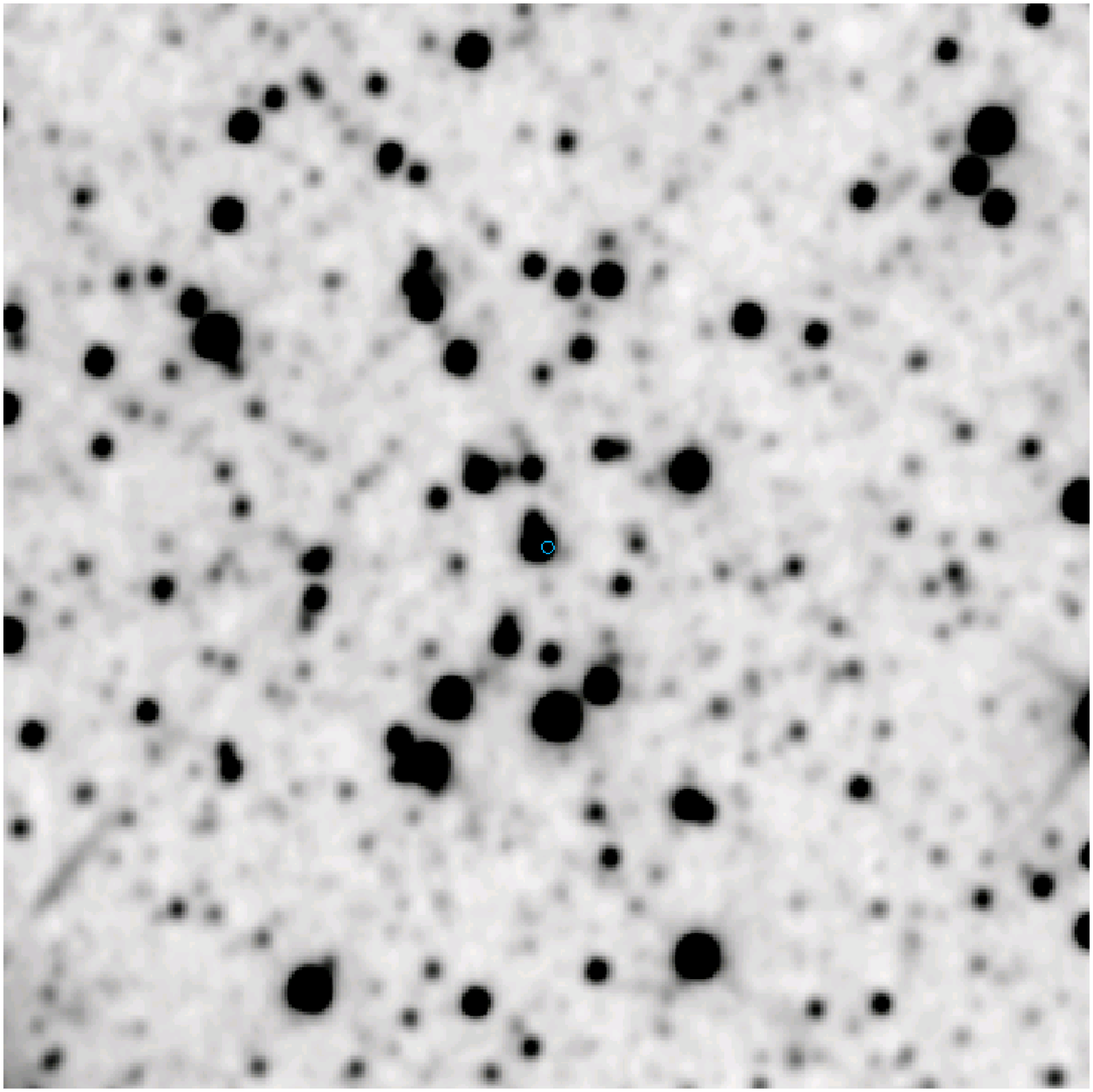}
\put(-105.0,155.0){\makebox(0.0,0.0)[5]{\fontsize{16}{16}\selectfont {\bf C 192 (ECC)}}}
\end{minipage}\hfill
\hspace{0.03cm}
\begin{minipage}[b]{0.328\linewidth}
\includegraphics[width=\textwidth]{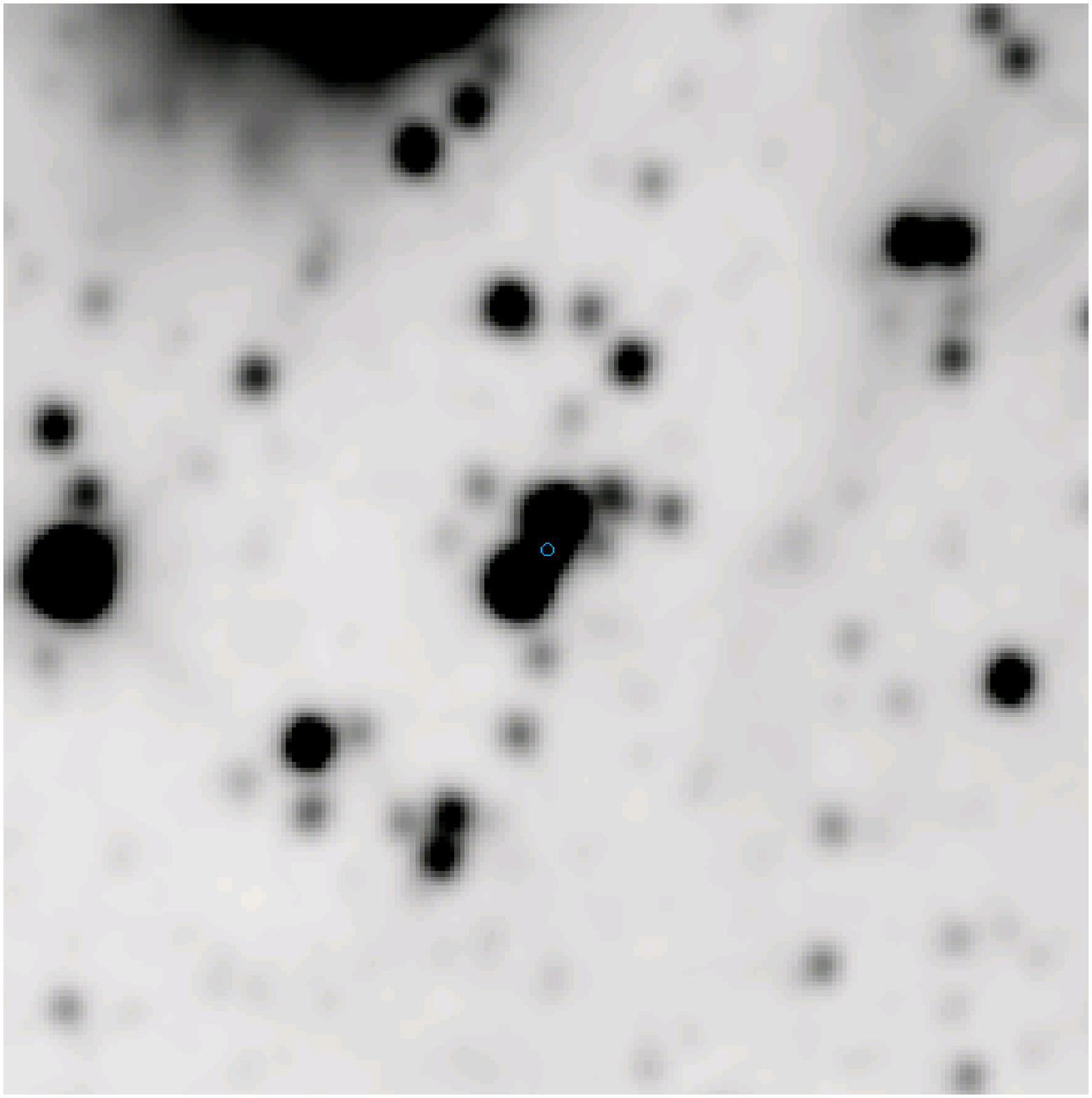}
\put(-112.0,155.0){\makebox(0.0,0.0)[5]{\fontsize{16}{16}\selectfont {\bf C 53 (EGr)}}}
\end{minipage}\hfill
\hspace{0.03cm}
\begin{minipage}[b]{0.328\linewidth}
\includegraphics[width=\textwidth]{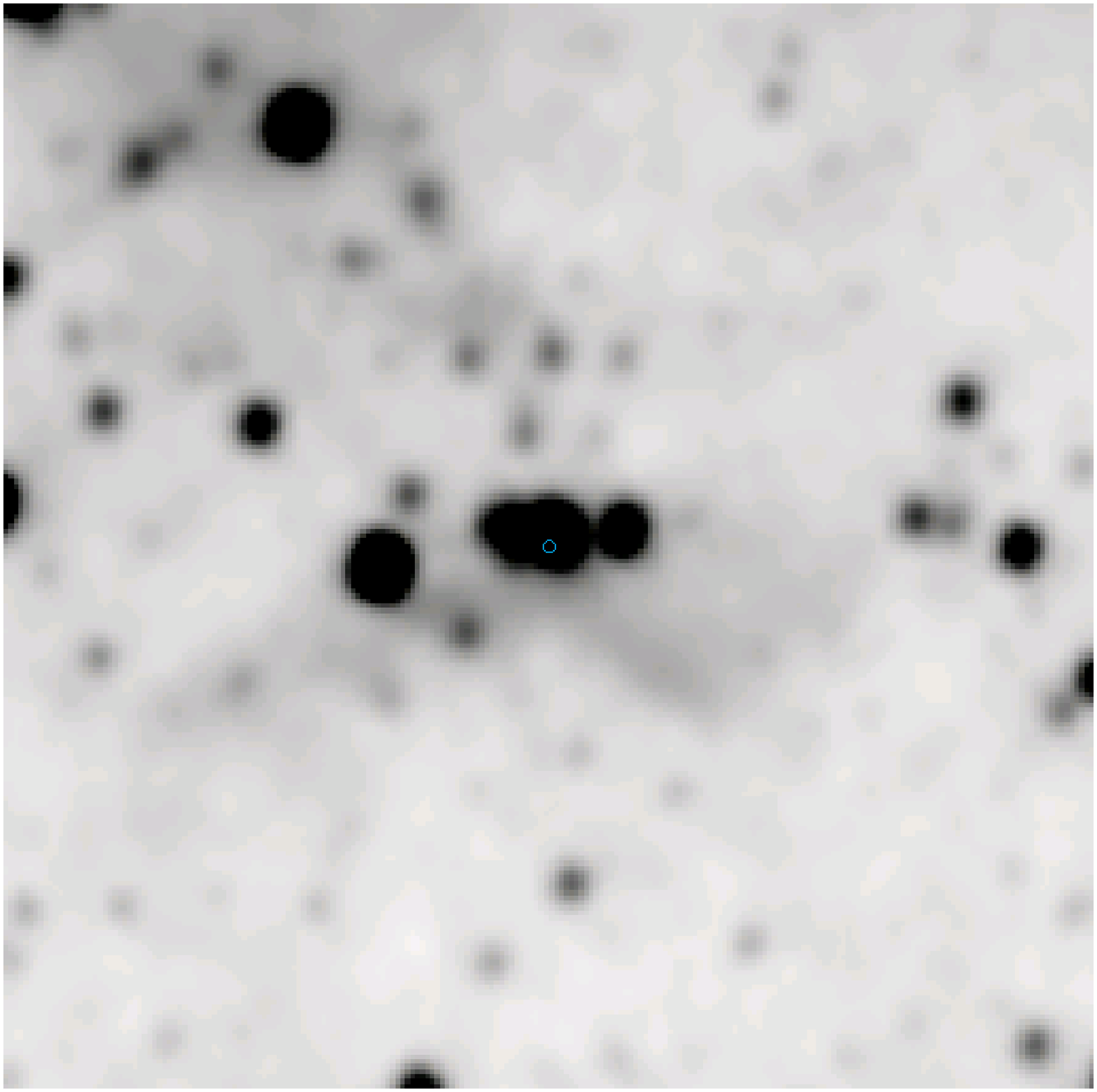}
\put(-110.0,155.0){\makebox(0.0,0.0)[10]{\fontsize{16}{16}\selectfont {\bf C 43 (EGr)}}}
\end{minipage}\hfill
\caption[]{Examples of additional sample classifications. Top panels: W2 WISE images ($10'\times10'$)  centred on the OC C 79 and on the OCCs C 70, and C 193. Bottom: W2 images ($10'\times10'$) centred on the ECC C 192 and W2 images ($5'\times5'$) centred on the EGrs C 53, and C 43.}
\label{nova}
\end{figure*}

\section{New star clusters and candidates}
\label{sec:2}

We searched for new clusters, stellar groups, and candidates in the WISE\footnote{The Wide-field Infrared Survey Explorer} \citep{Wright10} image atlas. WISE is a survey covering the entire sky in the infrared bands W1, W2, W3, and W4  centred at 3.4, 4.6, 12 and $22{\mu}$m.  The project was funded by NASA and operated by the Jet Propulsion Laboratory (JPL). The data processing and analysis was done by the Infrared Processing and Analysis Center (IPAC). Processed images and  data are online available via NASA/IPAC Infrared Science Archive (IRSA).

\begin{figure*}[ht]

\begin{minipage}[b]{0.328\linewidth}
\includegraphics[width=\textwidth]{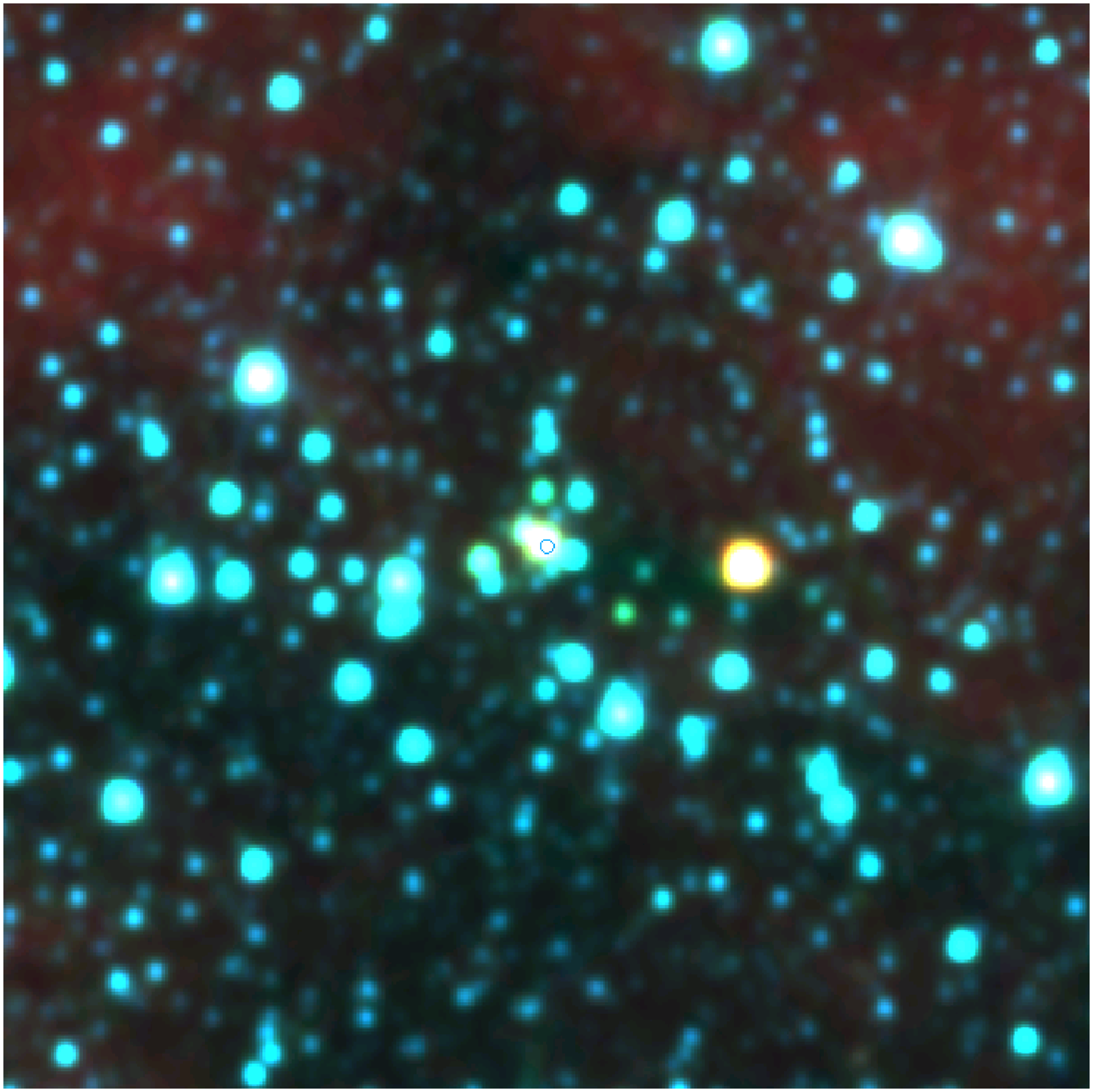}
\put(-140.0,155.0){\makebox(0.0,0.0)[5]{\fontsize{14}{14}\selectfont \color{red}C 245}}
\end{minipage}\hfill
\hspace{0.03cm}
\vspace{0.01cm}
\begin{minipage}[b]{0.328\linewidth}
\includegraphics[width=\textwidth]{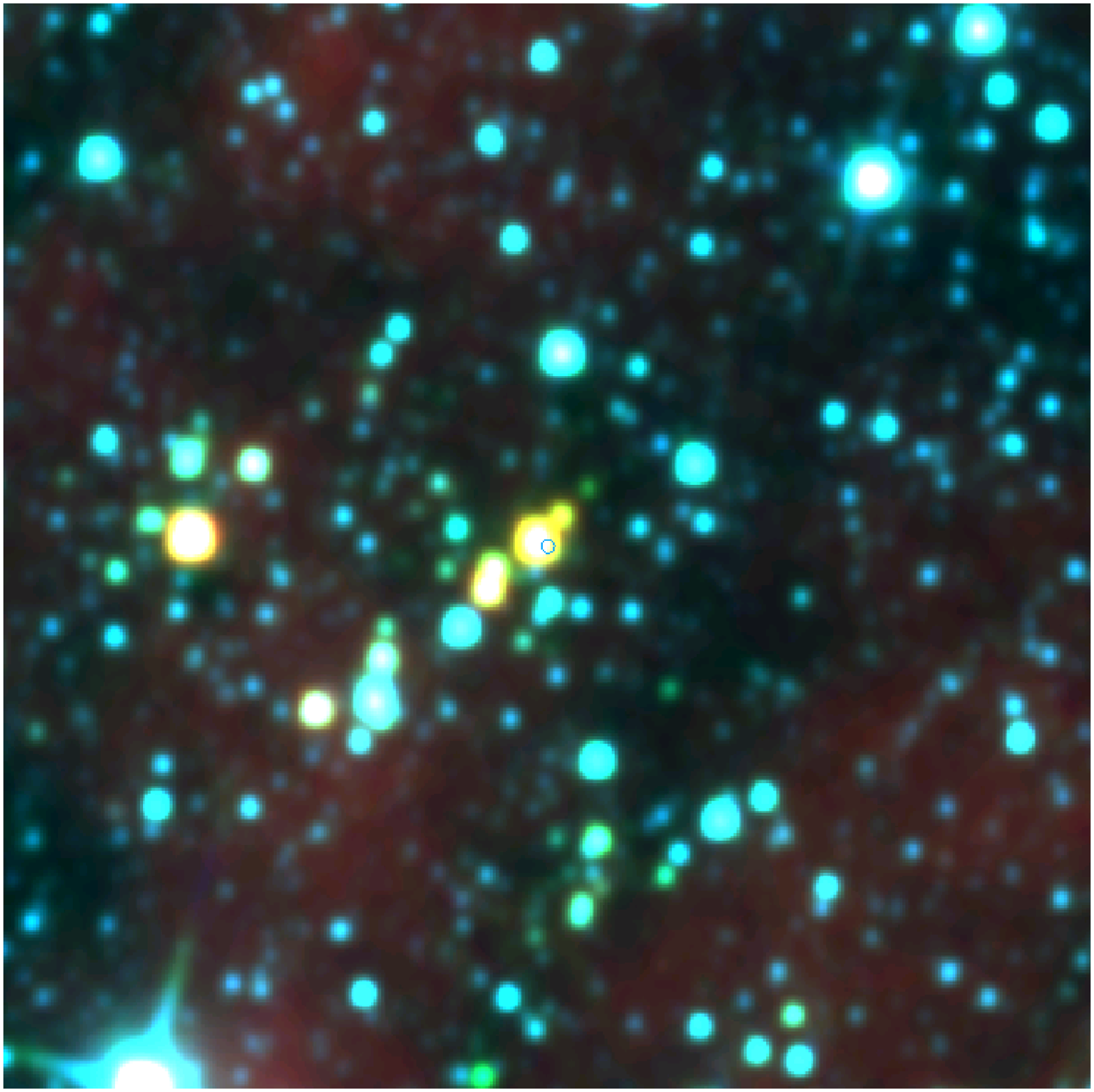}
\put(-140.0,155.0){\makebox(0.0,0.0)[5]{\fontsize{14}{14}\selectfont \color{red}C 253}}
\end{minipage}\hfill
\vspace{0.01cm}
\begin{minipage}[b]{0.328\linewidth}
\includegraphics[width=\textwidth]{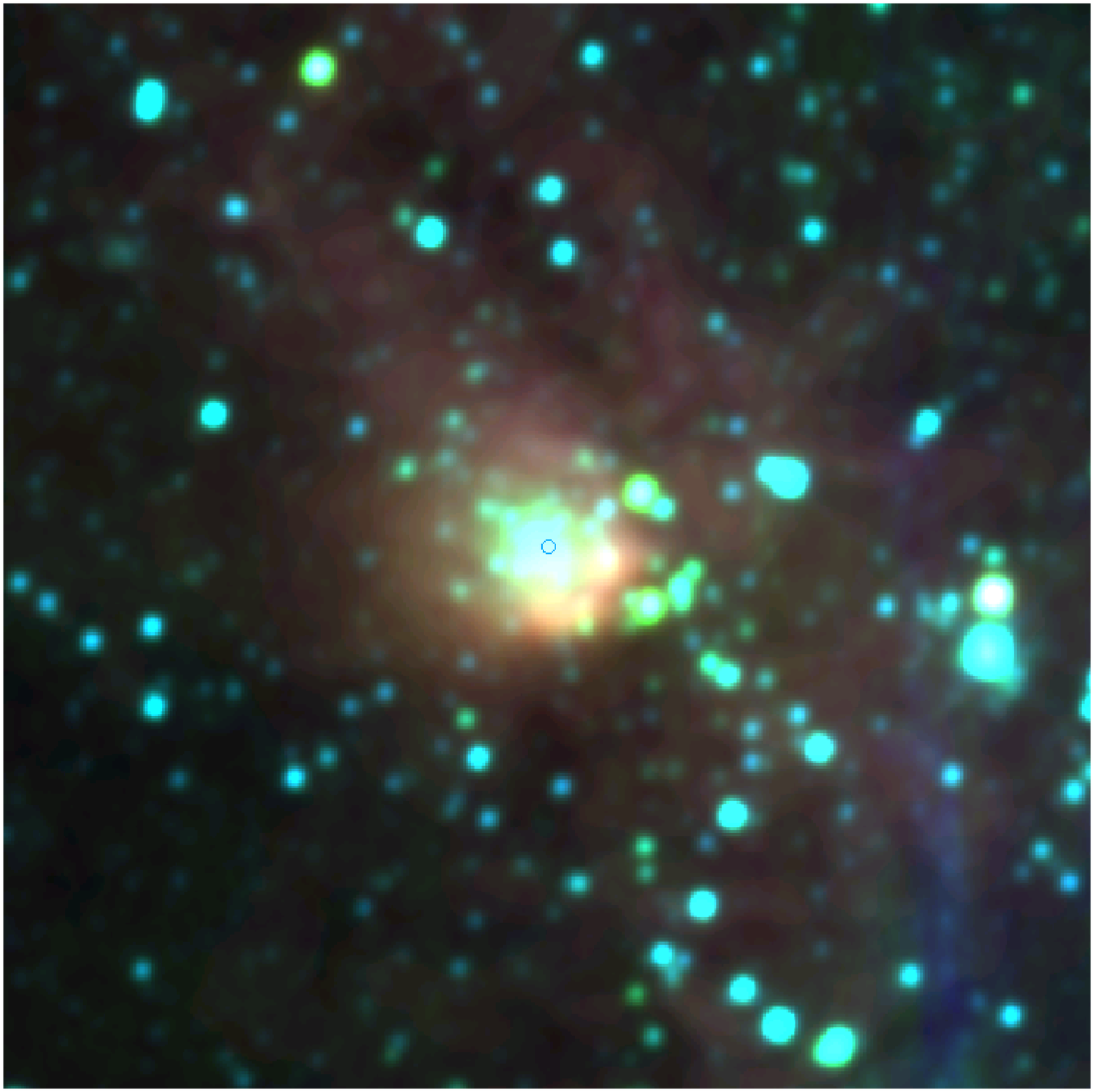}
\put(-140.0,155.0){\makebox(0.0,0.0)[5]{\fontsize{14}{14}\selectfont \color{red}C 40}}
\end{minipage}\hfill
\hspace{0.03cm}
\vspace{0.01cm}
\begin{minipage}[b]{0.328\linewidth}
\includegraphics[width=\textwidth]{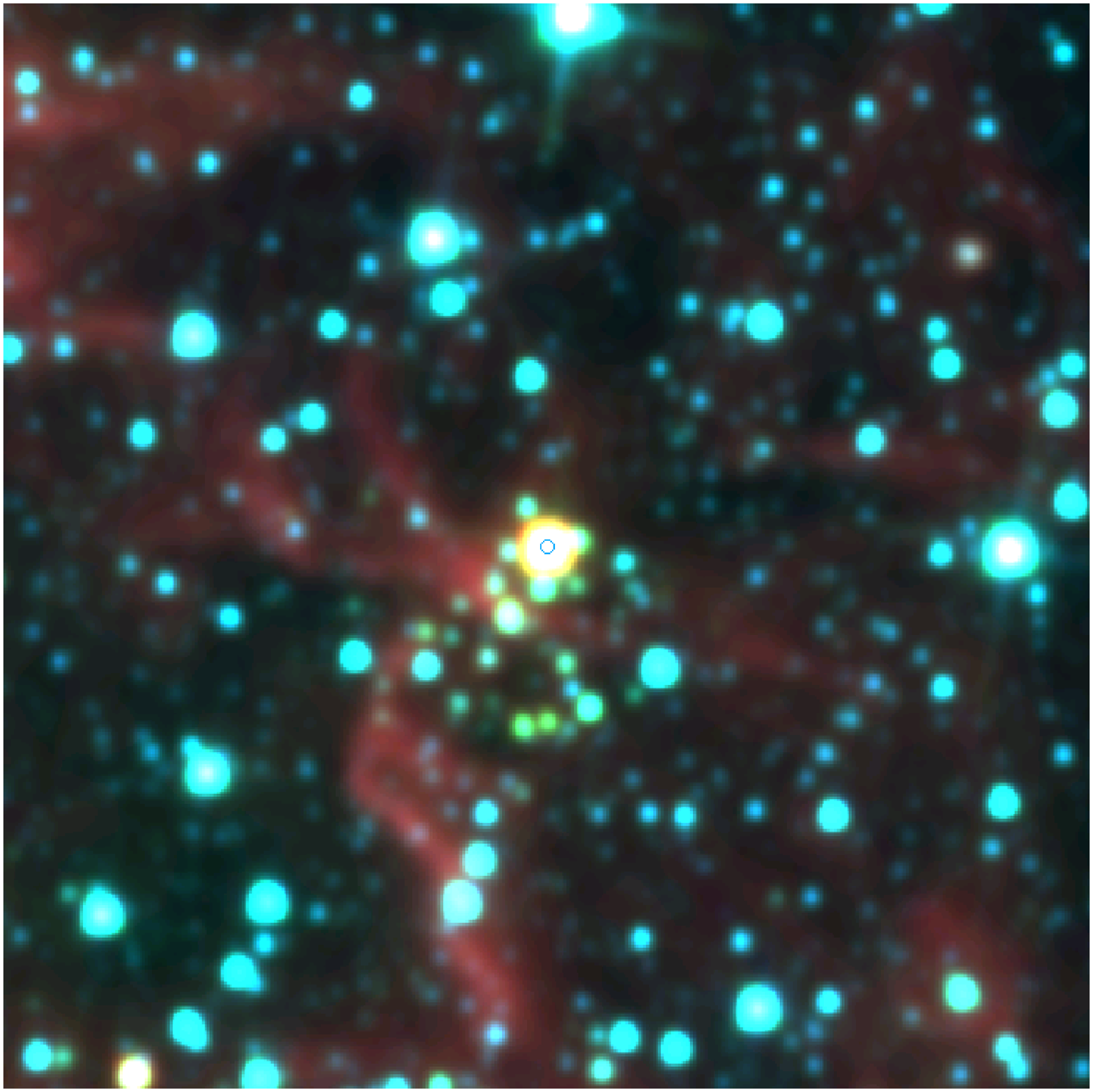}
\put(-140.0,155.0){\makebox(0.0,0.0)[5]{\fontsize{14}{14}\selectfont \color{red}C 292}}
\end{minipage}\hfill
\hspace{0.03cm}
\begin{minipage}[b]{0.328\linewidth}
\includegraphics[width=\textwidth]{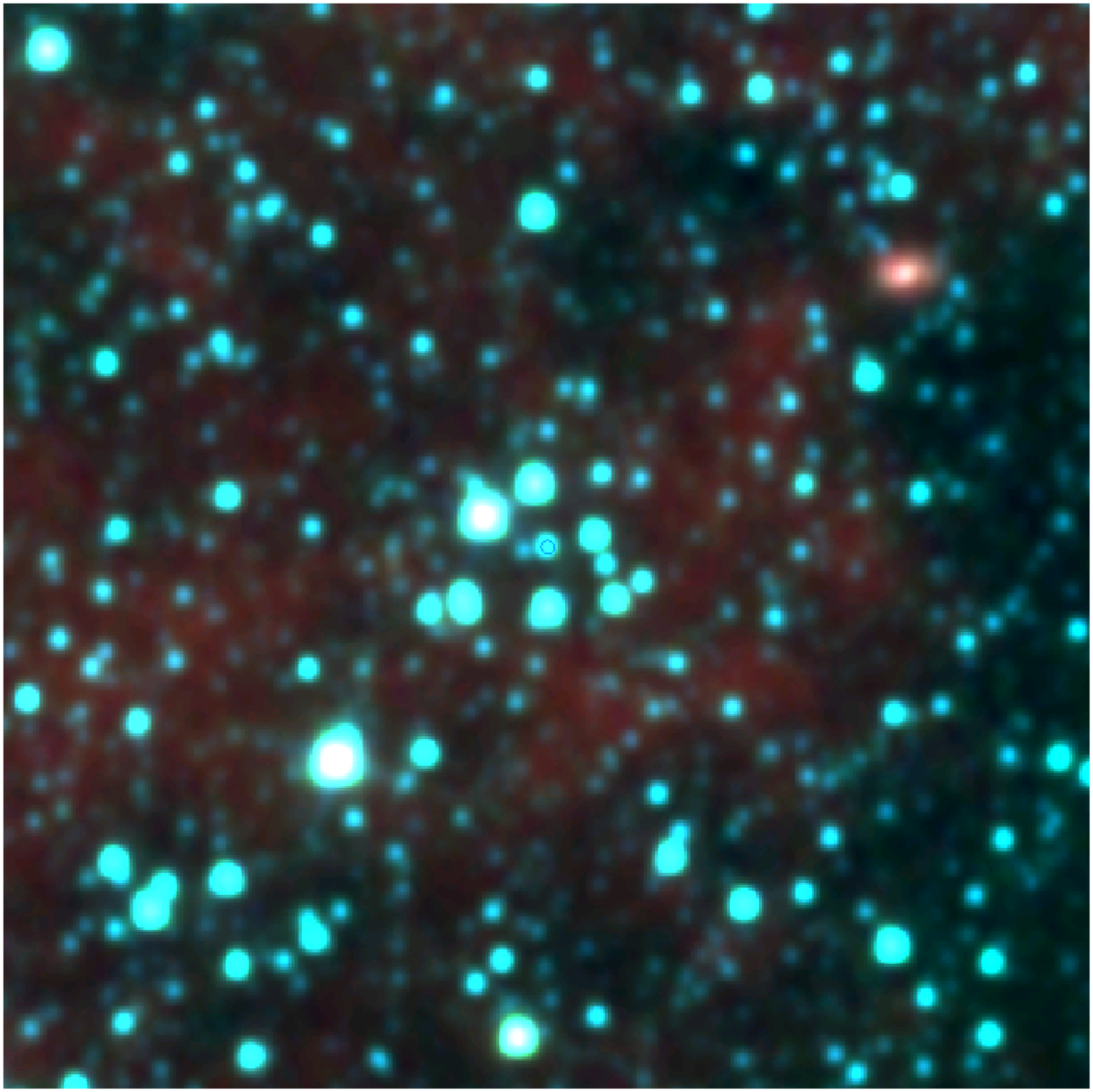}
\put(-140.0,155.0){\makebox(0.0,0.0)[5]{\fontsize{14}{14}\selectfont \color{red}C 241}}
\end{minipage}\hfill
\hspace{0.03cm}
\begin{minipage}[b]{0.328\linewidth}
\includegraphics[width=\textwidth]{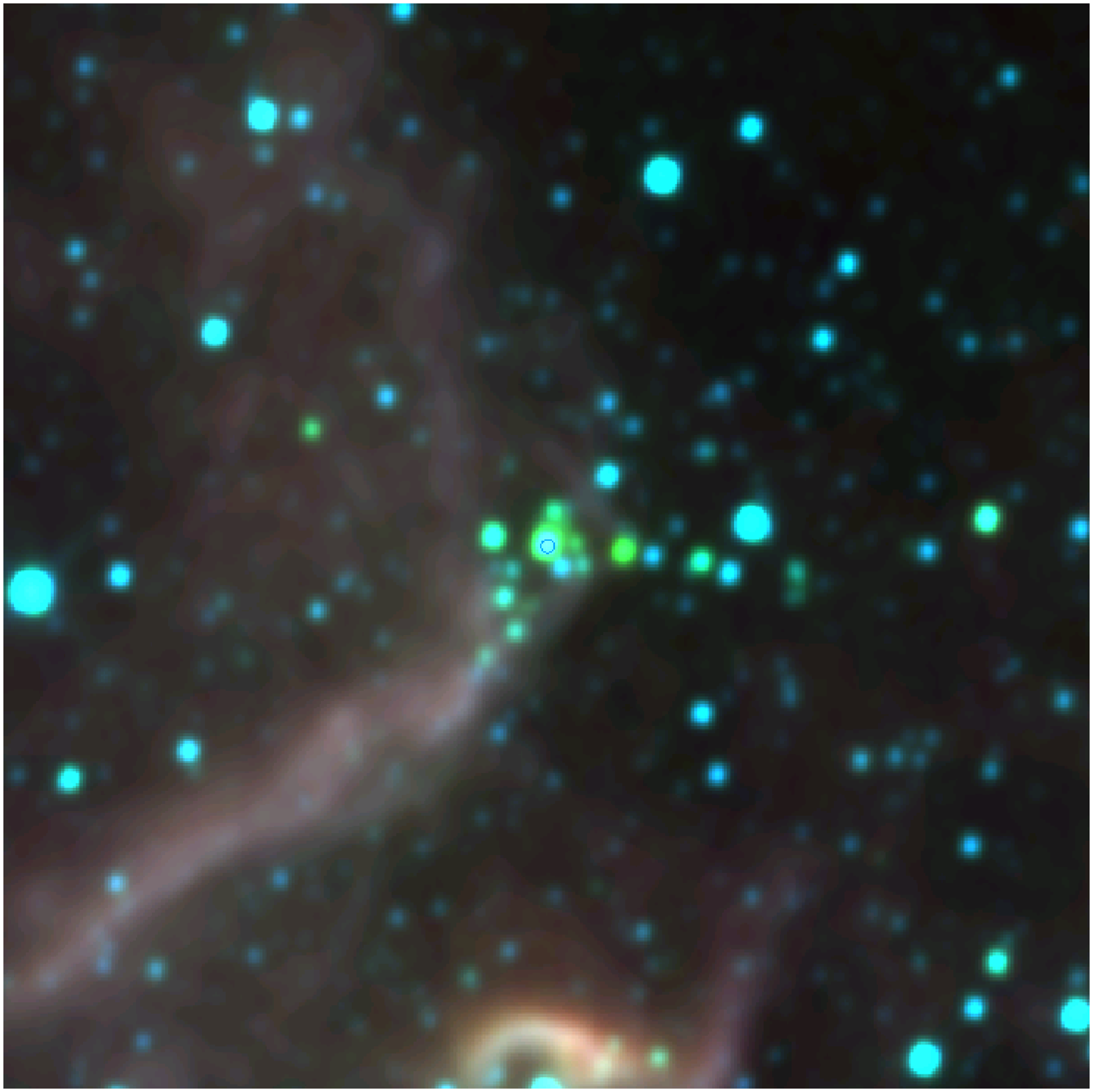}
\put(-140.0,155.0){\makebox(0.0,0.0)[5]{\fontsize{14}{14}\selectfont \color{red}C 39}}
\end{minipage}\hfill
\caption[]{RGB examples  of various cluster/dust distributions in  sample objects.
 Top panels: WISE RGB images ($10'\times10'$)  centred on C 245, C 253 and C 40. Bottom panels: the same for C 292, C 241, and C 39. C 292 is in the central region of a bubble. These objects are ECs.}
\label{f3}
\end{figure*}

\begin{figure*}[hp]
\begin{center}
   \includegraphics[scale=0.2435,angle=0,viewport=0 0 980 980,clip]{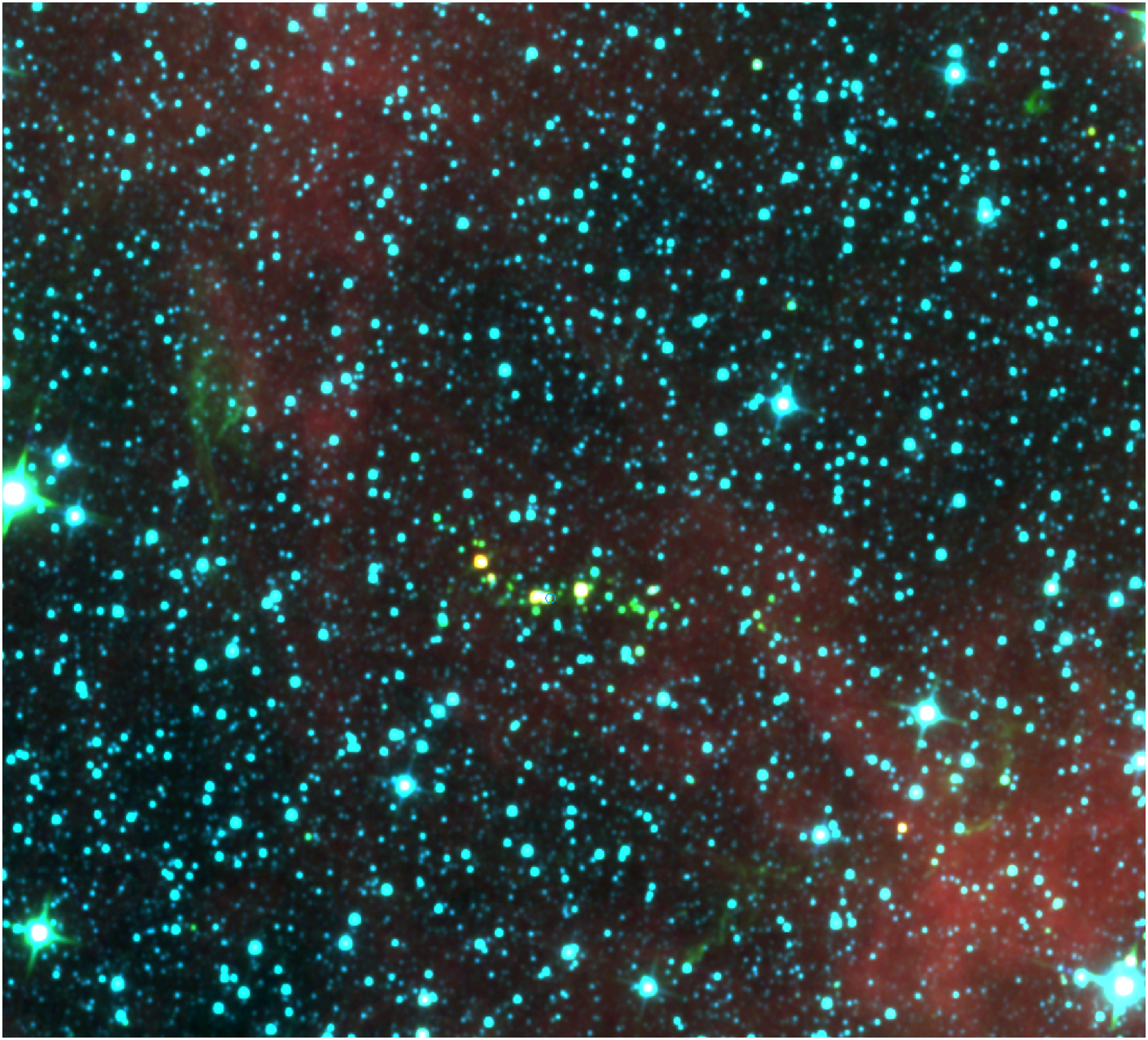}
\put(-210.0,230.0){\makebox(0.0,0.0)[5]{\fontsize{16}{16}\selectfont \color{red}{\bf C 260}}}
\hfill
   \includegraphics[scale=0.3435,angle=0,viewport=0 0 810 700,clip]{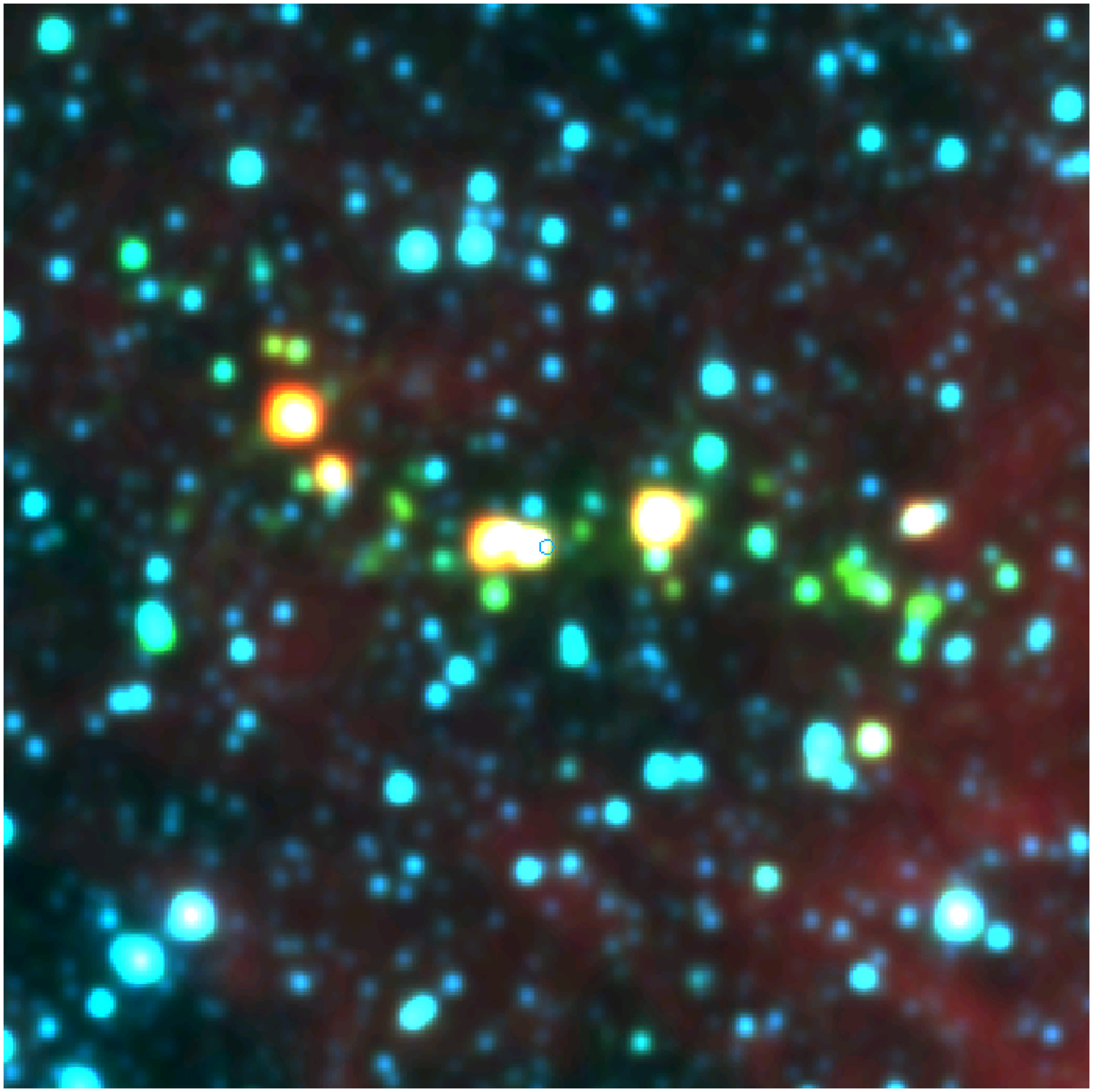}
\put(-210.0,230.0){\makebox(0.0,0.0)[5]{\fontsize{16}{16}\selectfont \color{red}{\bf C 260}}}
\hfill
   \includegraphics[scale=0.3465,angle=0,viewport=230 60 920 750,clip]{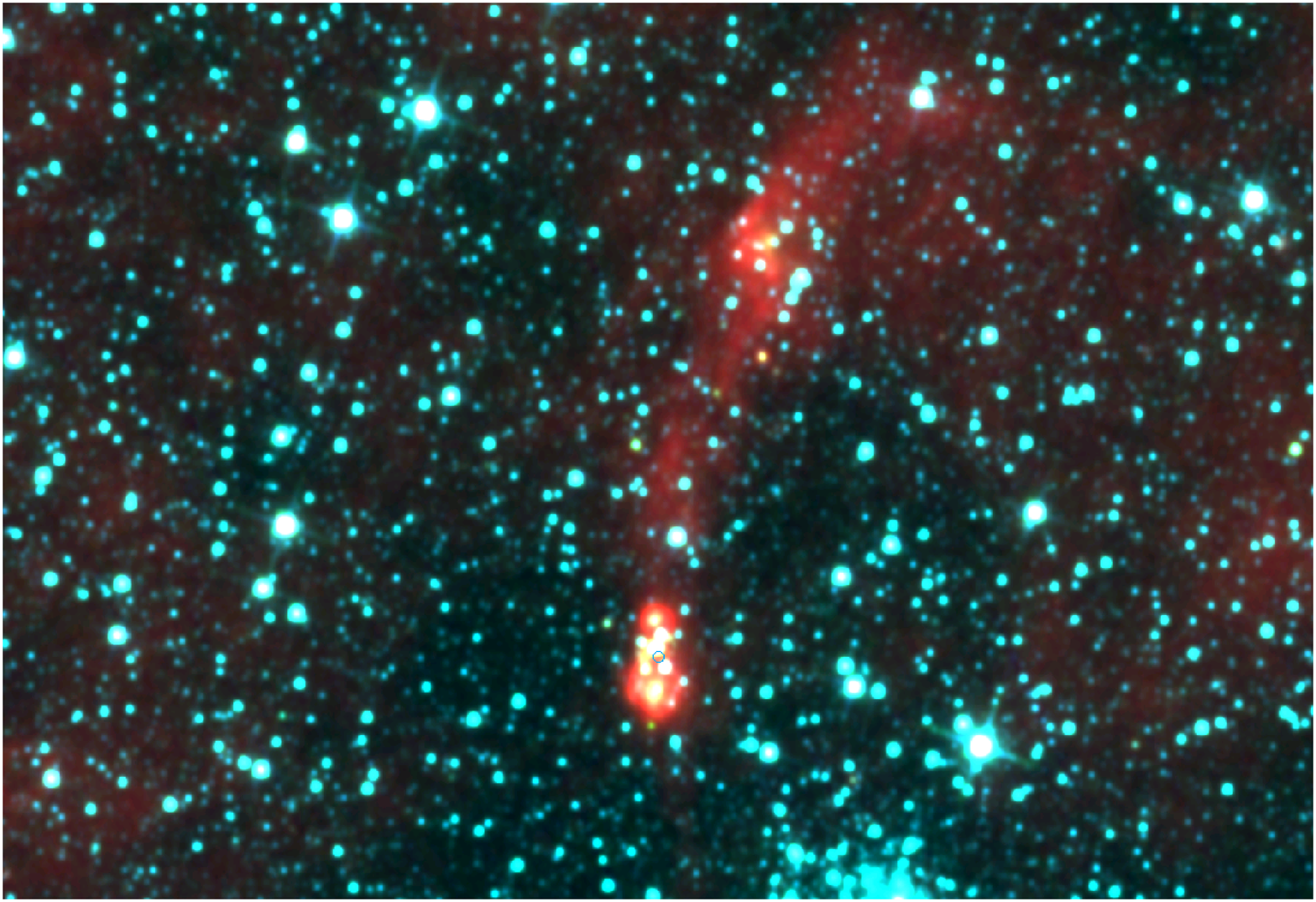}
\put(-170.0,230.0){\makebox(0.0,0.0)[5]{\fontsize{16}{16}\selectfont \color{red}{\bf C 217 and C 213}}}
\hfill
   \includegraphics[scale=0.3445,angle=0,viewport=0 0 810 700,clip]{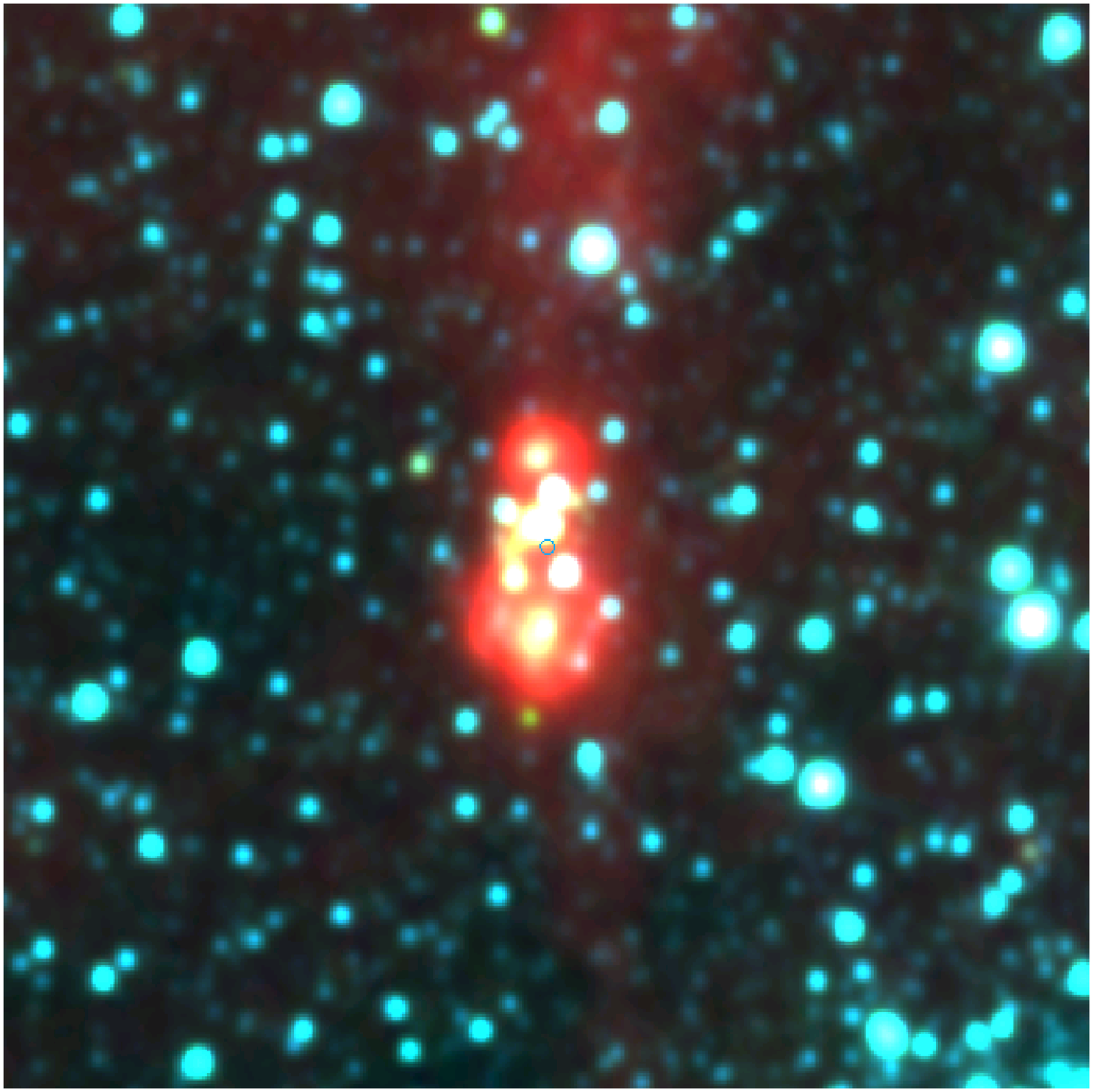}
\put(-210.0,230.0){\makebox(0.0,0.0)[5]{\fontsize{16}{16}\selectfont \color{red}{\bf C 217}}}\hfill
\end{center}
\caption[]{Cluster in filamentary dust structures. Top panels: WISE RGB images centred on \astrobj{C 260} ($40'\times40'$), and \astrobj{C 260} ($10'\times10'$). Bottom panels: the same for \astrobj{C 217} and \astrobj{C 213} ($30'\times30'$), and \astrobj{C 217} ($10'\times10'$). These images show evidence of filamentary cluster formation. \astrobj{C 260} shows unusual s-shaped structure.}

\label{g}
\end{figure*}

\begin{figure*}[hp]
\begin{minipage}[b]{0.495\linewidth}
\includegraphics[width=\textwidth]{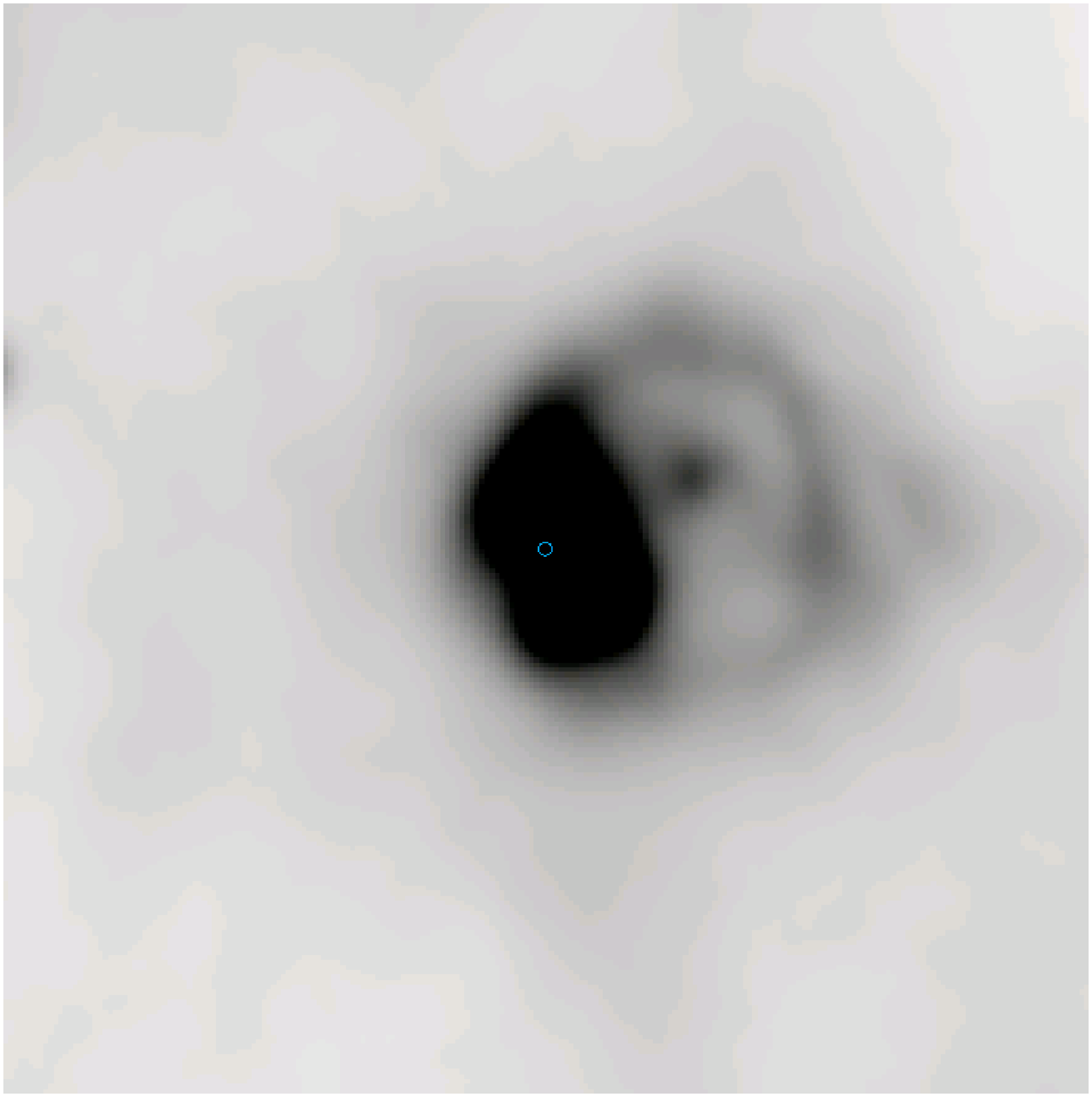}
\put(-210.0,230.0){\makebox(0.0,0.0)[5]{\fontsize{16}{16}\selectfont {\bf C 201}}}
\end{minipage}\hfill
\hspace{0.05cm}
\vspace{0.03cm}
\begin{minipage}[b]{0.4953\linewidth}
\includegraphics[width=\textwidth]{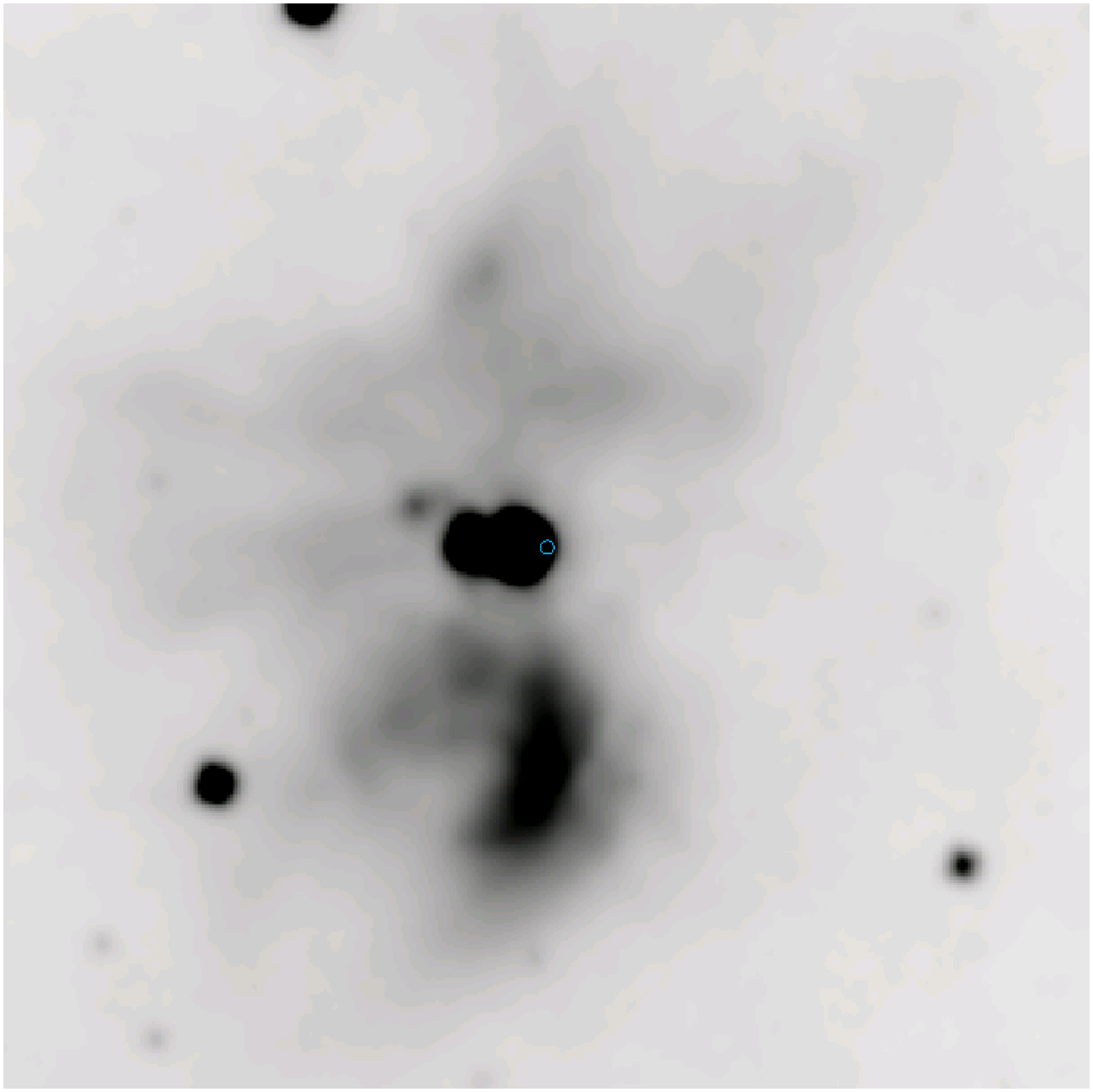}
\put(-40.0,230.0){\makebox(0.0,0.0)[5]{\fontsize{16}{16}\selectfont {\bf C 178}}}
\end{minipage}\hfill
\vspace{0.03cm}
\begin{minipage}[b]{0.495\linewidth}
\includegraphics[width=\textwidth]{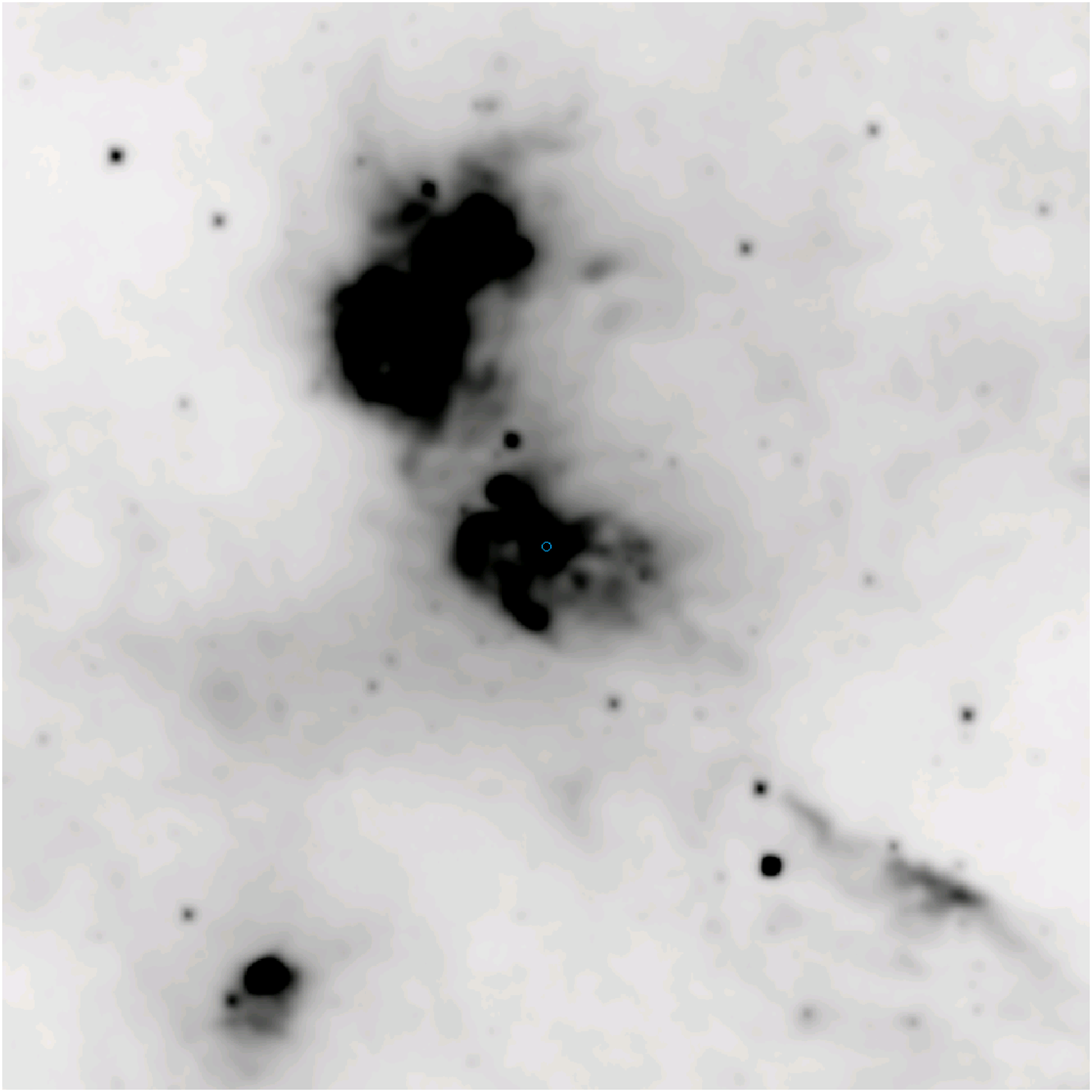}
\put(-210.0,230.0){\makebox(0.0,0.0)[5]{\fontsize{16}{16}\selectfont {\bf C 219}}}
\end{minipage}\hfill
\hspace{0.05cm}
\begin{minipage}[b]{0.495\linewidth}
\includegraphics[width=\textwidth]{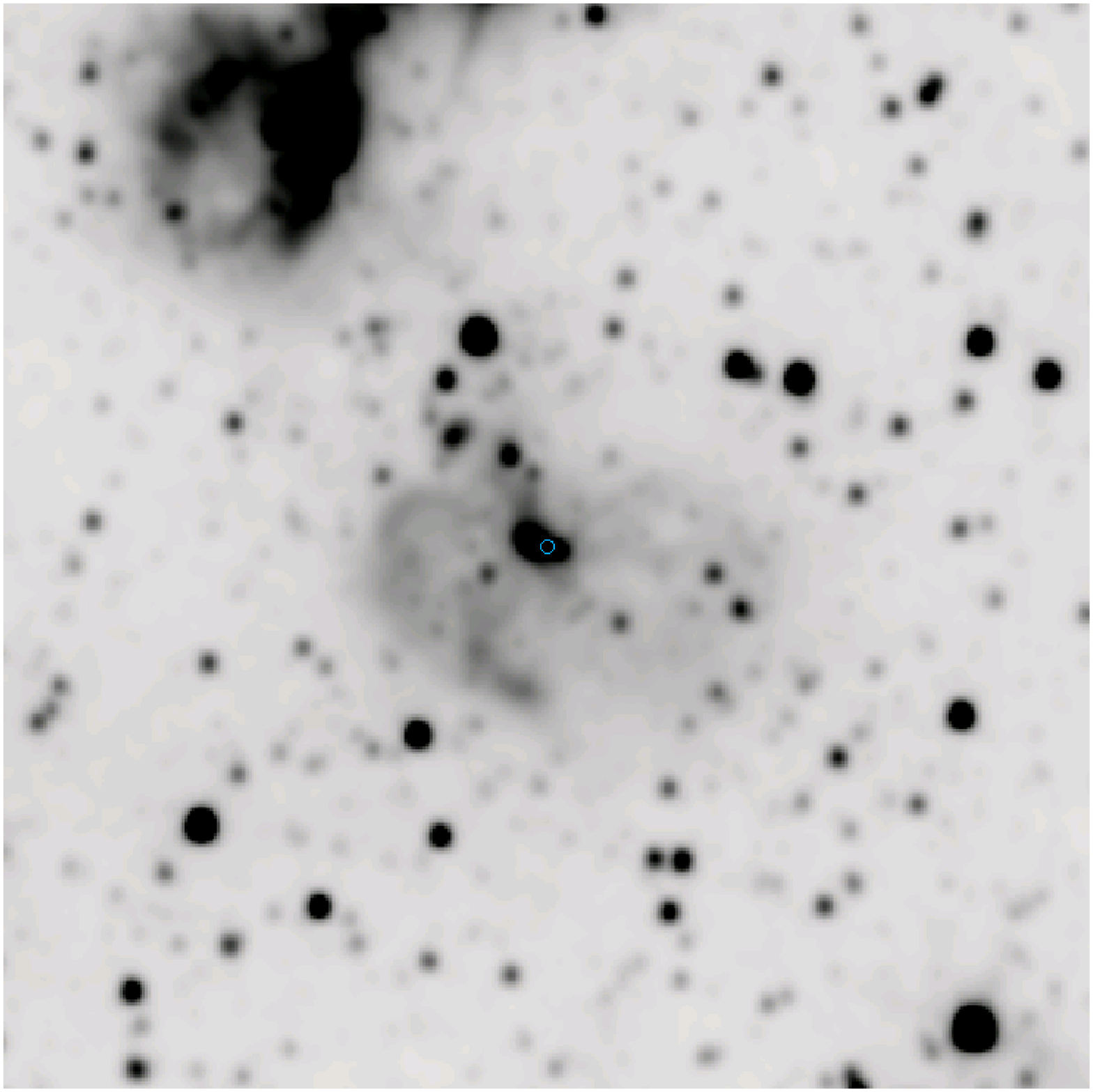}
\put(-40.0,230.0){\makebox(0.0,0.0)[5]{\fontsize{16}{16}\selectfont {\bf C 219}}}
\end{minipage}\hfill
\caption[]{Examples of objects with remarkable structure  in the sample. Top panels: W3 WISE images centred on C 201 ($5'\times5'$), and C 178 ($10'\times10'$). Bottom panels: the W3 WISE image for C 219 ($20'\times20'$) and W2 WISE image for C 219 ($10'\times10'$).}
\label{f4}
\end{figure*}

\begin{figure*}[!htp]
\begin{minipage}[b]{0.328\linewidth}
\includegraphics[width=\textwidth]{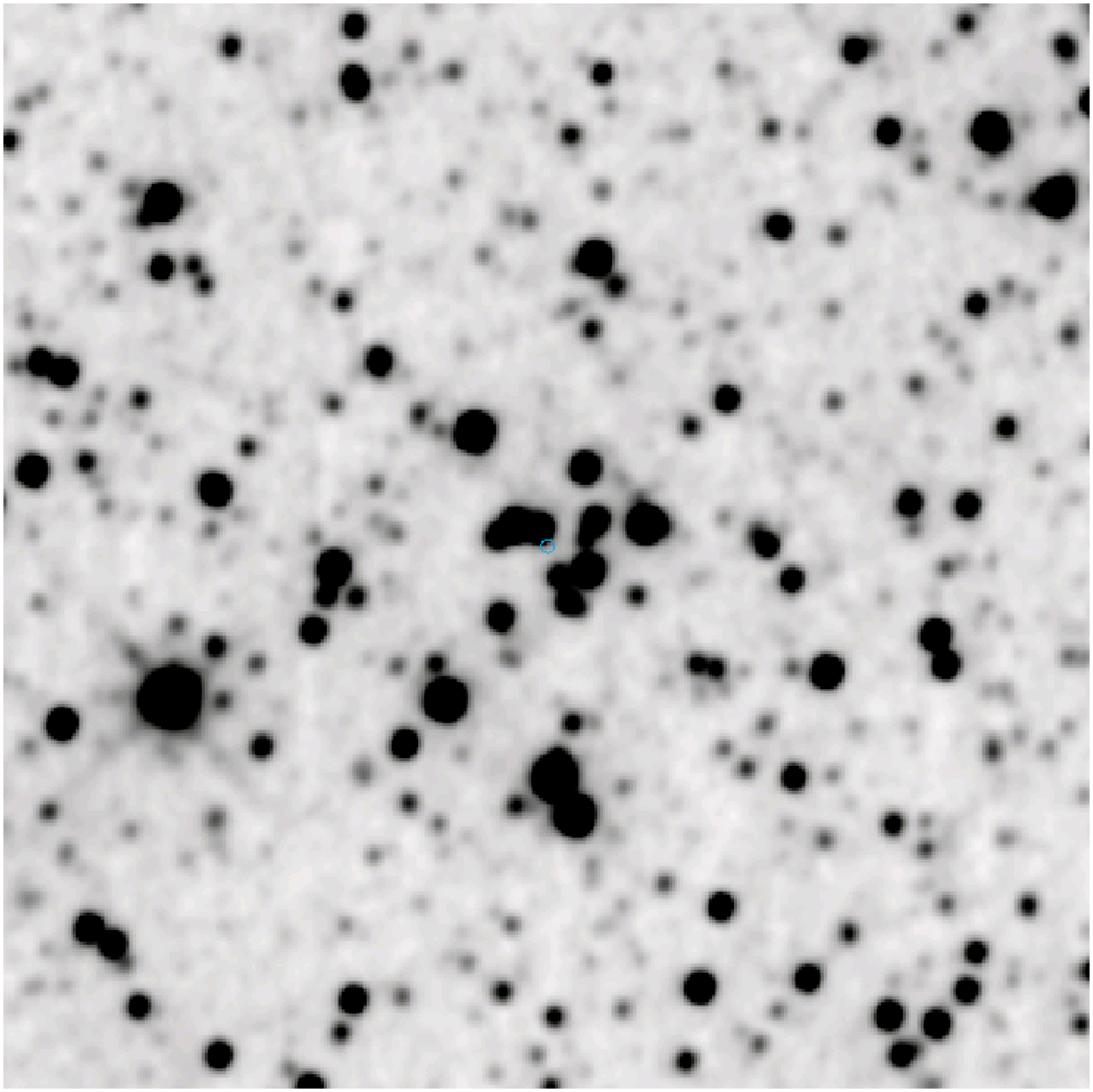}
\put(-135.0,155.0){\makebox(0.0,0.0)[5]{\fontsize{14}{14}\selectfont {\bf C 14}}}
\end{minipage}\hfill
\hspace{0.03cm}
\vspace{0.03cm}
\begin{minipage}[b]{0.328\linewidth}
\includegraphics[width=\textwidth]{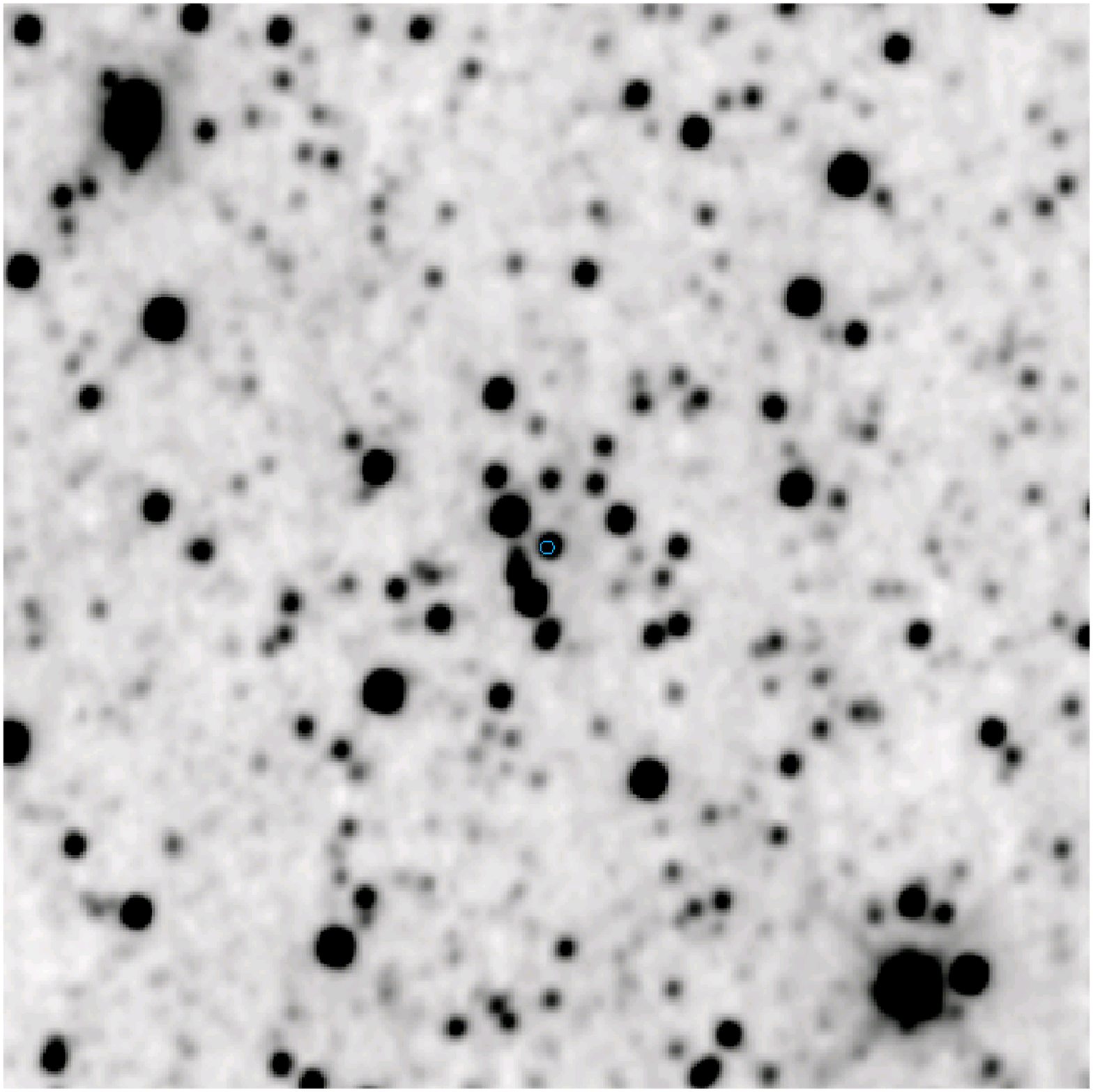}
\put(-135.0,155.0){\makebox(0.0,0.0)[5]{\fontsize{14}{14}\selectfont {\bf C 74}}}
\end{minipage}\hfill
\hspace{0.03cm}
\vspace{0.03cm}
\begin{minipage}[b]{0.328\linewidth}
\includegraphics[width=\textwidth]{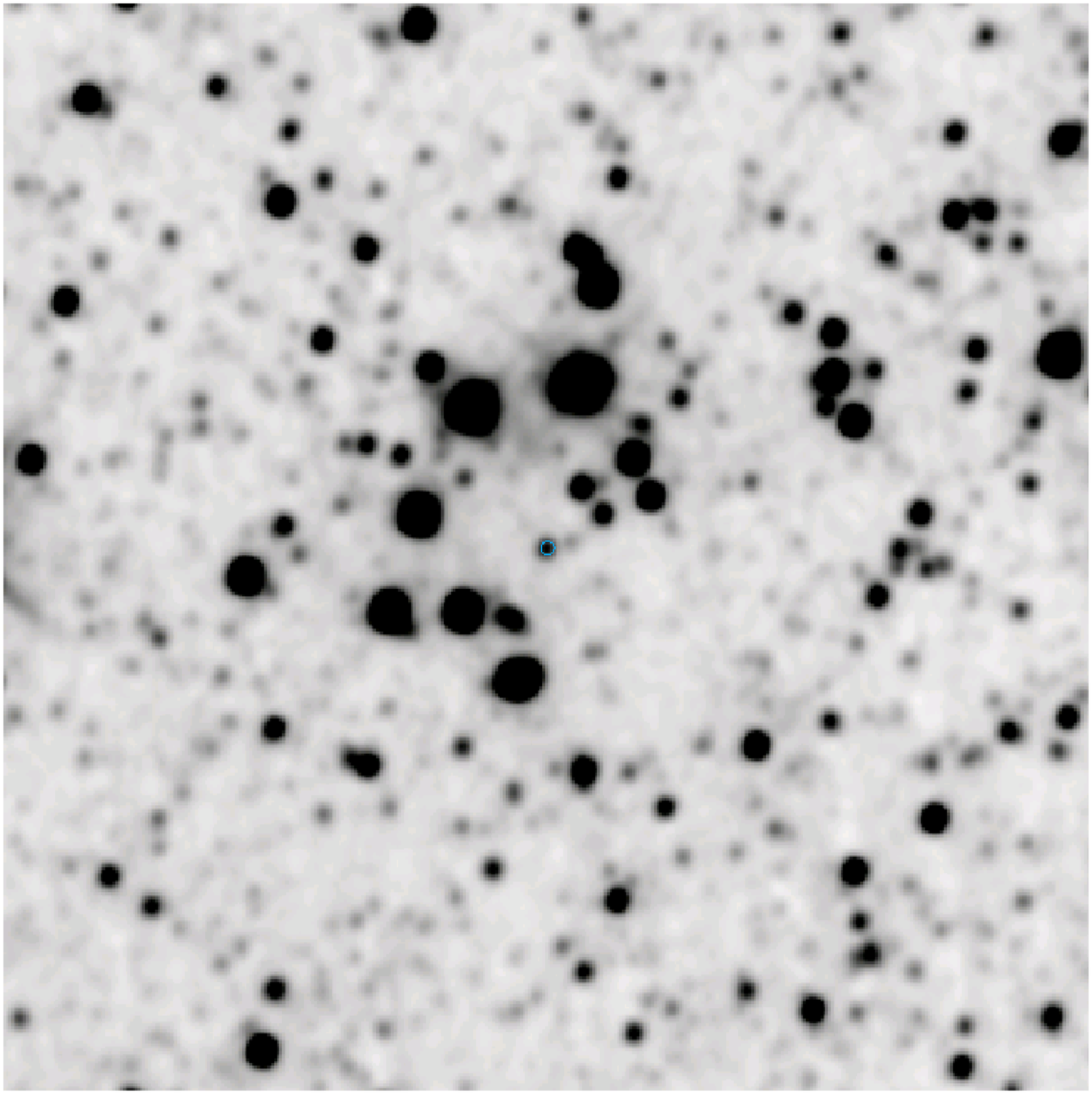}
\put(-135.0,155.0){\makebox(0.0,0.0)[5]{\fontsize{14}{14}\selectfont {\bf C 133}}}
\end{minipage}\hfill
\vspace{0.03cm}
\begin{minipage}[b]{0.328\linewidth}
\includegraphics[width=\textwidth]{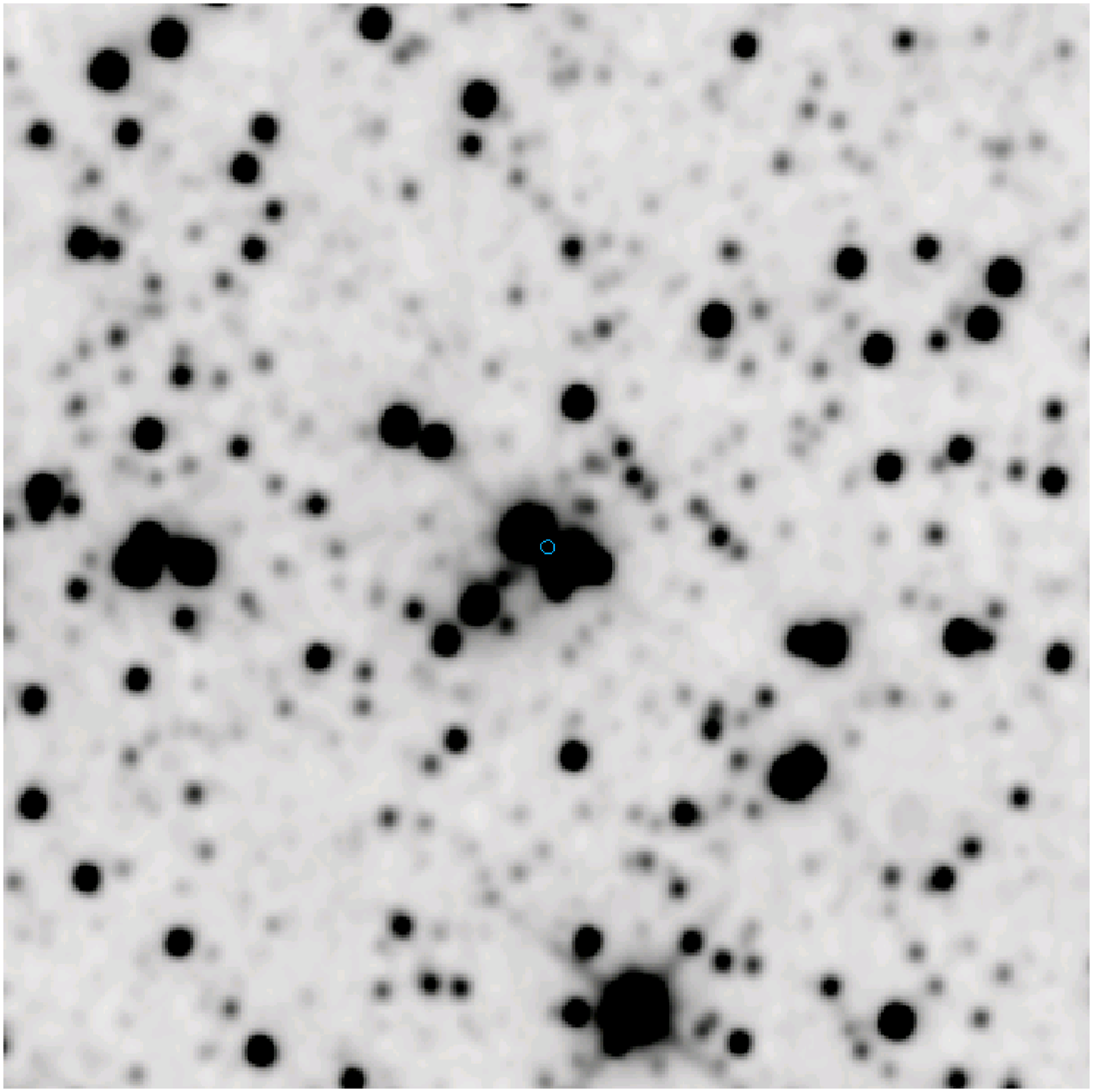}
\put(-135.0,155.0){\makebox(0.0,0.0)[5]{\fontsize{14}{14}\selectfont {\bf C 209}}}
\end{minipage}\hfill
\hspace{0.03cm}
\begin{minipage}[b]{0.328\linewidth}
\includegraphics[width=\textwidth]{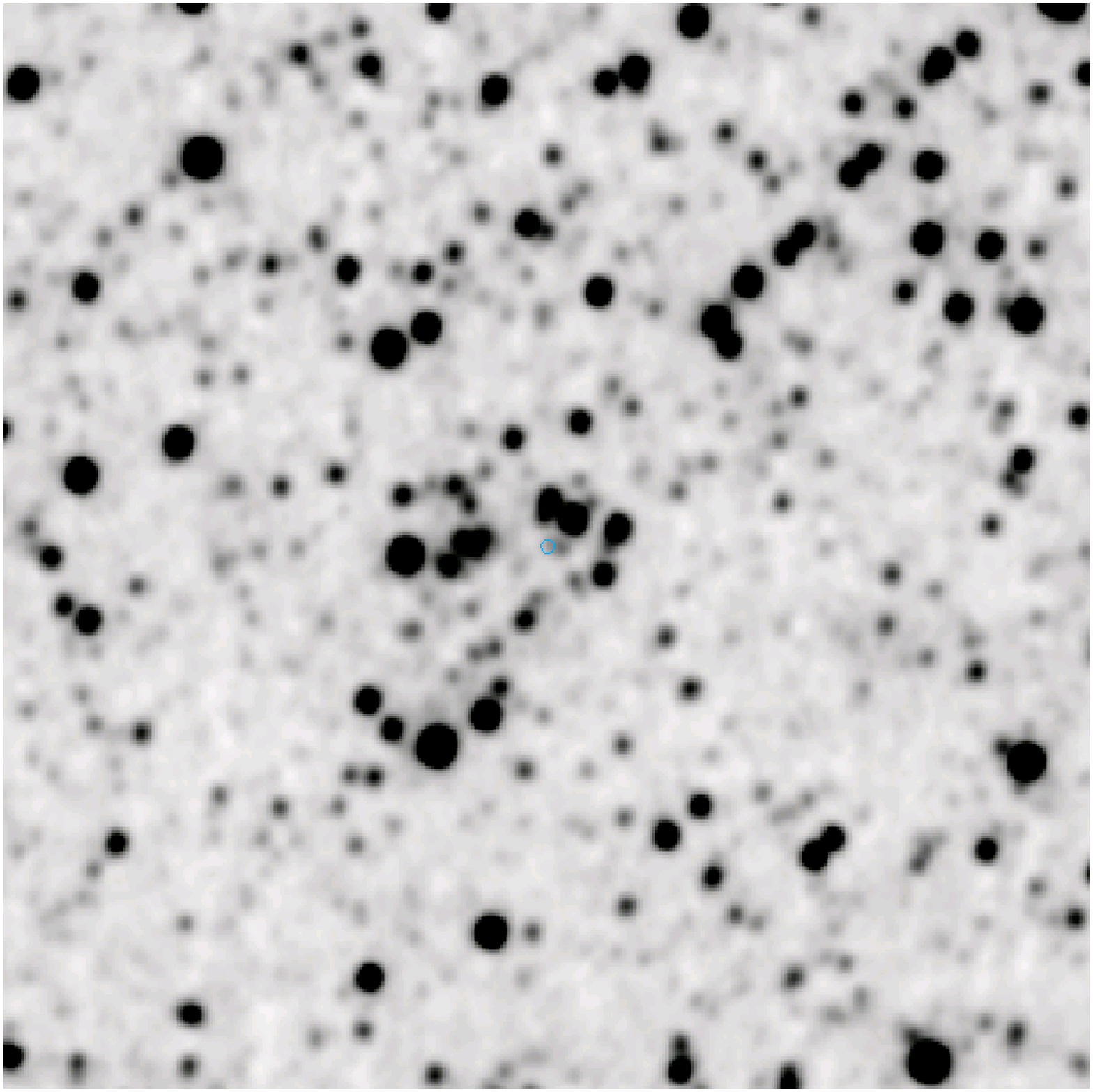}
\put(-135.0,155.0){\makebox(0.0,0.0)[5]{\fontsize{14}{14}\selectfont {\bf C 18}}}
\end{minipage}\hfill
\hspace{0.03cm}
\begin{minipage}[b]{0.328\linewidth}
\includegraphics[width=\textwidth]{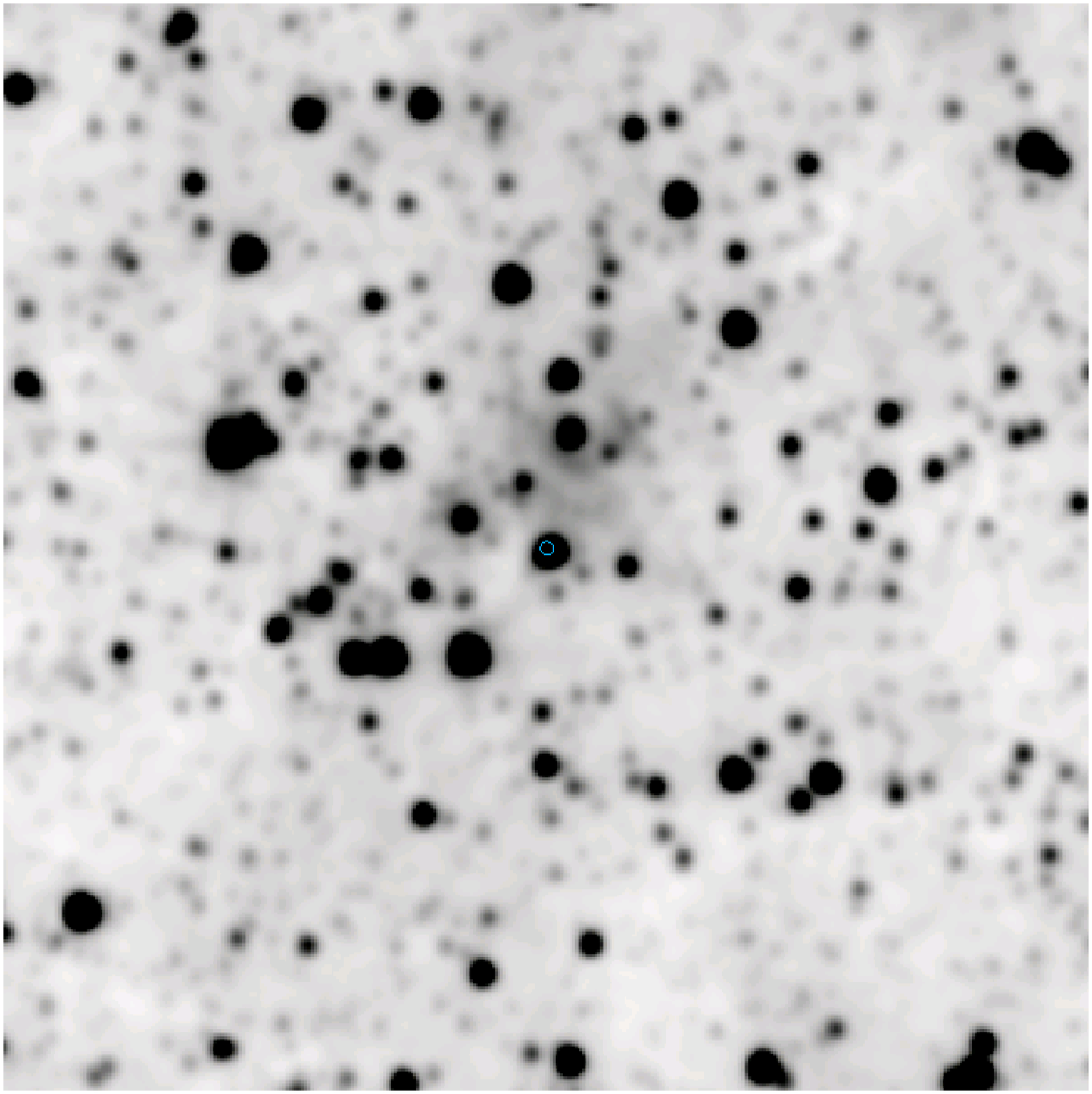}
\put(-135.0,155.0){\makebox(0.0,0.0)[5]{\fontsize{14}{14}\selectfont {\bf C 429}}}
\end{minipage}\hfill
\caption[]{Examples of clusters in the test sub-sample.
 Top panels: W2 images ($10'\times10'$)  centred on C 14, C 74, and C 133. Bottom panels: the same for C 209, C 18, and C 429.}
\label{f5}
\end{figure*}

WISE provides very efficient detections of star forming clusters \citep{Majaess13}.  We searched for new clusters, stellar groups and candidates in dust nebulae in the 3rd Quadrant and extensions to the 2nd and 4th Quadrants (Table~\ref{tab1}). WISE images for a representative sample of the newly found clusters are shown in Figs.~\ref{f2} to \ref{f5}. 

Recently, \citet{Majaess13} employed WISE in a search for very young clusters.
His method differs from the present one in the sense that  he employs a hybrid $JHK_s-W_1W_2W_3W_4$ photometry to identify concentrations of young stellar objects (YSOs). Our search focuses on the stellar density and gas/dust distribution to identify star clusters, stellar groups, and candidates. A similar early procedure was used by \citet{Hodapp94} who analysed 164 nebulae with molecular outflow sources, 54 of them resulting associated with embedded clusters. Also similarly, \citet{Bica03a} and \citet{Dutra03}, in a survey around the Galactic disk, provided 346 infrared clusters, stellar groups, and candidates.

\begin{table}[p]
\centering

\caption{The new list of star clusters or candidates.}
{\tiny
\label{tab1}
\renewcommand{\tabcolsep}{1.8mm}
\renewcommand{\arraystretch}{1.5}
\begin{tabular}{lrrrrrrrr}
\hline
\hline
Target &$\ell$&$b$&$\alpha(2000)$&$\delta(2000)$&Size&Type&Avend.\\
&$(^{\circ})$&$(^{\circ})$&(h\,m\,s)&$(^{\circ}\,^{\prime}\,^{\prime\prime})$&$(\,^{\prime}\,)$&& \\
($1$)&($2$)&($3$)&($4$)&($5$)&($6$)&($7$)&($8$)\\
\hline
\astrobj{C 1} &147.52&10.03&4:44:32&61:12:03&$8\times8$&EC&n\\
\astrobj{C 2} &147.88&9.46&4:42:50&60:33:46&$9\times9$&EC&n\\
\astrobj{C 3} &147.99&9.79&4:45:28&60:41:25&$10\times10$&EC&n\\
\astrobj{C 4} &149.83&-24.50&2:47:55&32:13:07&$8\times8$&OCC&n\\
\astrobj{C 5} &151.83&15.69&5:42:49&60:51:31&$7\times7$&EC&n\\
\astrobj{C 6} &151.84&16.06&5:45:34&61:01:36&$8\times8$&EC&n\\
\astrobj{C 7} &151.96&16.11&5:46:19&60:56:45&$12\times12$&ECC&n\\
\astrobj{C 8} &151.97&15.52&5:42:05&60:39:31&$5\times5$&EC&n\\
\astrobj{C 9} &152.12&16.14&5:47:08&60:49:12&$6\times6$&EGr&n\\
\astrobj{C 10} &152.30&15.76&5:45:02&60:29:36&$6\times6$&EGr&n\\
\astrobj{C 11} &160.11&-18.43&3:41:22&31:54:37&$4\times4$&EC&n\\
\astrobj{C 12} &160.28&-18.42&3:41:58&31:48:45&$6\times6$&EC&n\\
\astrobj{C 13} &164.15&0.56&5:03:50&42:22:15&$5\times5$&EC&n\\
\astrobj{C 14} &169.52&-1.79&5:10:50&36:39:45&$6\times6$&EC&n\\
\astrobj{C 15} &170.13&-1.17&5:15:04&36:31:54&$7\times7$&EGr&n\\
\astrobj{C 16} &171.17&0.45&5:24:41&36:36:16&$5\times5$&EC&y\\
\astrobj{C 17} &171.43&-0.01&5:23:30&36:07:49&$7\times7$&EC&n\\
\astrobj{C 18} &171.77&1.45&5:30:30&36:39:53&$7\times7$&EC&n\\
\astrobj{C 19} &172.08&-2.26&5:16:18&34:18:54&$4\times4$&EGr&y\\
\astrobj{C 20} &172.42&2.44&5:36:23&36:39:41&$5\times5$&EC&n\\
\hline
\end{tabular}
\vspace{-0.2cm}
\begin{list}{Table Notes.}
\item We suggest the designation Camargo for the objects, since one of us (D.C.) discovered them. Cols. $2-5$: Central coordinates. Col. $6$: cluster size. Col. $7$: object type - OC means open cluster, OCC open cluster candidate, EC embedded cluster, ECC embedded cluster candidate, and EGr stellar group. Col. $8$: cross-identifications with \citet{Avedisova02} SFRs or candidates within 5' of the central coordinates, y and n means yes and not. The full table is available online.
\end{list}
}
\end{table}

The object search started with inspection of $2^{\circ}\times2^{\circ}$ WISE images near the Galactic plane.  We used the WISE standard RGB colour images primarily looking for dust emission nebulae at different temperatures in the  wavelength bands W1, W2, W3, and W4. Subsequently, we were guided by stellar concentrations. 

The survey provided 437 new star clusters, stellar groups, and candidates.
The discovered objects were classified in five classes: open clusters (OCs), open cluster candidates (OCCs), embedded clusters (ECs), embedded cluster candidates (ECCs), and embedded stellar groups (EGrs). The classification follows a stellar/dust density estimation on the W1, W2, W3, W4, and RGB WISE images. Populous dust-free objects are classified as OCs and the OCCs are in general poor or unresolved objects that require deeper photometry or higher resolution. Prominent objects that are still cradled in their natal molecular clouds are ECs and the less populous candidates are ECCs. Star-forming low density groups are EGrs. Fig.~\ref{f2} shows a representative sample of ECs and Fig.~\ref{nova} shows samples of OC, OCC, ECC, and EGr. 

We verified  that  the present objects do not have previous identifications in the literature. The comparison catalogue that we used is essentially complete for all published  star clusters in the Galaxy and was made by one of us, E.B., and will be presented in a forthcoming paper. Additionally, we used e.g. SIMBAD, the catalogue of infrared star clusters, stellar groups and candidates by \citet{Bica03a}, the  cluster and candidate
catalogue by \citet{Froebrich07}, the recent protocluster results  in WISE by \citet{Majaess13}, and many others. We also cross correlated our cluster dust emission nebulae with the planetary nebulae catalogue of \citet{Acker92}. This way we could also classify some clump candidates as probable star-forming nebulae (SFNs) in the present list Table \ref{tab1}.

During our WISE search for stellar clusters and candidates, we came across
with 49 objects from the literature (Table \ref{cross}). They belong  mostly
to our EC classification, except the OC NGC 2301, and the OCCs FSR 1144 and FSR 1149.
The FSR object recognition  as ECs was derived by us from the  WISE images.
This cross-identification effort is important to cross relate classifications
in other lists, catalogues and individual papers.

\citet{Avedisova02} presented  3300 star forming regions (SFRs), of which about 500 are candidates. She complemented this with available photometric data and presented information on diffuse and reflecting nebulae, dark and molecular clouds. Her data do not refer directly to star clusters. It is fundamental to cross-identify her results with ours found in the WISE bands, in view of establishing or not the presence of stellar clusters. These comparisons might help answer  questions such as \textit{i}) are   very young clusters systematically found  outside or inside Avendisova's SFRs? \textit{ii}) Clues about  the threshold   of cluster/individual star formation?  The cross-identifications within 5' of the central coordinates
provided  95 coincidences between \citet{Avedisova02}
and the present list of 437 entries. This amounts to $21.7\%$ of our sample.
We indicate in Table~\ref{tab1} the entries that coincided with Avendisova ones.

\begin{table*}[p]
\centering

\caption{Objects from the literature detected during our WISE search.}
{\footnotesize
\label{cross}
\renewcommand{\tabcolsep}{2.7mm}
\renewcommand{\arraystretch}{1.2}
\begin{tabular}{lrrrrrrrr}
\hline
\hline
$\,\ell$&$b\,\,$&$\alpha(2000)$&$\delta(2000)$&Diameter&Type&Designation&Reference\\
$(^{\circ})$&$(^{\circ})$&(h\,m\,s)&$(^{\circ}\,^{\prime}\,^{\prime\prime})$&$(\,^{\prime}\,)$&& \\

\hline
160.48& -17.81&  3:44:33&  32:09:19&   10&    EC&     IC 348        &\citet{Lada03}\\
169.93&  -0.60&  5:16:49&  37:01:16&    7&    EC&     Majaess 54    &\citet{Majaess13}\\
170.66&  -0.27&  5:20:16&  36:37:21&    6&    EC&     KKC 6         &\citet{Kumar06}\\
170.72&  -0.11&  5:21:07&  36:39:45&    7&    EC&     Majaess 56    &\citet{Majaess13}\\
170.82&   0.01&  5:21:53&  36:38:52&    5&    EC&     Majaess 57    &\citet{Majaess13}\\
171.87&   0.45&  5:26:37&  36:01:46&    8&    EC&     Czernik 21    &\citet{Maciejewski07}\\
173.08&  -3.46&  5:14:22&  32:48:13&    7&    EC&     Dolidze 16    &\citet{Camargo10}\\
195.62&  -2.91&  6:08:32&  13:58:09&    8&    EC&     NGC 2169      &\citet{Krone10}\\
211.88&  -1.32&  6:44:49&   0:20:18&    6&    EC&     PHN 2         &\citet{Puga09}\\
212.27&  -1.08&  6:46:22&   0:04:46&    7&    EC&     PHN 3         &\citet{Puga09}\\
212.54&   0.28&  6:51:46&   0:28:21&   11&    OC&     NGC 2301      &\citet{Sharma06}\\
214.07& -19.65&  5:42:49&  -9:47:22&   10&    EC&     Majaess 64    &\citet{Majaess13}\\
216.84&   0.81&  7:01:27&  -3:06:41&  4.5&    EC&     de Wit 2      &\citet{Wit04}\\
217.49&  -0.02&  6:59:43&  -4:04:00&    3&    EC&     Ivanov 9      &\citet{Ivanov02}\\
217.63&  -0.18&  6:59:24&  -4:15:54&    4&    EC&     BDS 89        &\citet{Bica03b}\\
218.15&  -0.64&  6:58:41&  -4:56:13&    5&    EC&     FSR 1129      &\citet{Froebrich07}\\
219.47& -10.57& 6:25:14& -10:33:32 &   5 &   EC &    BDS 91        &\citet{Bica03b}\\
219.49&  -1.72&  6:57:15&  -6:37:12&   10&    OCC&    FSR 1144      &\citet{Froebrich07}\\
220.07&  -0.11&  7:04:06&  -6:24:02&   10&    OCC&    FSR 1149      &\citet{Froebrich07}\\
220.79&  -1.71&  6:59:42&  -7:46:29&    5&    EC &    NGC 2316      &\citet{Hodapp94}\\
224.01&  -1.81&  7:05:16& -10:40:10&    5&    EC &    FSR 1178      &\citet{Froebrich07}\\
224.23&  -1.47&  7:06:56& -10:43:45&    7&    EC &    Collinder 466 &\citet{Bukowiecki11}\\
224.44&  -2.36&  7:04:08& -11:18:58&    4&    EC &    Soares-Bica 1 &\citet{Soares02}\\
224.40&  -2.73&  7:02:42& -11:27:03&    5&    EC &    FSR 1184      &\citet{Froebrich07}\\
224.50&  -2.41&  7:04:03& -11:23:40&    4&    EC &    BRC 27        &\citet{Soares02}\\
224.79&  -1.73&  7:07:03& -11:20:16&  4.5&    EC &    FSR 1190      &\citet{Froebrich07}\\
225.48&  -2.57&  7:05:18& -12:19:44&    4&    EC &    BDS 96        &\citet{Bica03b}\\
259.27&  -2.61&  8:22:22& -41:36:14&    3&    EC &    Majaess 100   &\citet{Majaess13}\\
259.29&  -3.04&  8:20:32& -41:51:47&    3&    EC &    Majaess 98    &\citet{Majaess13}\\
259.33&  -2.55&  8:22:49& -41:37:10&    4&    EC &    Majaess 101   &\citet{Majaess13}\\
259.50&  -2.57&  8:23:15& -41:46:06&    6&    EC &    Majaess 104   &\citet{Majaess13}\\
259.60&  -2.98&  8:21:45& -42:04:55&   10&    EC &    Majaess 99    &\citet{Majaess13}\\
259.61&  -2.70&  8:23:00& -41:55:45&    9&    EC &    Majaess 103   &\citet{Majaess13}\\
259.76&  -2.84&  8:22:52& -42:07:58&    5&    EC &    DBS 18        &\citet{Dutra03}\\
260.76&   0.66&  8:41:07& -40:52:05&    6&    EC &    MLG 1         &\citet{Massi00}\\
260.93&   0.11&  8:39:20& -41:19:53&  3.5&    EC &    MLG 3         &\citet{Massi00}\\
263.53&  -0.35&  8:46:04& -43:39:43&    6&    EC &    DBS 24        &\citet{Dutra03}\\
264.19&   0.18&  8:50:40& -43:50:44&    8&    EC &    FSR 1424      &\citet{Froebrich07}\\
264.23&  -0.85&  8:46:20& -44:31:16&    5&    EC &    MLG 15        &\citet{Massi00}\\
264.32&  -0.18&  8:49:33& -44:10:47&  1.0&    EC &    MLG 16        &\citet{Massi00}\\
264.47&  -0.27&  8:49:41& -44:21:31&  4.5&    EC &    BH 54         &\citet{van75}\\
264.67&  -0.28&  8:50:21& -44:30:43&  1.7&    EC &    MLG 18        &\citet{Massi03}\\
264.73&   0.26&  8:52:55& -44:12:39&    5&    EC &    FSR 1432      &\citet{Froebrich07}\\
264.99&   0.99&  8:56:56& -43:56:30&    7&    EC &    DBS 29        &\citet{Dutra03}\\
266.16&   1.14&  9:01:54& -44:43:32&    6&    EC &    DBS 30        &\citet{Dutra03}\\
270.67&  -2.45&  9:03:43& -50:28:32&   13&    EC &    Majaess 109   &\citet{Majaess13}\\
276.96&  -2.04&  9:34:42& -54:40:26&    6&    EC &    BH 75         &\citet{van75}\\
281.76&  -2.02& 10:01:22& -57:43:10&  4.5&    EC &    DBS 40        &\citet{Dutra03}\\
281.84& -1.61& 10:03:40& -57:26:38&    6 &   EC  &   Majaess 116   &\citet{Majaess13}\\
\hline
\end{tabular}
}
\end{table*}
Out of the 437 objects, 297 ($\sim67.9\%$) were classified as ECs, 23 ($\sim5.3\%$) are ECCs, 90 ($\sim20.8\%$) are EGrs, and the remaining 27 ($\sim6.2\%$) are OCCs. These classes may suffer from completeness effects, especially the less dense EGrs.

\begin{table*}[hp]
{\footnotesize
\begin{center}
\caption{Derived fundamental parameters for a test sample of 14 star clusters from Table~\ref{tab1}.}
\renewcommand{\tabcolsep}{1.6mm}
\renewcommand{\arraystretch}{1.4}
\begin{tabular}{lrrrrrrrrrr}
\hline
\hline
Cluster &phase&$\alpha(2000)$&$\delta(2000)$&$A_V$&Age&$d_{\odot}$&$R_{GC}$&$x_{GC}$&$y_{GC}$&$z_{GC}$\\
&&$(^{\circ}\,^{\prime}\,^{\prime\prime})$&(mag)&(Myr)&(kpc)&(kpc)&(kpc)&(kpc)&(kpc)\\
($1$)&($2$)&($3$)&($4$)&($5$)&($6$)&($7$)&($8$)&($9$)&($10$)&($11$) \\
\hline
C 14 &EC&5:10:50&36:39:45 &$3.57\pm0.2$&$3\pm2$&$2.8\pm0.26$&$9.95\pm0.26$&$-9.93\pm0.26$&$0.50\pm0.05$&$-0.09\pm0.01$\\
C 18 &EC&5:30:30&36:39:53 &$2.28\pm0.2$&$2\pm1$&$5.66\pm0.54$&$12.85\pm0.53$&$-12.82\pm0.53$&$0.81\pm0.08$&$-0.14\pm0.01$\\
C 64 &EGr&6:08:49&13:09:38&$4.27\pm0.2$&$1\pm1$&$3.8\pm0.4$&$10.9\pm0.5$&$-10.9\pm0.5$&$-1.1\pm0.1$&$-0.22\pm0.02$\\
C 74 &EC&6:29:55&12:32:30&$4.27\pm0.2$&$3\pm2$&$3.83\pm0.27$&$10.9\pm0.34$&$-10.83\pm0.34$&$-1.27\pm0.12$&$0.07\pm0.01$\\
C 133 &EC&6:54:54&-3:21:39 &$2.28\pm0.2$&$10\pm5$&$3.26\pm0.3$&$10.0\pm0.25$&$-9.85\pm0.25$&$-1.93\pm0.18$&$-0.04\pm0.01$\\
C 149 &EC&6:59:46&-3:36:33 &$1.49\pm0.2$&$2\pm1$&$9.1\pm2.0$&$15.5\pm1.8$&$-14.45\pm1.8$&$-5.47\pm0.57$&$0.03\pm0.01$\\
C 209 &EC&6:59:12&-11:58:49 &$4.46\pm0.2$&$2\pm1$&$4.1\pm0.5$&$10.53\pm0.28$&$-10.13\pm0.28$&$-2.86\pm0.27$&$-0.27\pm0.03$\\
C 245 &EC&8:25:16&-43:26:30 &$3.67\pm0.2$&$1\pm1$&$3.94\pm0.40$&$8.6\pm0.3$&$-7.8\pm0.1$&$-3.9\pm0.4$&$-0.22\pm0.02$\\
C 260 &EC&8:30:24&-44:27:09 &$2.28\pm0.2$&$2\pm1$&$4.50\pm0.43$&$9.0\pm0.22$&$-7.8\pm0.1$&$-4.45\pm0.42$&$-0.24\pm0.02$\\
C 292 &EC&9:14:28&-47:24:01 &$3.47\pm0.2$&$1\pm1$&$3.86\pm0.37$&$8.21\pm0.17$&$-7.25\pm0.01$&$-3.86\pm0.37$&$0.06\pm0.01$\\
C 353 &EC&9:35:39&-54:11:40 &$3.47\pm0.2$&$1\pm1$&$4.23\pm0.40$&$7.9\pm0.2$&$-6.7\pm0.1$&$-4.2\pm0.4$&$-0.12\pm0.01$\\
C 394 &EC&9:58:12&-56:08:24 &$1.98\pm0.2$&$1\pm1$&$5.12\pm0.50$&$8.1\pm0.3$&$-6.3\pm0.1$&$-5.1\pm0.5$&$-0.09\pm0.01$\\
C 399 &EC&9:58:12&-56:27:03 &$5.46\pm0.2$&$1\pm1$&$3.0\pm0.30$&$7.3\pm0.2$&$-6.7\pm0.1$&$-2.94\pm0.3$&$-0.07\pm0.01$\\
C 429 &EGr&10:04:07&-57:37:50 &$3.27\pm0.2$&$1\pm1$&$5.0\pm0.50$&$7.9\pm0.3$&$-6.2\pm0.1$&$-4.9\pm0.5$&$-0.15\pm0.01$\\
\hline
\end{tabular}
\begin{list}{Table Notes.}
\item Col. 3: $A_V$ in the cluster central region. Col. 4: age, from 2MASS photometry. Col. 5: distance from the Sun. Col. 6: $R_{GC}$ calculated using $R_{\odot}=7.2$ kpc for the distance of the Sun to the Galactic centre \citep{Bica06}. Cols. 7 - 9: Galactocentric components.
\end{list}
\label{tab2}
\end{center}
}
\end{table*}

\begin{figure}[!ht]
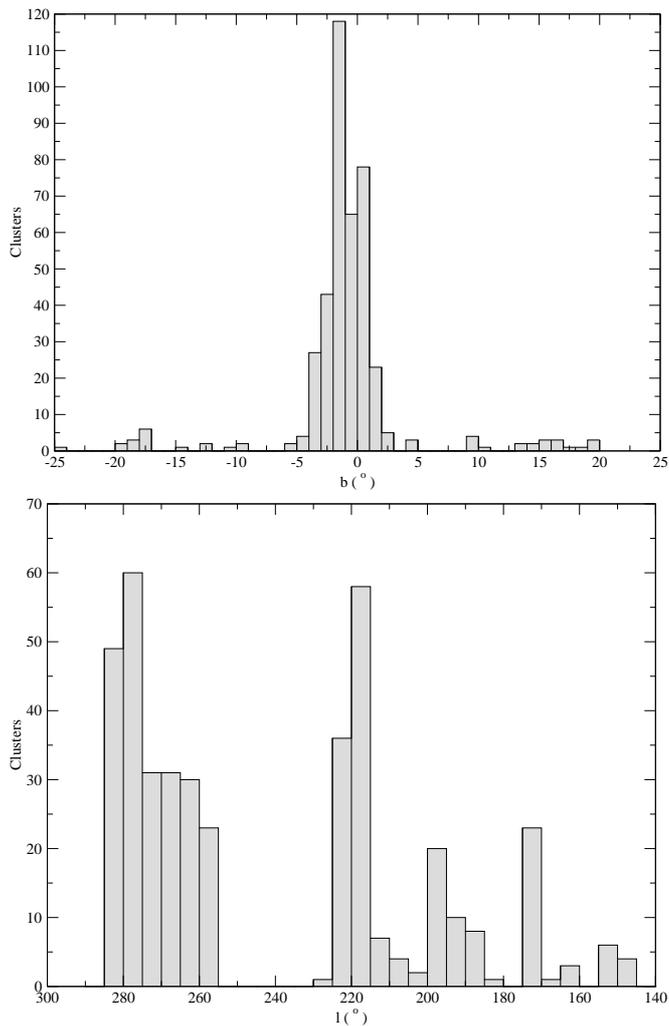

\begin{minipage}[b]{0.98\linewidth}
\includegraphics[width=\textwidth]{bhistogram.eps}
\end{minipage}\hfill
\vspace{0.05cm}
\begin{minipage}[b]{0.98\linewidth}
\includegraphics[width=\textwidth]{lhistogram.eps}
\end{minipage}\hfill
\caption[]{Distribution of the new clusters and candidates in Galactic latitude
(top panel) and longitude (bottom panel).}
\label{histogram}
\end{figure}

\begin{figure}[!hb]
\resizebox{\hsize}{!}{\includegraphics{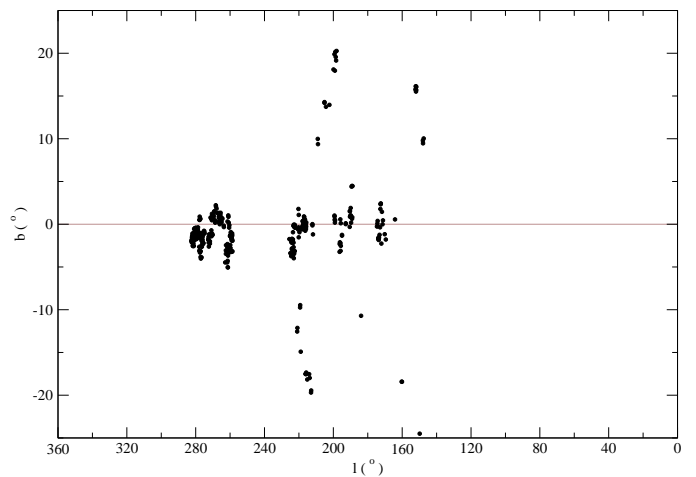}}
\caption[]{Galactic distribution of the new clusters and candidates.}
\label{lb}
\end{figure}

\section{Cluster structure and dust distribution}
\label{sec:3}

One of us (D. C.) carried out the cluster search and discovered  437 objects. The first 20 entries are listed  in Table \ref{tab1}. The entire list is available in the online version of the paper.
Table \ref{tab1} shows  the ranges of $\ell$ and  $b$ , so that most of the present survey was carried out in the 3rd Quadrant. Fig.~\ref{histogram} shows histograms of  $\ell$ and $b$, while the Fig.~\ref{lb} shows the $\ell\times{b}$ distribution of the new clusters and candidates.
The present objects are mostly projected within small Galactic latitude  values, as expected 
for a peaked  thin disk recent stellar cluster  formation. 
As a cautionary remark,  the dips  along longitude (Figs.~\ref{histogram} and \ref{lb}) reflect more where we searched clusters than any intrinsic distribution. The sample does not intend to be complete in any sense. The distribution in Fig.~\ref{lb} of populated areas and excursions that we made to search for the clusters suggest  that a large number of ECs and other classes are yet to be found.

Star clusters show a variety of structures that depend on their evolutionary phase. They are formed as embedded clusters and their morphology in the earliest stage is tightly associated with the formation process, reflecting the morphology of the gas in their natal molecular clouds \citep{Lada03, Cartwright04}. Well known embedded cluster-forming regions are radially concentred or present a fractal-like structure with small ECs \citep{Schmeja08, Sanchez10, Lomax11, Camargo11, Camargo12}. These substructures are quickly erased by dynamical effects forming a massive cluster by coalescence or being completely dissolved in the field. On the other hand, some of them may evolve to individual clusters, forming a group of clusters or  aggregate \citep{Feigelson11, Camargo11, Camargo12}. In the early ages substructured clusters and EC aggregates appear to be more able to survive the primordial gas expulsion than those centrally concentred \citep{Kruijssen12a}.   ECs may present an expanding bubble generated by OB stars \citep{Lee12}, some of which may develop subsequently a second generation of stars, and possibly a second generation of ECs, by means of sequential star formation.

\begin{figure}[!ht]
\resizebox{\hsize}{!}{\includegraphics{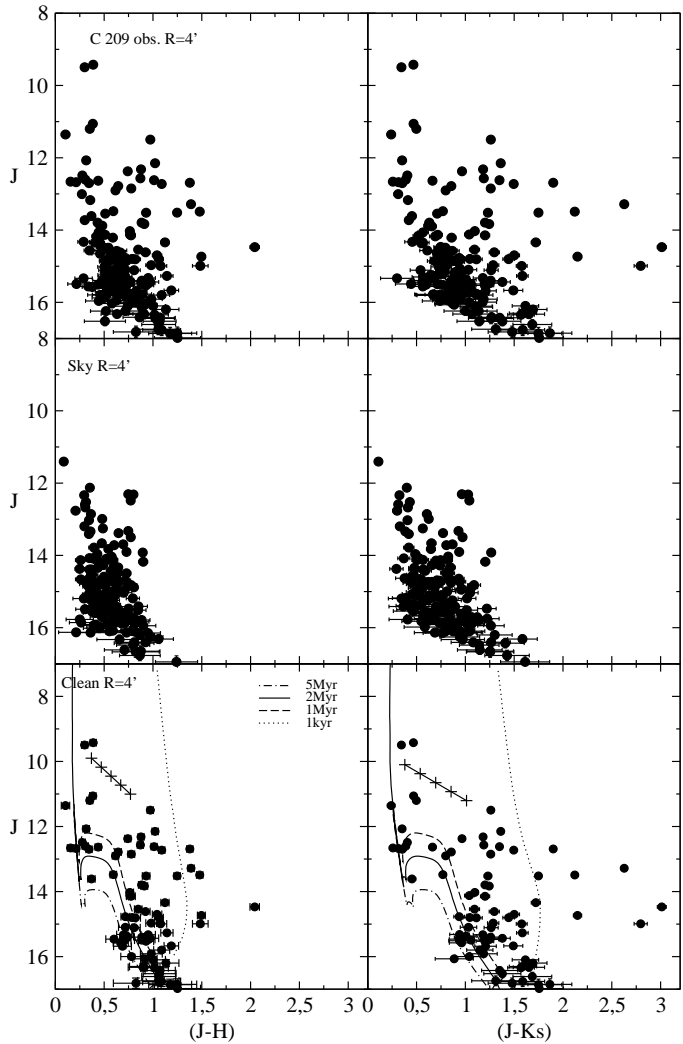}}
\caption[]{Band combinations in the colour-magnitude diagrams.
 2MASS CMDs extracted from the central region of C 209. Top panels: observed CMDs $J\times(J-H)$ (left) and $J\times(J-K_s)$ (right). Middle panels: equal area comparison field. Bottom panels: field star decontaminated CMDs fitted with MS and PMS Padova isochrones.  We also present the reddening vector for $A_V=0$ to 5.}
\label{cmd1}
\end{figure}

The stellar systems analysed in this work present diversified morphologies.
ECs have a core as a rule. The core may be stellar, dusty or dusty/stellar, suggesting that a collapse has occurred. EGrs have in general little central concentration. ECs and EGrs appear to be in general related to dust emission in bubbles or winds. 

In Fig.~\ref{f2}  we show representative WISE images of ECs found in this study (C 62, C 72, C 258, C 44, C 284, and C 288). These clusters present evident stellar-density contrast with respect to the background. C 62 and C 44 are ECs with prominent dust/stellar cores.

Fig.~\ref{f3} shows standard RGB images of representative objects in the sample (C 245, C 253, C 40, C 292, C 241, and C 39).

Fig.~\ref{g} shows interesting filamentary cluster-forming clouds. The s-shaped distribution of stars in C 260 may be linked to the remaining gas in the filamentary structure. The pair C 217 and C 213 are forming in a gas filament. C 217  presents an elongated stellar distribution aligned with the filament.  It appears that the elongated shape of ECs in Fig.~\ref{g} is related to the primordial nebular structure.

\begin{figure}[!ht]
\resizebox{\hsize}{!}{\includegraphics{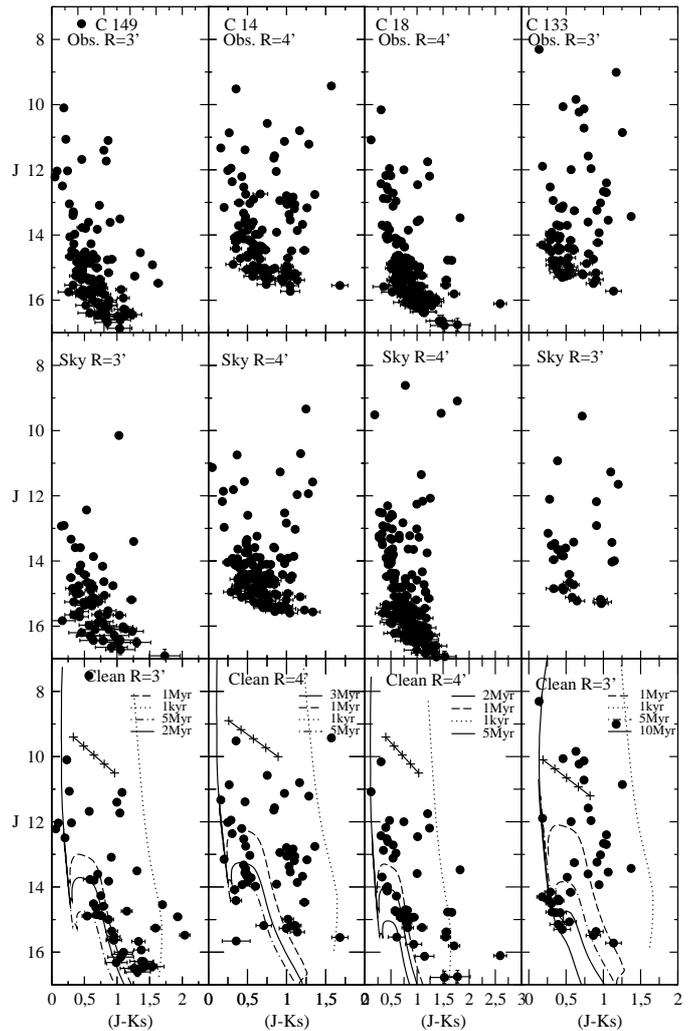}}
\caption[]{Colour-magnitude diagrams for the test sub-sample.
 2MASS CMDs for the clusters C 149, C 14, C 18, and C 133. Top panels: observed CMDs $J\times(J-K_s)$. Middle panels: equal area comparison field. Bottom panels: field star decontaminated CMDs fitted with MS and PMS Padova isochrones. We show the reddening vector for $A_V=0$ to 5.}
\label{cmd2}
\end{figure}

In Fig.~\ref{f4} we  show some remarkable objects.
 C 201 presents a prominent structure resembling a diamond ring. C 178 is an EC-forming, which appears to be undergoing powerful wind shocks. C 219 together with C 218 and C 222 (in the same panel) form a probable group of clusters. They are aligned, probably as a consequence of the gas dynamics. The dust in C 219 shows a twisted filamentary structure.

\begin{figure}[!ht]
\resizebox{\hsize}{!}{\includegraphics{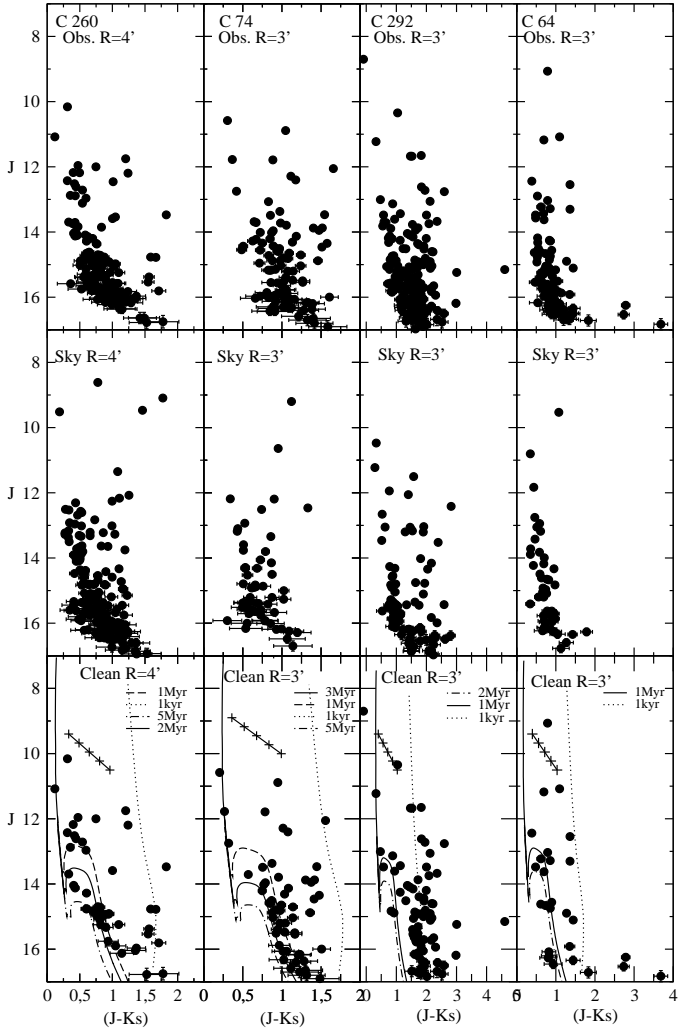}}
\caption[]{Same as Fig.~\ref{cmd2} for the ECs C 260, C 74, C 292, and the EGr C 64.}
\label{cmd3}
\end{figure}

The WISE RGB images of some presently discovered clusters, e.g., C 39, C 40, C 253, C 292, and  C 260 (Figs.~\ref{f3} and \ref{g}) show Extended Green Objects (EGOs), which are considered to be  young stellar object (YSO) candidates \citep[][and references therein]{Churchwell06, Churchwell07, Zavagno07, Cyganowski08, Deharveng10}.  EGOs are often associated with infrared dark clouds (IRDCs) that are the precursors of cluster-forming clumps \citep{Carey00, Rathborne10}.  

We conclude that dust enshrouded stellar clusters and aggregates in this work show a variety of projected stellar and dust distributions, and inferences can be made about star forming processes and stages.

\section{2MASS photometry and analytical tools}
\label{sec:4}

In this Section we present the tools to  test  whether   a representative sub-sample of  14  WISE  candidates (Table 2) are physical  stellar clusters.   We employ 2MASS\footnote{The Two Micron All Sky Survey, available at \textit{www..ipac.caltech.edu/2mass/releases/allsky/}} photometry \citep{Skrutskie06} in the $J$, $H$ and $K_{s}$ bands, extracted in circular concentric regions centred on the coordinates given in Table~\ref{tab2}. 2MASS gives an adequate spatial and photometric uniformity required for wide extractions in view of a consistent field star decontamination. The analytical tools have been widely applied to stellar clusters in our previous papers, and below we refer to them for details on the methods.

\begin{figure}[!ht]
\resizebox{\hsize}{!}{\includegraphics{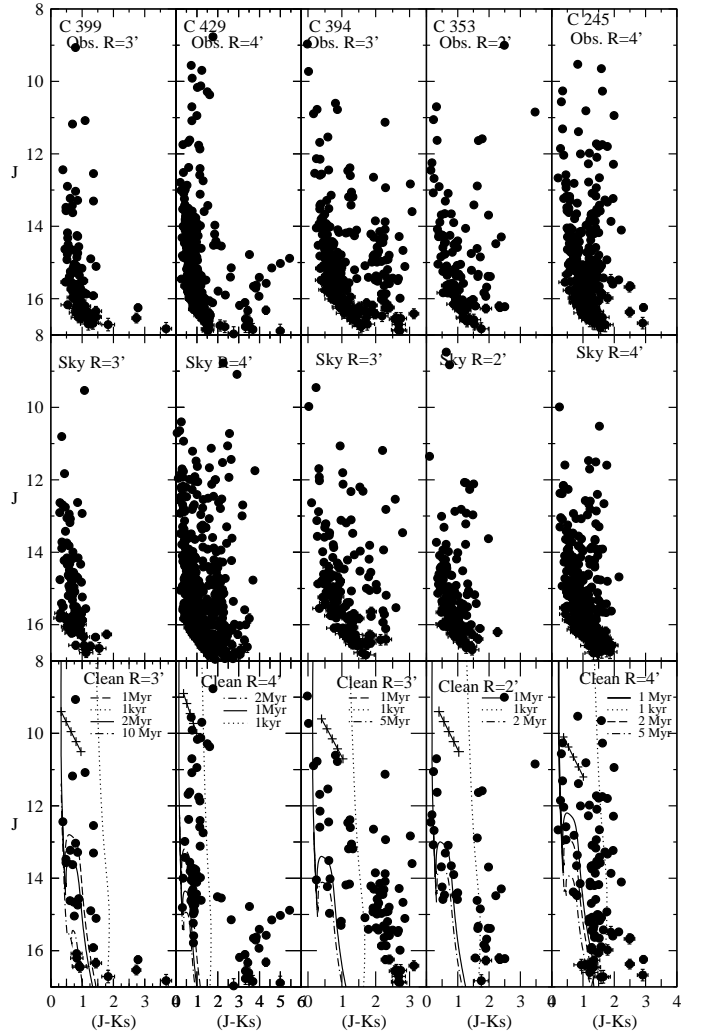}}
\caption[]{Same as Fig.~\ref{cmd2} for the ECs C 245, C 353, C 394, C 399, and the EGr C 429.}
\label{cmd4}
\end{figure}

Aiming to uncover the intrinsic CMD morphology from the foreground/background stars, we applied a field-star decontamination procedure \citep{Bonatto07a}.
The decontamination algorithm developed by one of us (C.B.) is fully described in \citet{Bonatto07b} and \citet{Bica08}.  In short, the CMD is divided into a 3D grid of cells with axes along the $J$ magnitude and the ($J-H$) and ($J-K_s$) colours, then the expected number-density of field stars in a given cell is computed based on the number of comparison field stars. Subsequently, the expected number of field stars is subtracted from each cell, generating the decontaminated CMD.

Fundamental parameters are derived using the field decontaminated CMD fitted with MS and PMS Padova isochrones \citep{Bressan12}. The derived parameters are the observed distance modulus $(m-M)_{J}$ and reddening $E(J-H)$, which converts to $E(B-V)$ and $A_{V}$ with the relations $A_{J}/{A_{V}}=0.276$, $A_{H}/{A_{V}}=0.176$, $A_{K_{s}}/{A_{V}}=0.118$, $A_{J}=2.76\times{E(J-H)}$  and $E(J-H)=0.33\times{E(B-V)}$ \citep{Dutra02}. We assume a constant total-to-selective absorption ratio $R_{V}=3.1$ \citep{Cardelli89}. Decontaminated CMDs are shown in Figs.~\ref{cmd1} to \ref{cmd4}. The fundamental parameters are estimated by applying shifts to the isochrone set from the zero distance modulus and reddening to the best fit.

\begin{figure}[!ht]
\resizebox{\hsize}{!}{\includegraphics{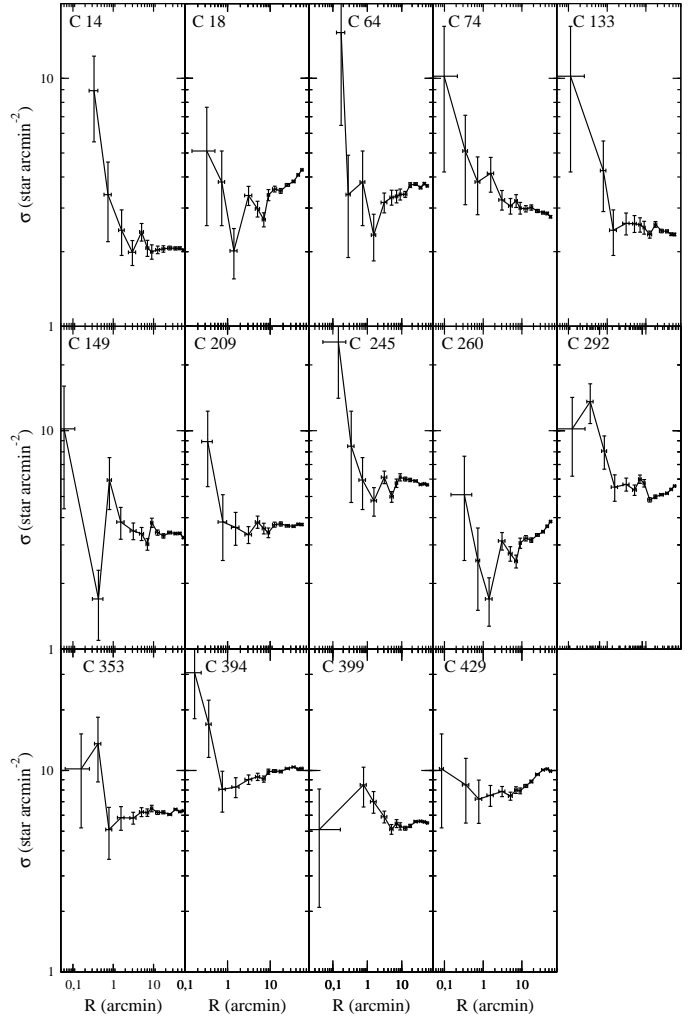}}
\caption[]{Radial density profiles for the embedded clusters and stellar groups in the test sub-sample.}
\label{rdp}
\end{figure}

As a rule, a star cluster structure is studied by means of the stellar radial density profile (RDP). Usually, star clusters have   RDPs following a power law profile \citep{King62, Wilson75, Elson87}. However, ECs are often heavily embedded in their embryonic molecular cloud, which may absorb the near background and part of the cluster member stars. As a result, some ECs may present RDPs with  decreasing density towards the cluster centre or bumps and dips  \citep{Lada03, Camargo11, Camargo12, Camargo13}. Some RDP irregularities appear to be intrinsic to the embedded evolutionary stage and related to a fractal-like structure \citep{Lada03, Camargo11, Camargo12, Camargo13}. We thus argue that overdensities  in general are not enough  to describe ECs in the earliest stages (Fig.~\ref{rdp}).

\section {CMD and RDP analyses of 14 test objects}
\label{sec:5}

Fig.~\ref{f5} shows WISE images of a representative sub-sample of objects analysed in this section. In Figs.~\ref{cmd1} to \ref{cmd3} are show their CMDs. The upper panels give CMDs extracted from a circular area centred on the coordinates of each object (Table~\ref{tab1}). The middle panels correspond to the comparison fields of equal area, and the bottom panels give  the decontaminated CMDs. The derived fundamental parameters (\textit{age, reddening, distance}) are in Table~\ref{tab2}.

Most of the present ECs and EGrs  are in an initial evolutionary phase, and cannot be fitted by King's profile (Fig.~\ref{rdp}). Normally, we apply a colour magnitude filter to the observed photometry to enhance the RDP contrast relative to the background \citep[e.g.][and references therein]{Bonatto07a, Bonatto09, Bonatto10, Bonatto11, Camargo09, Camargo10}, but the decontaminated CMDs of the present clusters show extremely reddened stars so that the filter is not  effective by discarding such stars.  

Fig.~\ref{cmd1} shows for illustrative purposes the \jh~ and \jk~ CMDs for C 209. This EC presents high absorption  ($A_V\approx4.5$ mag).
The decontaminated CMDs show extremely reddened stars mainly in the \jk~ colour.

Probably, differential reddening contributed to the IR excess, but part of it can be intrinsic to their evolutionary stage, because the most reddened stars as a rule occupy  the cluster core on the WISE images. Most of them are probably EGOs. In this sense, WISE is sensitive to the youngest stars in Class 0, which are still accreting material \citep{Wright10}.

In Fig.~\ref{cmd2} we show \jk~ CMDs for C 149, C 14, C 18, and C 133, respectively. C 149 was initially classified as ECC, but based on the decontaminated CMD we now confirm it as EC. The best-fitting Padova MS and PMS isochrones suggest an age of $2\pm1$ Myr, a distance from the Sun of $d_{\odot}=9.1$ kpc, and a distance from the Galactic centre of $R_{GC}=15.45\pm1.8$ kpc. C 14 is  recognised as a cluster by the presence of a  core and combination of MS and PMS sequences in the decontaminated CMD. The estimated age is $3\pm2$ Myr for  $d_{\odot}=2.8\pm0.26$ kpc. The RDP of C 14 (Fig.~\ref{rdp}) presents a high density contrast to the field suggesting a cluster. The decontaminated CMD of the EC C 18 presents extremely reddened stars, again suggesting a cluster initial evolutionary stage. We derived an age of $2\pm1$ Myr and $d_{\odot}=5.6\pm0.5$ kpc. The RDP of C 18 is also  irregular with the background density centrally decreasing.  The prominent C 133 has decontaminated CMD and RDP suggesting a cluster. We derive for C 133 an age of $10\pm5$ and $d_{\odot}=3.3\pm0.3$ kpc. In the early age range ECs may be centrally concentrated with $R\sim1$ pc \citep{Parmentier12, Camargo13}. 

Fig.~\ref{cmd3} provides CMDs of C 260, C 74, C 292, and C 64. The WISE image of C 260 (Fig.~\ref{f4}) shows very reddened stars, but they are  absent in the 2MASS photometry. Nonetheless, the decontaminated CMD shows features of an EC. The best fit age using MS and PMS isochrones is $2\pm1$ Myr, which implies in a $d_{\odot}=4.5\pm0.4$ kpc and $R_{GC}=9\pm0.2$ kpc. The decontaminated  CMD of   C 74 is typical of  ECs with a poorly populated MS and a well-defined PMS. The cluster presents an important absorption  of $A_V\sim4.27$ mag, age of $3\pm2$ Myr, and $d_{\odot}=3.83\pm0.27$ kpc. The RDP of  C 74 is constrasted  with respect to the background. C 292 is an EC with age of $1\pm1$ Myr and $d_{\odot}=3.86\pm0.37$ kpc. The decontaminated CMD shows a prominent PMS with very reddened stars. Despite a dip in the inner RDP region, the structure of  C 292 points to a cluster. The EGr C 64 also presents extremely reddened stars (Fig.~\ref{cmd3}). The decontaminated CMD points to an early age and the RDP is irregular, but shows a central peak.

Fig.~\ref{cmd4} shows  the CMDs for C 245, C 353, C 394, C 399, and the EGr C 429. The best fit of MS and PMS Padova isochrones provides an age of $1\pm1$ for these objects.

Some clusters appear to show an age spread, arising possibly from the merge of two generations of  ECs or cluster-forming clumps (e.g. C 14 and C 133, Fig.~\ref{cmd2}). 

We conclude from this chapter's analyses of CMDs and density profiles  that this  representative subsample is made up of  star clusters,  in general in their  cold gas and dust embryonic cloud. This in turn reinforces expectations that the present list of 437 objects  will become  a major source for  future studies of star cluster early evolutionary stages.

\section{Concluding remarks}
\label{sec:6}

We report the discovery on WISE  images  of 437  Galactic star clusters, stellar groups, and candidates, mostly in the 3rd Galactic Quadrant. They are as a rule deeply embedded in their natal molecular clouds. Out of the 437 objects, 297 ($\sim67.9\%$) were classified as embedded clusters, 23 ($\sim5.3\%$) as candidates, 90 ($\sim20.8\%$) as embedded stellar groups, and the remaining 27 ($\sim6.2\%$) as open cluster candidates. 
We analysed in  detail a subsample of 14 clusters by means of  2MASS photometry to probe their nature. We derived CMD fundamental cluster parameters for  them. Radial density profiles do  not follow a King's law, as expected for such young clusters. Most of them  are very young ($\sim1$ Myr) , and  in  part  present decontaminated CMDs with extremely reddened stars, comparable to  those in infrared dark clouds that are the precursors of  ECs.

The present  catalogue of newly found embedded clusters and alike objects will certainly become a major object source   for future  studies of star cluster formation and early evolution.

\vspace{0.8cm}

\textit{Acknowledgements}:  We thank an anonymous referee for important comments and suggestions. This publication makes use of data products from the Two Micron All Sky Survey and Wide-field Infrared Survey Explorer. The first is a joint project of the University of Massachusetts and the Infrared Processing and Analysis Centre/California Institute of Technology. The second, is a joint project of the University of California, Los Angeles, and the Jet Propulsion Laboratory/California Institute of Technology. Both funded by the National Aeronautics and Space Administration and the National Science Foundation. We acknowledge support from INCT-A and CNPq (Brazil).

\appendix
\section{Online material}

We carried out a search that provided 437 new star clusters, stellar
groups, and candidates. The first 20 entries are listed in Table~\ref{tab1} and Table~\ref{tbl-1} provide the full list of newly found star clusters.

\onecolumn

\begin{deluxetable}{lrrrrrrrr}

\tablecolumns{9}
\tablewidth{0pc}

\tabletypesize{\scriptsize}

\tablecaption{The list of discovered star clusters or candidates. \label{tbl-1}}

\tablehead{

\colhead{Target} & \colhead{$\ell$}   & \colhead{$b$}    & \colhead{$\alpha(2000)$} &
\colhead{$\delta(2000)$}    & \colhead{Size}   & \colhead{Type}    & \colhead{Avend.}  &  \colhead{Comments}\\

\colhead{} & \colhead{$(^{\circ})$}   & \colhead{$(^{\circ})$}    & \colhead{(h\,m\,s)} &
\colhead{$(^{\circ}\,^{\prime}\,^{\prime\prime})$}    & \colhead{$(\,^{\prime}\,)$}   & \colhead{}    & \colhead{}  &  \colhead{}\\

\colhead{($1$)} & \colhead{($2$)}   & \colhead{($3$)}    & \colhead{($4$)} &
\colhead{($5$)}    & \colhead{($6$)}   & \colhead{($7$)}    & \colhead{($8$)}  &  \colhead{($9$)}\\[-0.3cm]
}

\startdata

\astrobj{Camargo 1} &147.52&10.03& $4:44:32$& $61:12:03$&$8\times8$&EC&n& loose core, prominent\\
\astrobj{Camargo 2} &147.88&  9.46& $4:42:50$& $60:33:46$&$9\times9$&EC&n& loose core, impressive\\
\astrobj{Camargo 3} &147.99&  9.79&  $4:45:28$&  $60:41:25$&  $10\times10$&EC&n& loose core\\
\astrobj{Camargo 4} &149.83&  -24.50&  $2:47:55$&  $32:13:07$&  $8\times8$&OCC&n& \\
\astrobj{Camargo 5} &151.83&  15.69&  $5:42:49$&  $60:51:31$&  $7\times7$&EC&n& prominent\\
\astrobj{Camargo 6} &151.84&  16.06&  $5:45:34$&  $61:01:36$&  $8\times8$&EC&n& loose core\\
\astrobj{Camargo 7} &151.96&  16.11&  $5:46:19$&  $60:56:45$&  $12\times12$&ECC&n&  \\
\astrobj{Camargo 8} &151.97&  15.52&  $5:42:05$&  $60:39:31$&  $5\times5$&EC&n& core, impressive\\
\astrobj{Camargo 9} &152.12&  16.14&  $5:47:08$&  $60:49:12$&  $6\times6$&EGr&n&\\
\astrobj{Camargo 10} &152.30&  15.76&  $5:45:02$&  $60:29:36$&  $6\times6$&EGr&n&\\
\astrobj{Camargo 11} &160.11&  -18.43&  $3:41:22$&  $31:54:37$&  $4\times4$&EC&n& core \\
\astrobj{Camargo 12} &160.28&  -18.42&  $3:41:58$&  $31:48:45$&  $6\times6$&EC&n& core, prominent \\
\astrobj{Camargo 13} &164.15&  0.56&  $5:03:50$&  $42:22:15$&  $5\times5$&EC&n& impresssive dusty\\
\astrobj{Camargo 14} &169.52&  -1.79&  $5:10:50$&  $36:39:45$&  $6\times6$&EC&n& core\\
\astrobj{Camargo 15} &170.13&  -1.17&  $5:15:04$&  $36:31:54$&  $7\times7$&EGr&n&\\
\astrobj{Camargo 16} &171.17&  0.45&  $5:24:41$&  $36:36:16$&  $5\times5$&EC&y& impressive\\
\astrobj{Camargo 17} &171.43&  -0.01&  $5:23:30$&  $36:07:49$&  $7\times7$&EC&n& core\\
\astrobj{Camargo 18} &171.77&  1.45&  $5:30:30$&  $36:39:53$&  $7\times7$&EC&n& impressive\\
\astrobj{Camargo 19} &172.08&  -2.26&  $5:16:18$&  $34:18:54$&  $4\times4$&EGr&y& wind, cirrus, impressive\\
\astrobj{Camargo 20} &172.42&  2.44&  $5:36:23$&  $36:39:41$&  $5\times5$&EC&n& impressive\\
\astrobj{Camargo 21} &172.53&   2.35&  $5:36:17$&  $36:31:07$&    $7\times7$&EC&y& prominent\\
\astrobj{Camargo 22} &172.69&   1.78&  $5:34:21$&  $36:04:14$&    $4\times4$&EC&n& prominent, core\\  
\astrobj{Camargo 23} &172.86&  -0.36&  $5:26:03$&  $34:45:11$&    $6\times6$&EC&n& core, impressive winds\\
\astrobj{Camargo 24} &173.14&  -1.24&  $5:23:15$&  $34:01:35$&    $2\times2$&EGr&y& shocks\\
\astrobj{Camargo 25} &173.51&  -1.58&  $5:22:55$&  $33:31:52$&    $6\times6$&EC&y& impressive, shocks\\
\astrobj{Camargo 26} &172.53&   2.35&  $5:36:17$&  $36:31:07$&    $6\times6$&EC&y& wind\\
\astrobj{Camargo 27} &173.56&  -1.52&  $5:23:19$&  $33:31:42$&    $2\times2$&EC&y& shocks, in Do 18b\\
\astrobj{Camargo 28} &173.58&  -1.57&  $5:23:09$&  $33:28:42$&    $2\times2$&EC&y& \\
\astrobj{Camargo 29} &173.59&  -1.62&  $5:22:58$&  $33:26:40$&    $2\times2$&EGr&y& \\
\astrobj{Camargo 30} &173.61&  -1.72&  $5:22:39$&  $33:22:18$&    $5\times5$&EC&y& core\\
\astrobj{Camargo 31} &173.68&  -1.84&  $5:22:21$&  $33:14:39$&    $3\times3$&EGr&n& winds\\
\astrobj{Camargo 32} &173.85&  -1.67&  $5:23:28$&  $33:11:52$&    $3\times3$&EC&y& winds\\
\astrobj{Camargo 33} &173.86&  -1.73&  $5:23:18$&  $33:09:19$&    $10\times10$&EGr&y& winds\\
\astrobj{Camargo 34} &174.26&  -0.23&  $5:30:18$&  $33:39:47$&    $4\times4$&ECC&n& faint core, distant?\\
\astrobj{Camargo 35} &174.28&   0.39&  $5:32:53$&  $33:59:05$&    $6\times6$&EC&n& core\\
\astrobj{Camargo 36} &174.41&   0.08&  $5:31:57$&  $33:42:02$&    $6\times6$&EC&n& core, impressive\\
\astrobj{Camargo 37} &174.66&  -0.26&  $5:31:14$&  $33:18:48$&    $6\times6$&EC&n& impressive, diffuse core\\
\astrobj{Camargo 38} &183.87& -10.71&  $5:15:18$&  $19:52:42$&    $7\times7$&OCC&n& \\
\astrobj{Camargo 39} &188.86&   4.46&  $6:22:20$&  $23:24:47$&    $5\times5$&EC&n& impressive, wind\\
\astrobj{Camargo 40} &189.43&   4.40&  $6:23:15$&  $22:52:45$&    $6\times6$&EC&y& impressive, dusty core \\
\astrobj{Camargo 41} &189.00&   0.82&  $6:08:46$&  $21:33:52$&    $3\times3$&EGr&y& \\ 
\astrobj{Camargo 42} &189.14&   0.64&  $6:08:23$&  $21:21:27$&    $7\times7$&EC&n& winds, impressive\\
\astrobj{Camargo 43} &189.15&   4.48&  $6:22:59$&  $23:10:08$&    $4\times4$&EGr&y& dusty core, 2nd generation?\\
\astrobj{Camargo 44} &189.68&   0.18&  $6:07:47$&  $20:39:32$&    $5\times5$&EC&y& impressive, core\\
\astrobj{Camargo 45} &189.87&   1.04&  $6:11:23$&  $20:54:19$&    $8\times8$&EC&n& impressive, core\\
\astrobj{Camargo 46} &189.93&   1.90&  $6:14:46$&  $21:16:01$&   $10\times10$&OCC&n&  core\\
\astrobj{Camargo 47} &190.13&   0.98&  $6:11:46$&  $20:39:22$&    $7\times7$&EC&n& winds, impressive\\
\astrobj{Camargo 48} &190.24&   0.92&  $6:11:41$&  $20:31:43$&    $7\times7$&EC&n& impressive, wind\\
\astrobj{Camargo 49} &190.29&   0.87&  $6:11:38$&  $20:27:23$&    $7\times7$&EC&n& loose\\
\astrobj{Camargo 50} &190.32&   1.56&  $6:14:15$&  $20:45:20$&    $8\times8$&EC&n& core \\
\astrobj{Camargo 51} &190.44&   1.50&  $6:14:17$&  $20:37:33$&   $10\times10$&EGr&n& \\
\astrobj{Camargo 52} &190.50&  -0.31&  $6:07:40$&  $19:42:03$&    $7\times7$&EC&n& bubble, impressive, core\\
\astrobj{Camargo 53} &192.76&   0.01&  $6:13:25$&  $17:52:40$&    $4\times4$&EGr&y& core\\
\astrobj{Camargo 54} &192.79&   0.11&  $6:13:52$&  $17:53:48$&    $5\times5$&EC&y& prominent\\
\astrobj{Camargo 55} &194.97&  -1.25&  $6:13:13$&  $15:20:13$&  $3.5\times2.5$&EC&y& impressive\\
\astrobj{Camargo 56} &194.97&  -1.34&  $6:12:56$&  $15:17:25$&   $6\times6$&EC&n& shock  prominent\\
\astrobj{Camargo 57} &195.65&  -2.53&  $6:09:57$&  $14:07:24$&    $6\times6$&EC&n& asymmetric?\\
\astrobj{Camargo 58} &195.68&   0.11&  $6:19:36$&  $15:21:21$&    $7\times7$&EC&y& core, impressive\\
\astrobj{Camargo 59} &195.77&  -3.12&  $6:08:04$&  $13:43:55$&    $7\times7$&OCC&n& core \\
\astrobj{Camargo 60} &195.86&  -2.22&  $6:11:29$& $14:05:09$&     $4\times4$&EGr&n& few stars, prominent\\
\astrobj{Camargo 61} &195.94&   0.57&  $6:21:48$& $15:20:37$&      $5\times5$&EGr&n& \\
\astrobj{Camargo 62} &196.19&  -2.11&  $6:12:31$&  $13:51:17$&    $5\times5$&EC&y& impressive, core\\
\astrobj{Camargo 63} &196.30&  -2.31&  $6:12:01$&  $13:39:36$&    $8\times8$&EC&n&\\   
\astrobj{Camargo 64} &196.36&  -3.23&  $6:08:49$&  $13:09:38$&    $6\times6$&EGr&y& winds\\
\astrobj{Camargo 65} &198.11&  20.27&  $7:40:56$&  $21:48:37$&    $5\times5$&OCC&n& \\
\astrobj{Camargo 66} &198.36&  19.15&  $7:36:51$&  $21:09:56$&    $6\times6$&OCC&n& small core\\
\astrobj{Camargo 67} &198.64&  19.57&  $7:38:57$&  $21:05:17$&    $8\times8$&OCC&n&\\
\astrobj{Camargo 68} &198.86&  20.16&  $7:41:38$&  $21:06:52$&    $3\times3$&ECC&n&\\
\astrobj{Camargo 69} &198.96&   0.46&  $6:27:13$&  $12:37:22$&    $5\times5$&EC&n& core\\ \hline \tablebreak
\astrobj{Camargo 70} &199.09&  17.96&  $7:33:19$&  $20:04:03$&    $8\times8$&OCC&n& small core\\
\astrobj{Camargo 71} &199.10&   0.19&  $6:26:30$&  $12:22:04$&    $7\times7$&EC&n& impressive\\
\astrobj{Camargo 72} &199.22&   0.53&  $6:27:58$&  $12:25:38$&    $6\times6$&EC&n& impressive\\      
\astrobj{Camargo 73} &199.27&   0.90&  $6:29:25$&  $12:33:03$&    $7\times7$&EGr&n& \\
\astrobj{Camargo 74} &199.34&   1.01&  $6:29:55$&  $12:32:30$&    $6\times6$&EC&n& impressive\\
\astrobj{Camargo 75} &199.35&  19.88&  $7:41:15$&  $20:34:57$&    $5\times5$&SFN&n& dusty core? star forming?\\
\astrobj{Camargo 76} &199.90&  18.10&  $7:35:09$&  $19:24:44$&   $10\times10$&OCC&n& \\
\astrobj{Camargo 77} &202.29&  13.96&  $7:23:12$&  $15:38:04$&    $8\times8$&EGr&n& \\
\astrobj{Camargo 78} &204.30& 13.74& $7:25:47$&  $13:46:09$&    $6\times6$&OCC&n& loose core\\
\astrobj{Camargo 79} &208.92&   9.37&  $7:17:44$&   $7:47:50$&    $7\times7$&OC&n& loose core, impressive\\
\astrobj{Camargo 80} &209.05&   9.98&  $7:20:10$&   $7:57:19$&    $4\times4$&ECC&n& \\
\astrobj{Camargo 81} &211.58&  -1.13&  $6:44:56$&   $0:41:16$&    $7\times7$&EGr&n&  wind\\
\astrobj{Camargo 82} &211.66&  -1.05&  $6:45:23$&   $0:38:55$&    $8\times8$&EGr&y& wind\\ 
\astrobj{Camargo 83} &211.78&  -1.50&  $6:43:59$&   $0:20:37$&    $6\times4$&EC&n& \\
\astrobj{Camargo 84} &211.79&  -1.35&  $6:44:32$&   $0:23:49$&    $6\times6$&EC&n& \\
\astrobj{Camargo 85} &211.82&  -1.44&  $6:44:17$&   $0:19:51$&    $6\times6$&EC&n& \\
\astrobj{Camargo 86} &211.85&  -1.19&  $6:45:14$&   $0:25:00$&    $8\times8$&EC&y& wind, prominent\\
\astrobj{Camargo 87} &211.89&  -1.28&  $6:44:59$&   $0:20:47$&    $6\times6$&EGr&y& \\
\astrobj{Camargo 88} &211.90&  -1.20&  $6:45:16$&   $0:22:20$&    $4\times4$&EC&y& \\
\astrobj{Camargo 89} &211.92&  -0.98&  $6:46:07$&   $0:27:17$&    $8\times8$&EGr&n& wind\\
\astrobj{Camargo 90} &211.92&  -1.53&  $6:44:09$&   $0:12:11$&    $6\times6$&EC&n& winds, prominent\\
\astrobj{Camargo 91} &211.95&  -1.42&  $6:44:36$&   $0:13:47$&    $7\times7$&EC&y& impressive morphology\\
\astrobj{Camargo 92} &211.99&  -1.13&  $6:45:40$&   $0:19:51$&    $7\times7$&EC&y&\\
\astrobj{Camargo 93} &212.03&  -0.40&  $6:48:22$&   $0:37:19$&    $7\times7$&EC&y& prominent dusty core\\
\astrobj{Camargo 94} &212.04&  -1.08&  $6:45:58$&   $0:18:15$&    $6\times6$&EGr&y& winds\\
\astrobj{Camargo 95} &212.10&  -1.28&  $6:45:23$&   $0:09:19$&    $7\times7$&EGr&n& scattered\\
\astrobj{Camargo 96} &212.11&  -0.97&  $6:46:28$&   $0:17:30$&    $7\times7$&EC&n&\\
\astrobj{Camargo 97} &212.11&  -1.11&  $6:46:00$&   $0:13:42$&    $7\times7$&EGr&y& wind\\
\astrobj{Camargo 98} &212.14&  -0.14&  $6:49:30$&   $0:38:46$&    $6\times6$&EC&n& prominent, wind,  core\\
\astrobj{Camargo 99} &212.14&   0.01&  $6:50:02$&   $0:42:34$&    $6\times6$&EC&n& prominent\\
\astrobj{Camargo 100} &212.15&  -1.42&  $6:44:57$&   $0:03:05$&    $8\times8$&EGr&y& impressive, winds\\
\astrobj{Camargo 101} &212.18&  -1.34&  $6:45:18$&   $0:03:30$&    $6\times6$&EC&n& prominent, winds\\
\astrobj{Camargo 102} &212.22&  -1.02&  $6:46:31$&   $0:10:05$&    $5\times5$&EC&y& bubble\\
\astrobj{Camargo 103} &212.26&  -1.27&  $6:45:42$&   $0:01:08$&    $7\times7$&EC&n& small loose core\\
\astrobj{Camargo 104} &212.28&  -0.62&  $6:48:03$&   $0:17:56$&    $5\times5$&EC&y& impressive\\  
\astrobj{Camargo 105} &212.35&  -0.92&  $6:47:07$&   $0:06:07$&    $6\times6$&EC&y& winds\\
\astrobj{Camargo 106} &212.42&  -1.13&  $6:46:30$&  $-0:03:35$&    $6\times6$&EC&y& bubble\\
\astrobj{Camargo 107} &212.45&  -0.98&  $6:47:04$&  $-0:01:08$&    $7\times7$&EC&y& long core\\
\astrobj{Camargo 108} &212.86& -19.45& $5:41:30$&  $-8:40:14$&    $8\times8$&EC&n& impressive\\
\astrobj{Camargo 109} &212.95& -19.69& $5:40:49$&  $-8:51:21$&    $7\times7$&OCC&n&\\
\astrobj{Camargo 110} &212.53&  -0.74&  $6:48:05$&   $0:00:58$&    $6\times6$&EC&n& prominent\\
\astrobj{Camargo 111} &212.64&  -0.79&  $6:48:07$&  $-0:06:12$&    $6\times6$&EC&n&\\
\astrobj{Camargo 112} &213.75& -17.98& $5:48:21$&  $-8:47:32$&    $7\times7$&OCC&n&\\
\astrobj{Camargo 113} &214.05& -17.54&  $5:50:28$&  $-8:51:36$&    $5\times5$&OCC&n&\\
\astrobj{Camargo 114} &215.16& -18.16&  $5:50:05$& $-10:04:34$&    $7\times7$&EGr&n&\\
\astrobj{Camargo 115} &215.73&   0.20&  $6:57:15$&  $-2:24:12$&   $7\times7$&EC&n&\\
\astrobj{Camargo 116} &215.75& -17.35& $5:54:01$& $-10:14:29$&  $4.5\times4.5$&EC&n&\\
\astrobj{Camargo 117} &215.79&  -0.27&  $6:55:42$&  $-2:40:02$&   $7\times7$&EGr&n&\\
\astrobj{Camargo 118} &215.86& -17.54& $5:53:30$& $-10:24:38$&    $3\times2$&EGr&n& 2nd generation, wind \\
\astrobj{Camargo 119} &215.87& -17.49& $5:53:42$& $-10:23:52$&    $8\times8$&OCC&n&\\
\astrobj{Camargo 120} &215.91&  -0.56&  $6:54:52$& $-2:54:27$&    $10\times10$&ECC&n&\\
\astrobj{Camargo 121} &216.00&  -0.13&  $6:56:35$&  $-2:47:11$&    $4\times4$&ECC&n&\\
\astrobj{Camargo 122} &216.02&  -0.38&  $6:55:42$&  $-2:55:20$&   $7\times7$&OCC&n&\\
\astrobj{Camargo 123} &216.06&  -0.12&  $6:56:44$&  $-2:50:19$&   $6\times6$&EC&n& core\\ 
\astrobj{Camargo 124} &216.11&  -0.25&  $6:56:22$&  $-2:56:24$&   $5\times5$&EGr&n& PN A 18 3.5´ NW\\
\astrobj{Camargo 125} &216.14&  -0.24&  $6:56:25$&  $-2:57:44$&    $9\times9$&EC&n&\\
\astrobj{Camargo 126} &216.15&  -0.47&  $6:55:38$&  $-3:04:46$&   $10\times10$&OCC&n&\\
\astrobj{Camargo 127} &216.18&   0.04&  $6:57:30$&  $-2:52:12$&   $10\times10$&ECC&y&\\
\astrobj{Camargo 128} &216.18&   0.12&  $6:57:48$&  $-0:50:26$&    $5\times5$&EGr&n&\\
\astrobj{Camargo 129} &216.22& -17.51& $5:54:13$& $-10:42:37$&    $6\times6$&OCC&n&\\
\astrobj{Camargo 130} &216.23&  -0.50&  $6:55:40$&  $-3:09:46$&    $8\times8$&EC&n& prominent dusty core\\
\astrobj{Camargo 131} &216.23&  -0.11&  $6:57:03$&  $-2:59:06$&    $8\times8$&EC&n&\\
\astrobj{Camargo 132} &216.26&  -0.24&  $6:56:38$&  $-3:04:10$&    $7\times7$&ECC&n& prominent\\
\astrobj{Camargo 133} &216.32&  -0.76&  $6:54:54$&  $-3:21:39$&   $10\times10$&EC&n& prominent\\
\astrobj{Camargo 134} &216.36&  -0.28&  $6:56:43$&  $-3:10:45$&    $7\times7$&EC&n& wind\\
\astrobj{Camargo 135} &216.41&   0.10&  $6:58:08$&  $-3:02:50$&    $8\times8$&ECC&n&\\
\astrobj{Camargo 136} &216.41&   0.18&  $6:58:26$&  $-3:00:52$&    $9\times9$&ECC&n&\\
\astrobj{Camargo 137} &216.45&   0.52&  $6:59:43$&  $-2:53:20$&    $9\times9$&EC&n&\\
\astrobj{Camargo 138} &216.51&  -0.06&  $6:57:45$&  $-3:12:28$&   $10\times10$&ECC&n& wind\\ \hline \tablebreak
\astrobj{Camargo 139} &216.64&   0.26&  $6:59:08$&  $-3:10:50$&  $2.5\times2.5$&EC&n&\\ 
\astrobj{Camargo 140} &216.66&  -0.46&  $6:56:36$&  $-3:31:55$&    $8\times8$&ECC&n&\\
\astrobj{Camargo 141} &216.68&   0.54&  $7:00:12$&  $-3:05:33$&    $7\times7$&EC&n&\\
\astrobj{Camargo 142} &216.89&   0.28&  $6:59:40$&  $-3:23:42$&    $3\times2.5$&EC&n&\\
\astrobj{Camargo 143} &216.92&   0.03&  $6:58:49$&  $-3:31:55$&    $2\times1.5$&SFN&n& dust ring with bright spots\\
\astrobj{Camargo 144} &216.92&  -0.34&  $6:57:30$&  $-3:42:03$&   $10\times10$&EC&n& prominent\\
\astrobj{Camargo 145} &216.92&   0.82&  $7:01:40$&  $-3:10:37$&    $2\times2$&EC&n&\\   
\astrobj{Camargo 146} &217.03&  -0.14&  $6:58:26$&  $-3:42:19$&    $4\times4$&ECC&y&\\
\astrobj{Camargo 147} &217.05&   0.91&  $7:02:12$&  $-3:14:57$&    $5\times5$&EC&y&\\
\astrobj{Camargo 148} &217.09&   0.34&  $7:00:15$&  $-3:32:40$&    $8\times8$&EC&n& prominent\\
\astrobj{Camargo 149} &217.09&   0.20&  $6:59:46$&  $-3:36:33$&    $3\times2$&EC&n& shock, few stars\\
\astrobj{Camargo 150} &217.11&   0.43&  $7:00:36$&  $-3:31:06$&    $2\times2$&EC&n& prominent\\
\astrobj{Camargo 151} &217.11&  -0.19&  $6:58:24$&  $-3:48:18$&   $10\times10$&EC&y&\\
\astrobj{Camargo 152} &217.15&  -0.03&  $6:59:03$&  $-3:46:04$&    $5\times3.5$&EC&n&\\
\astrobj{Camargo 153} &217.19&  -0.07&  $6:58:58$&  $-3:49:07$&    $2\times2$&EC&y& core of a larger cluster?\\
\astrobj{Camargo 154} &217.21&   0.48&  $7:00:58$&  $-3:35:14$&    $5\times5$&EC&n&\\
\astrobj{Camargo 155} &217.27&  -0.34&  $6:58:08$&  $-4:00:45$&   $10\times10$&EC&n&\\
\astrobj{Camargo 156} &217.32&   0.32&  $7:00:35$&  $-3:45:23$&    $5\times5$&EC&n&\\
\astrobj{Camargo 157} &217.40&   0.09&  $6:59:56$&  $-3:56:03$&    $6\times6$&EGr&n&\\
\astrobj{Camargo 158} &217.42&   0.32&  $7:00:46$&  $-3:51:15$&  $1.5\times1.5$&EC&y&\\
\astrobj{Camargo 159} &217.42&   0.28&  $7:00:38$&  $-3:51:41$&    $2\times2$&EC&y& shock induced?\\
\astrobj{Camargo 160} &217.45&   0.35&  $7:00:57$&  $-3:51:34$&  $2.5\times2.5$&EC&y& shock induced? impressive\\
\astrobj{Camargo 161} &217.62&  -0.15&  $6:59:29$&  $-4:14:25$&   $2\times1$&EC&y& bubble, subcluster or 2nd generation of BDS 89\\
\astrobj{Camargo 162} &217.67&   1.14&  $7:04:10$&  $-3:41:26$&    $5\times5$&EC&n& small cl. and large dust cores\\
\astrobj{Camargo 163} &217.90&   0.35&  $7:01:46$&  $-4:15:25$&  $2.5\times2.5$&EC&n&\\
\astrobj{Camargo 164} &217.96&  -0.51&  $6:58:49$&  $-4:42:12$&    $5\times5$&EC&n&\\ 
\astrobj{Camargo 165} &218.10&  -0.36&  $6:59:35$&  $-4:46:01$&    $4\times4$&EC&y&\\
\astrobj{Camargo 166} &218.15&  -0.57&  $6:58:57$&  $-4:53:56$&    $5\times5$&EC&y& include 1´ SE possible subcluster, rel. [G84d]1-3\\
\astrobj{Camargo 167} &218.16&  -0.58&  $6:58:55$&  $-4:54:53$&    $1\times1$&EC&y& subcluster or in pair\\
\astrobj{Camargo 168} &218.16&  -0.29&  $6:59:58$&  $-4:46:54$&    $6\times6$&EC&y&\\
\astrobj{Camargo 169} &218.74&  -0.47&  $7:00:24$&  $-5:22:42$&  $5.5\times5.5$&EGr&n&\\
\astrobj{Camargo 170} &218.93& -14.91&  $6:08:19$&  $-11:57:17$&   $8\times8$&OCC&n& core\\
\astrobj{Camargo 171} &219.24&  -9.46&  $6:28:53$&  $-9:52:08$&    $5\times5$&EC&n& prominent\\
\astrobj{Camargo 172} &219.32&  -9.74&  $6:28:00$&  $-10:03:43$&   $10\times10$&EC&y& prominent\\
\astrobj{Camargo 173} &219.68&  -0.94&  $7:00:26$&  $-6:26:02$&   $4.5\times4.5$&OCC&n&\\
\astrobj{Camargo 174} &220.10&   1.09&  $7:08:27$&  $-5:52:40$&   $6\times6$&ECC&n&\\
\astrobj{Camargo 175} &220.14&  -1.53&  $6:59:10$&  $-7:06:31$&   $7\times7$&OCC&n&\\
\astrobj{Camargo 176} &220.25&  -0.36&  $7:03:33$&  $-6:40:30$&  $4.5\times4.5$&EGr&n&\\
\astrobj{Camargo 177} &220.30&   1.79&  $7:11:19$&  $-5:43:49$&   $5\times5$&EC&n& core\\
\astrobj{Camargo 178} &220.46&  -0.61&  $7:03:02$&  $-6:58:26$&   $8\times8$&EC&y& inc shock\\
\astrobj{Camargo 179} &220.89& -12.13&  $6:21:56$& $-12:29:22$&   $6\times6$&EC&n& prominent\\
\astrobj{Camargo 180} &221.00& -12.55&  $6:20:34$& $-12:46:16$&    $5\times5$&OCC&n& core \\
\astrobj{Camargo 181} &221.29&   0.07&  $7:07:02$&  $-7:24:10$&    $3\times2$&SFN&n& small dust neb, bright in W3, star forming?\\
\astrobj{Camargo 182} &222.28&  -0.41&  $7:07:08$&  $-8::30:16$&   $8\times8$&OCC&n&\\
\astrobj{Camargo 183} &222.43&  -3.14&  $6:57:35$&  $-9:52:41$&    $3\times3$&EGr&n&\\
\astrobj{Camargo 184} &222.43&  -0.12&  $7:08:28$&  $-8:30:00$&    $4\times4$&EC&n& shock origin, 1st generation\\
\astrobj{Camargo 185} &222.47&  -0.06&  $7:08:44$&  $-8:30:20$&    $3\times2$&EC&n& shock induced,2nd generation\\
\astrobj{Camargo 186} &222.74&  -0.32&  $7:08:19$&  $-8:51:58$&    $8\times8$&OCC&n& core\\   
\astrobj{Camargo 187} &222.74&  -3.42&  $6:57:09$& $-10:16:55$&    $4\times4$&EC&n&\\
\astrobj{Camargo 188} &222.95&  -0.05&  $7:09:40$& $-8:55:59$&     $5\times5$&OCC&n&core\\
\astrobj{Camargo 189} &222.98&  -3.98&  $6:55:32$& $-10:44:59$&  $1.5\times1.5$&EC&n&\\
\astrobj{Camargo 190} &223.12&  -2.69&  $7:00:30$&  $-10:17:37$&  $4.5\times4.5$&ECC&n&bubble?\\
\astrobj{Camargo 191} &223.13&  -2.15&  $7:02:28$&  $-10:03:09$&    $5\times5$&EC&n& prominent\\
\astrobj{Camargo 192} &223.16&  -3.51&  $6:57:34$&  $-10:42:00$&   $7\times7$&ECC&n&\\
\astrobj{Camargo 193} &223.20&  -3.16&  $6:58:56$&  $-10:34:46$&   $5\times5$&OCC&n& prominent core\\
\astrobj{Camargo 194} &223.24&  -0.18&  $7:09:46$&  $-9:14:50$&    $5\times5$&EC&n& prominent\\
\astrobj{Camargo 195} &223.45&  -0.94&  $7:07:25$&  $-9:46:57$&    $5\times5$&EC&n& shock, impressive, winds\\
\astrobj{Camargo 196} &223.51&  -1.78&  $7:04:31$&  $-10:13:25$&   $5\times5$&EC&n&\\
\astrobj{Camargo 197} &223.51&  -3.75&  $6:57:22$&  $-11:07:04$&   $1.5\times1.5$&EGr&n&\\
\astrobj{Camargo 198} &223.54&  -1.86&  $7:04:15$&  $-10:16:57$&   $4.5\times4.5$&EC&n&\\
\astrobj{Camargo 199} &223.62&  -1.89&  $7:04:18$&  $-10:22:32$&   $6\times6$&EGr&n& winds \\
\astrobj{Camargo 200} &223.71&  -1.90&  $7:04:25$&  $-10:27:17$&   $4\times4$&EC&n&\\
\astrobj{Camargo 201} &223.71&  -2.20&  $7:03:20$&  $-10:35:20$&   $2\times2$&EC&y& diamond ring, shock impressive\\
\astrobj{Camargo 202} &223.85&  -1.87&  $7:04:48$&  $-10:33:43$&    $8\times8$&ECC&n&\\
\astrobj{Camargo 203} &224.01&  -1.70&  $7:05:43$&  $-10:37:58$&   $5\times4$&EC&n& prominent\\
\astrobj{Camargo 204} &224.09&  -2.94&  $7:01:22$&  $-11:16:02$&   $4.5\times4.5$&EC&n& bubble\\
\astrobj{Camargo 205} &224.09&  -2.84&  $7:01:44$&  $-11:13:20$&   $4\times4$&EGr&n& shock\\ 
\astrobj{Camargo 206} &224.14&  -1.94&  $7:05:05$&  $-10:51:39$&    $5\times5$&EC&n& \\
\astrobj{Camargo 207} &224.34&  -2.14&  $7:04:44$&  $-11:07:14$&    $3\times3$&EC&n& shock\\ \hline \tablebreak
\astrobj{Camargo 208} &224.35&  -2.02&  $7:05:12$&  $-11:04:36$&   $6\times6$&EC&n& impressive\\          
\astrobj{Camargo 209} &224.48&  -3.74&  $6:59:12$&  $-11:58:49$&   $10\times10$&EC&n& prominent core\\
\astrobj{Camargo 210} &224.97&  -3.39&  $7:01:22$&  $-12:15:24$&   $4.5\times4.5$&EC&n& prominent core \\
\astrobj{Camargo 211} &225.55&  -1.74&  $7:08:27$&  $-12:01:06$&   $10\times10$&EGr&n&\\
\astrobj{Camargo 212} &258.61&  -1.93&  $8:23:20$&  $-40:39:58$&   $5\times5$&EC&n& impressive, dusty  core\\
\astrobj{Camargo 213} &258.67&  -3.16&  $8:18:07$&  $-41:25:02$&    $5\times5$&EC&n& prominent\\
\astrobj{Camargo 214} &258.74&  -1.31&  $8:26:23$&  $-40:25:21$&   $5\times5$&EGr&n& bubble, wind\\
\astrobj{Camargo 215} &258.79&  -1.09&  $8:27:32$&  $-40:20:04$&   $6\times6$&EC&n& core\\
\astrobj{Camargo 216} &258.81&  -1.80&  $8:24:30$&  $-40:45:23$&   $5\times5$&EGr&n& bubble\\
\astrobj{Camargo 217} &258.82&  -3.22&  $8:18:20$&  $-41:34:45$&    $4\times3$&EC&n& shock, wind, impressive\\
\astrobj{Camargo 218} &259.05&  -1.55&  $8:26:19$&  $-40:48:39$&   $5\times5$&EC&n& impressive, ring core\\
\astrobj{Camargo 219} &259.08&  -1.62&  $8:26:06$&  $-40:52:35$&   $5\times5$&EC&n& spiral wind,impressive\\
\astrobj{Camargo 220} &259.08&  -1.90&  $8:24:53$&  $-41:02:21$&   $5\times5$&EC&n&\\
\astrobj{Camargo 221} &259.09&  -1.78&  $8:25:27$&  $-40:58:54$&   $6\times6$&ECC&n& wind, impressive\\        
\astrobj{Camargo 222} &259.12&  -1.74&  $8:25:44$&  $-40:58:32$&    $5\times5$&EC&n& prominent, wind\\
\astrobj{Camargo 223} &259.14&  -1.08&  $8:28:38$&  $-40:36:33$&    $5\times5$&EC&n& core\\
\astrobj{Camargo 224} &259.24&  -1.63&  $8:26:33$&  $-41:00:36$&    $3\times3$&EC&n& dusty core\\
\astrobj{Camargo 225} &259.31&  -1.62&  $8:26:49$&  $-41:04:11$&    $5\times5$&EGr&n&\\
\astrobj{Camargo 226} &259.56&  -0.92&  $8:30:36$&  $-40:51:38$&    $6\times6$&EC&n& prominent\\
\astrobj{Camargo 227} &259.57&  -1.03&  $8:30:11$&  $-40:55:50$&    $5\times5$&EC&n&\\  
\astrobj{Camargo 228} &259.57&  -1.44&  $8:28:25$& $-41:10:08$&     $5\times5$&EGr&y&\\
\astrobj{Camargo 229} &259.64&  -1.31&  $8:29:09$&  $-41:08:58$&    $7\times7$&EC&y& in Majaess 107, impressive, chain\\
\astrobj{Camargo 230} &259.71&  -1.27&  $8:29:32$& $-41:11:06$&    $5\times5$&EC&y& in Majaess 107, bubble, shock\\
\astrobj{Camargo 231} &259.73&  -1.32&  $8:29:25$& $-41:14:01$&    $5\times5$&EC&y& in Majess 107, winds\\
\astrobj{Camargo 232} &259.77&  -2.78&  $8:23:10$& $-42:06:07$&    $5\times5$&EGr&n& bubble\\
\astrobj{Camargo 233} &259.81&  -3.26&  $8:21:09$& $-42:24:35$&   $2.5\times2.5$&EC&n& impressive\\       
\astrobj{Camargo 234} &259.89&  -2.43&  $8:25:05$& $-42:00:09$&   $6\times6$&EGr&n& impressive, shock\\
\astrobj{Camargo 235} &259.90&  -1.39&  $8:29:38$& $-41:24:56$&    $2\times2$&OCC&n& far OC?, also in 2MASS\\
\astrobj{Camargo 236} &259.91&  -1.33&  $8:29:56$& $-41:22:44$&    $6\times6$&EGr&n&\\                                        
\astrobj{Camargo 237} &260.08&  -1.38&  $8:30:14$& $-41:32:45$&    $8\times8$&EGr&n&\\                                        
\astrobj{Camargo 238} &260.42&  -0.12&  $8:36:44$& $-41:04:26$&    $8\times8$&EC&n&        shock \\                           
\astrobj{Camargo 239} &260.54&  -0.40&  $8:35:54$& $-41:20:21$&    $6\times6$&EC&n&                                     \\ 
\astrobj{Camargo 240} &260.81&   0.19&  $8:39:17$& $-41:11:30$&    $7\times7$&EC&n&         prominent                  \\     
\astrobj{Camargo 241} &260.91&  -3.10&  $8:25:18$& $-43:13:19$&    $8\times8$&EC&n&            core                    \\     
\astrobj{Camargo 242} &260.95&   0.87&  $8:42:36$& $-40:53:23$&    $5\times5$&EGr&n&              core                 \\     
\astrobj{Camargo 243} &260.96&  -3.33&  $8:24:25$& $-43:23:45$&    $9\times9$&EC&n&             ring of  stars         \\     
\astrobj{Camargo 244} &261.04&   1.03&  $8:43:34$& $-40:51:33$&    $6\times6$&EC&n&        prominent                   \\     
\astrobj{Camargo 245} &261.09&  -3.23&  $8:25:16$& $-43:26:30$&    $6\times5$&EC&n&        prominent                   \\     
\astrobj{Camargo 246} &261.20&  -3.67&  $8:23:39$& $-43:47:09$&    $5\times5$&EGr&n&         core                      \\    
\astrobj{Camargo 247} &261.26&  -5.05&  $8:17:26$& $-44:36:48$&    $7\times7$&EC&n&         impressive wind            \\     
\astrobj{Camargo 248} &261.32&  -3.06&  $8:26:45$& $-43:32:07$&    $6\times6$&EC&y&            core                    \\     
\astrobj{Camargo 249} &261.33&  -5.04&  $8:17:45$& $-44:40:02$&    $7\times7$&EC&n&                                    \\     
\astrobj{Camargo 250} &261.42&  -4.28&  $8:21:31$& $-44:18:53$&    $6\times6$&EC&n&       core                         \\     
\astrobj{Camargo 251} &261.47&   0.32&  $8:41:59$& $-41:38:18$&    $8\times8$&EC&n&              impressive            \\     
\astrobj{Camargo 252} &261.54&  -2.32&  $8:30:48$& $-43:17:02$&    $8\times7$&EC&n&     sequential star formation? bubble, winds, impressive\\ 
\astrobj{Camargo 253} &261.56&  -2.51&  $8:29:59$& $-43:24:45$&    $5\times5$&EC&n&         prominent                   \\    
\astrobj{Camargo 254} &261.58&  -2.68&  $8:29:18$& $-43:31:18$&    $6\times6$&EC&n&         prominent                  \\  
\astrobj{Camargo 255} &261.59&  -2.47&  $8:30:17$& $-43:24:38$&    $3\times3$&EC&n&                                   \\      
\astrobj{Camargo 256} &261.65&  -4.47&  $8:21:25$& $-44:36:45$&   $10\times10$&ECC&n&                                 \\       
\astrobj{Camargo 257} &261.78&  -3.04&  $8:28:20$& $-43:53:40$&    $6\times6$&EC&n&       winds                       \\     
\astrobj{Camargo 258} &262.36&  -2.45&  $8:32:53$& $-44:01:29$&    $6\times6$&EC&n&         impressive                \\      
\astrobj{Camargo 259} &262.41&  -3.46&  $8:28:29$& $-44:38:54$&    $6\times4$&ECC&n&                                  \\      
\astrobj{Camargo 260} &262.45&  -3.06&  $8:30:24$& $-44:27:09$&    $8\times8$&EC&y&         impressive,chain in W4    \\      
\astrobj{Camargo 261} &262.84&  -4.47&  $8:25:14$& $-45:35:10$&    $6\times6$&EC&n&               prominent           \\      
\astrobj{Camargo 262} &263.61&  -0.33&  $8:46:27$& $-43:42:48$&    $8\times8$&EC& n&       prominent                  \\      
\astrobj{Camargo 263} &263.71&  -0.16&  $8:47:30$& $-43:41:18$&    $7\times7$&EC& n&      2 other nearby centers possible\\   
\astrobj{Camargo 264} &264.54&   0.61&  $8:53:44$& $-43:50:33$&    $7\times7$&EC& y&               prominent             \\   
\astrobj{Camargo 265} &264.54&   0.70&  $8:54:07$& $-43:47:11$&    $6\times6$&EC& n&                                     \\  
\astrobj{Camargo 266} &264.97&   0.26&  $8:53:46$& $-44:23:47$&    $6\times6$&EC& n&        impressive nuclear dust      \\   
\astrobj{Camargo 267} &265.51&   1.34&  $9:00:18$& $-44:06:08$&    $8\times8$&EC& n&      core, impressive               \\   
\astrobj{Camargo 268} &265.68&   1.07&  $8:59:47$& $-44:24:24$&    $8\times8$&EC& n&        prominent                    \\   
\astrobj{Camargo 269} &265.82&   0.92&  $8:59:39$& $-44:36:39$&    $7\times7$&EC& n&                                     \\   
\astrobj{Camargo 270} &265.84&   1.10&  $9:00:31$& $-44:30:48$&    $6\times6$&EC& y&       prominent, core               \\   
\astrobj{Camargo 271} &265.92&   0.24&  $8:57:10$& $-45:08:03$&    $6\times6$&EGr&n&     wind                            \\   
\astrobj{Camargo 272} &265.99&   1.04&  $9:00:50$& $-44:40:00$&    $6\times6$ &EC& n&                                    \\    
\astrobj{Camargo 273} &265.99&   0.50&  $8:58:29$& $-45:01:02$&   $13\times13$&EC& n&          prominent                 \\    
\astrobj{Camargo 274} &266.02&   0.00&  $8:56:28$& $-45:21:34$&   $10\times10$&EC& n&      wind, bubble, prominent, core, dust\\   
\astrobj{Camargo 275} &266.09&   0.39&  $8:58:26$& $-45:10:07$&   $6\times6$&EC&   n&           bubble                 \\    
\astrobj{Camargo 276} &266.18&   0.33&  $8:58:29$& $-45:16:32$&    $6\times6$&EC&  n&          bubble                   \\  \hline \tablebreak  
\astrobj{Camargo 277} &266.20&   0.76&  $9:00:25$& $-45:00:04$&    $6\times6$&EC&  n&        core                        \\   
\astrobj{Camargo 278} &266.23&   0.88&  $9:01:00$& $-44:56:55$&    $7\times7$&EC&  y&       core                    \\        
\astrobj{Camargo 279} &266.27&   1.00&  $9:01:41$& $-44:53:52$&    $6\times6$&EGr& n&                               \\       
\astrobj{Camargo 280} &266.27&   1.10&  $9:02:09$& $-44:49:59$&    $5\times5$&EGr& n&          impressive mid IR core\\       
\astrobj{Camargo 281} &266.36&   0.75&  $9:00:57$& $-45:08:03$&    $7\times7$&EC&  n&      core                      \\       
\astrobj{Camargo 282} &266.42&   1.32&  $9:03:36$& $-44:47:43$&    $8\times8$&EC&  n&                                \\       
\astrobj{Camargo 283} &266.51&   0.22&  $8:59:15$& $-45:35:33$&    $6\times5$&EC&  n&        impressive              \\       
\astrobj{Camargo 284} &266.60&   0.90&  $9:02:29$& $-45:12:35$&    $7\times7$&EC&  n&      core,impressive           \\       
\astrobj{Camargo 285} &267.45&   1.33&  $9:07:34$& $-45:33:00$&    $6\times6$&EC&  n&     impressive, core           \\       
\astrobj{Camargo 286} &267.88&   1.83&  $9:11:21$& $-45:31:48$&    $5\times5$&EC&  n&                                \\       
\astrobj{Camargo 287} &268.17&   2.07&  $9:13:26$& $-45:34:29$&    $8\times8$&EC&  y&        small core              \\       
\astrobj{Camargo 288} &268.32&   2.22&  $9:14:39$& $-45:35:01$&    $5\times5$&EC&  n&      wind, impressive           \\       
\astrobj{Camargo 289} &268.62&   0.45&  $9:08:22$& $-47:01:04$&    $8\times8$&EGr& n&                                \\       
\astrobj{Camargo 290} &268.65&   0.18&  $9:07:19$& $-47:12:52$&    $7\times7$&EC&  n&        impressive              \\       
\astrobj{Camargo 291} &269.21&   1.51&  $9:15:14$& $-46:42:57$&    $6\times6$&EGr& n&        central concentration   \\       
\astrobj{Camargo 292} &269.61&   0.94&  $9:14:28$& $-47:24:01$&    $5\times5$&EC&  n&       impressive, winds         \\       
\astrobj{Camargo 293} &269.67&   1.07&  $9:15:18$& $-47:21:23$&    $6\times6$&EGr& n&        impressive, cirrus      \\       
\astrobj{Camargo 294} &269.79&   0.85&  $9:14:51$& $-47:35:21$&    $6\times6$&EC&  n&        impressive              \\       
\astrobj{Camargo 295} &269.87&   0.78&  $9:14:50$& $-47:41:40$&    $2\times2$&EGr& n&                                \\       
\astrobj{Camargo 296} &269.88&   1.14&  $9:16:26$& $-47:27:20$&    $6\times6$&EC&  n&              impressive        \\       
\astrobj{Camargo 297} &269.97&   0.84&  $9:15:34$& $-47:43:39$&    $6\times6$&EC&  n&          impressive            \\       
\astrobj{Camargo 298} &270.04&   0.76&  $9:15:28$& $-47:49:47$&    $6\times6$&EGr& n&             winds              \\       
\astrobj{Camargo 299} &270.20&   1.05&  $9:17:22$& $-47:45:09$&    $6\times6$&EC&  n&    winds                       \\       
\astrobj{Camargo 300} &270.25&  -1.26&  $9:07:28$& $ -49:22:01$&    $8\times8$ &EC& n&       impressive              \\         
\astrobj{Camargo 301} &270.33&   0.81&  $9:16:52$& $-48:00:37$&    $6\times6$ &EGr& y&              winds            \\        
\astrobj{Camargo 302} &270.37&   0.76&  $9:16:52$& $ -48:04:15$&    $3\times3$ &EGr&n&          wind                 \\         
\astrobj{Camargo 303} &270.37&  -1.24&  $9:08:01$& $-49:26:57$&    $7\times7$ &EC&  n&        impressive             \\        
\astrobj{Camargo 304} &270.39&  -1.13&  $9:08:35$& $ -49:23:12$&    $6\times6$ &EC& n&         prominent             \\         
\astrobj{Camargo 305} &270.45&  -1.21&  $9:08:30$& $ -49:29:18$&    $5\times5$ &EC& n&      impressive, shock        \\         
\astrobj{Camargo 306} &270.51&   0.74&  $9:17:20$& $ -48:11:08$&    $5\times5$ &EC& y&      resolved dusty core, impressive\\   
\astrobj{Camargo 307} &270.53&   1.22&  $9:19:29$& $ -47:51:59$&    $6\times6$ &EC& n&              bubble, impressive     \\   
\astrobj{Camargo 308} &270.55&   0.46&  $9:16:20$& $ -48:24:41$&   $10\times8$ &EC& y&                                     \\   
\astrobj{Camargo 309} &270.56&   0.82&  $9:17:55$& $ -48:10:03$&    $6\times6$ &EC& y&         impressive                  \\    
\astrobj{Camargo 310} &270.68&  -1.21&  $9:09:27$& $ -49:38:50$&    $6\times6$ &EC& n&     impressive, strong wind         \\   
\astrobj{Camargo 311} &280.77&  -1.06&  $9:59:49$& $ -56:21:56$&   $ 6\times6$ &EC& n&       dusty core                    \\   
\astrobj{Camargo 312} &270.77&   0.61&  $9:17:54$& $ -48:27:34$&    $5\times4$ &EC& y&                                     \\   
\astrobj{Camargo 313} &270.82&  -1.11&  $9:10:30$& $ -49:41:27$&    $7\times7$ &EC& y&         impressive, wind            \\   
\astrobj{Camargo 314} &270.82&  -1.40&  $9:09:10$& $ -49:53:03$&    $4\times4$ &EC& n&                                     \\   
\astrobj{Camargo 315} &270.88&  -0.72&  $9:12:30$& $ -49:28:00$&    $5\times5$ &EC& y&        impressive, dusty core       \\   
\astrobj{Camargo 316} &270.93&   0.70&  $9:18:57$& $ -48:30:56$&    $7\times7$ &EC& n&         core                        \\   
\astrobj{Camargo 317} &270.95&  -1.38&  $9:09:51$& $ -49:58:05$&    $6\times6$ &EGr&n&           bubble                    \\   
\astrobj{Camargo 318} &270.96&  -1.11&  $9:11:06$& $ -49:47:34$&    $8\times8$ &EC& n&       prominent                     \\   
\astrobj{Camargo 319} &271.12&   0.79&  $9:20:09$& $ -48:35:00$&    $8\times8$ &ECC&n&                                     \\   
\astrobj{Camargo 320} &271.33&   0.77&  $9:20:58$& $ -48:44:39$&    $8\times8$ &EC& n&       impressive                    \\   
\astrobj{Camargo 321} &271.72&  -2.00&  $9:10:17$& $ -50:57:01$&    $7\times7$ &EC& n&     prominent, wind                 \\   
\astrobj{Camargo 322} &271.73&  -2.27&  $9:09:03$& $ -51:08:45$&    $4\times4$ &EC& n&       dusty core                    \\   
\astrobj{Camargo 323} &272.25&  -1.19&  $9:16:17$& $ -50:46:29$&    $5\times5$ &EC& n&    impressive, wind, 2nd generation  \\   
\astrobj{Camargo 324} &272.35&  -2.61&  $9:10:08$& $ -51:50:00$&    $6\times6$ &EC& n&     binuclear, impressive, sequential\\   
\astrobj{Camargo 325} &272.40&  -2.16&  $9:12:29$& $ -51:33:10$&    $7\times7$ &EC& n&      dusty core, impressive         \\   
\astrobj{Camargo 326} &272.49&  -1.49&  $9:15:59$& $ -51:09:23$&    $7\times7$ &EC& n&        bubble                       \\   
\astrobj{Camargo 327} &272.58&  -1.73&  $9:15:16$& $ -51:23:15$&    $6\times6$ &EC& n&     impressive, dusty core          \\   
\astrobj{Camargo 328} &272.70&  -1.82&  $9:15:29$& $ -51:32:27$&    $6\times6$ &EC& n&      prominent                      \\  
\astrobj{Camargo 329} &274.92&  -1.06&  $9:29:06$& $ -52:33:40$&    $7\times7$ &EC& y&         dusty core                  \\   
\astrobj{Camargo 330} &275.08&  -0.85&  $9:30:48$& $ -52:31:16$&    $3\times3$ &EC& n&      subcluster                     \\   
\astrobj{Camargo 331} &275.09&  -0.88&  $9:30:45$& $ -52:33:10$&   $12\times12$ &EC&n&       incl. subcluster                \\    
\astrobj{Camargo 332} &275.07&  -1.09&  $9:29:43$& $ -52:41:11$&  $3.5\times3.5$ &EC&n&        dusty core shocks impressive\\    
\astrobj{Camargo 333} &275.142& -0.77&  $9:31:31$& $ -52:30:17$&   $10\times10$ &EGr& y&        impressive winds           \\    
\astrobj{Camargo 334} &275.14&  -1.07&  $9:30:09$& $ -52:43:35$&    $7\times7$ &ECC&  n&                                   \\   
\astrobj{Camargo 335} &275.17&  -0.98&  $9:30:40$& $ -52:40:49$&    $7\times7$ &EC&   y&   small core                      \\   
\astrobj{Camargo 336} &275.57&  -2.20&  $9:26:58$& $ -53:50:13$&    $5\times5$ &EC&   y&                                   \\   
\astrobj{Camargo 337} &275.65&  -2.19&  $9:27:27$& $ -53:53:04$&    $6\times6$ &EC&   n&   winds                           \\   
\astrobj{Camargo 338} &275.71&  -2.31&  $9:27:11$& $ -54:00:43$&    $6\times6$ &EGr&  n&       winds                       \\   
\astrobj{Camargo 339} &275.72&  -2.23&  $9:27:35$& $ -53:57:34$&   $10\times10$ &EC&  n&           cavity, winds            \\   
\astrobj{Camargo 340} &275.80&  -2.20&  $9:28:08$& $ -53:59:51$&    $7\times7$ &EC&   n&    impressive                     \\   
\astrobj{Camargo 341} &275.88&  -2.15&  $9:28:45$& $ -54:00:50$&    $7\times7$ &EGr&  n&                                   \\   
\astrobj{Camargo 342} &275.97&  -2.03&  $9:29:46$& $ -53:59:22$&    $8\times8$ &EC&   n&     prominent                     \\   
\astrobj{Camargo 343} &276.01&  -1.79&  $9:31:05$& $ -53:50:23$&    $7\times7$ &EC&   y&     impressive, dusty core         \\   
\astrobj{Camargo 344} &276.03&  -1.45&  $9:32:43$& $ -53:36:21$&    $3\times3$ &EC&   n&          include subcluster           \\   
\astrobj{Camargo 345} &276.04&  -1.00&  $9:34:51$& $ -53:17:11$&    $9\times9$ &EC&   n&   winds  impressive               \\  \hline \tablebreak 
\astrobj{Camargo 346} &276.07&  -1.44&  $9:33:02$& $ -53:37:36$&    $9\times9$ &EC&   n&     incl.  subcl. cavity        \\   
\astrobj{Camargo 347} &276.38&  -2.64&  $9:28:58$& $ -54:42:56$&    $5\times5$ &EC&   n&      object is a loose core?      \\   
\astrobj{Camargo 348} &276.51&  -2.45&  $9:30:29$& $ -54:40:14$&    $7\times7$ &EC&   n&           dusty core              \\   
\astrobj{Camargo 349} &276.58&  -3.88&  $9:23:54$& $ -55:44:35$&    $8\times8$ &EGr&  n&                                   \\   
\astrobj{Camargo 350} &276.60&  -1.91&  $9:33:27$& $ -54:20:01$&    $6\times6$ &EC&   n&          impressive               \\     
\astrobj{Camargo 351} &276.61&  -1.72&  $9:34:24$& $ -54:12:08$&    $7\times7$ &EGr&  n&                                   \\   
\astrobj{Camargo 352} &276.70&  -1.47&  $9:36:00$& $ -54:04:35$&    $8\times5$ &EGr&  n&                                   \\   
\astrobj{Camargo 353} &276.74&  -1.60&  $9:35:39$& $ -54:11:40$&    $8\times8$ &EC&   n&                                   \\   
\astrobj{Camargo 354} &276.97&  -4.02&  $9:25:10$& $ -56:06:45$&    $6\times6$ &EC&   n&     prominent                     \\   
\astrobj{Camargo 355} &277.06&  -3.16&  $9:29:54$& $ -55:33:19$&    $9\times9$ &EC&   n&                                   \\   
\astrobj{Camargo 356} &277.13&  -1.53&  $9:37:58$& $ -54:24:25$&    $5\times5$ &EGr&  n&                                   \\   
\astrobj{Camargo 357} &277.15&  -1.60&  $9:37:42$& $ -54:28:04$&    $7\times7$ &EC&   n&       prominent                   \\   
\astrobj{Camargo 358} &277.22&  -1.55&  $9:38:18$& $ -54:28:42$&    $6\times6$ &EC&   n&   prominent                       \\   
\astrobj{Camargo 359} &277.23&   0.64&  $9:48:00$& $ -52:49:47$&    $9\times9$ &EC&   n&   core, prominent                 \\   
\astrobj{Camargo 360} &277.36&  -1.36&  $9:39:56$& $ -54:25:46$&    $8\times8$ &EC&   n&        prominent                  \\   
\astrobj{Camargo 361} &277.50&  -1.23&  $9:41:13$& $ -54:25:23$&    $7\times6$ &EGr&  n&    1st generation, bubble          \\   
\astrobj{Camargo 362} &277.51&  -1.28&  $9:41:02$& $ -54:28:18$&    $6\times6$ &EGr&  n&  protocl, 2nd gener winds, dusty core\\  
\astrobj{Camargo 363} &277.52&  -3.32&  $9:31:28$& $ -55:59:33$&    $8\times8$ &EGr&  n&                                    \\  
\astrobj{Camargo 364} &277.56&   0.56&  $9:49:19$& $ -53:05:43$&    $8\times8$ &EGr&  n&                                    \\  
\astrobj{Camargo 365} &277.63&  -1.05&  $9:42:42$& $ -54:22:25$&    $6\times6$ &EC&   y&     dusty core                     \\  
\astrobj{Camargo 366} &277.68&   0.89&  $9:51:20$& $ -52:54:49$&    $9\times9$ &EGr&  n&                                    \\  
\astrobj{Camargo 367} &277.69&  -3.15&  $9:33:10$& $ -55:58:52$&   $10\times10$ &EGr& n&                                    \\   
\astrobj{Camargo 368} &277.71&   0.51&  $9:49:54$& $ -53:13:48$&   $10\times10$ &EGr& n&            winds                   \\   
\astrobj{Camargo 369} &277.80&   0.50&  $9:50:18$& $ -53:17:38$&    $6\times6$ &EC&   n& bubble, winds, impressive            \\   
\astrobj{Camargo 370} &277.84&  -3.08&  $9:34:20$& $ -56:01:30$&    $8\times8$ &EC&   n&      impressive                    \\  
\astrobj{Camargo 371} &277.94&  -1.43&  $9:42:34$& $ -54:51:48$&    $6\times6$ &EGr&  n&    protocl, winds, dusty core        \\  
\astrobj{Camargo 372} &277.95&  -2.65&  $9:36:58$& $ -55:47:20$&    $8\times8$ &EGr&  n&                                    \\  
\astrobj{Camargo 373} &278.35&  -0.89&  $9:47:09$& $ -54:43:10$&    $8\times8$ &EC&   y&   impressive, dusty core            \\  
\astrobj{Camargo 374} &278.60&  -0.99&  $9:48:02$& $ -54:57:18$&    $8\times8$ &EC&   n&      loose core                    \\  
\astrobj{Camargo 375} &278.63&  -0.37&  $9:50:56$& $ -54:29:41$&    $6\times6$ &EC&   y&   impressive stellar and dusty core\\  
\astrobj{Camargo 376} &278.78&  -0.65&  $9:50:31$& $ -54:48:25$&    $6\times6$ &EC&   n&    protocl., bubble, winds            \\  
\astrobj{Camargo 377} &279.04&  -1.13&  $9:49:49$& $ -55:20:45$&    $7\times7$ &EC&   n&                                    \\  
\astrobj{Camargo 378} &279.12&  -1.22&  $9:49:52$& $ -55:27:39$&    $6\times6$ &EGr&  n&                                    \\  
\astrobj{Camargo 379} &279.12&  -1.22&  $9:49:32$& $ -55:50:07$&   $10\times10$ &EGr& n&                                    \\   
\astrobj{Camargo 380} &279.40&  -1.68&  $9:49:19$& $ -55:59:45$&    $3\times2$ &EC&   n&     2nd generation, deeply absorbed \\  
\astrobj{Camargo 381} &279.42&  -0.96&  $9:52:39$& $ -55:27:10$&    $6\times6$ &EGr&  y&     bubble, impress, 1st and 2nd gener\\ 
\astrobj{Camargo 382} &279.43&  -1.69&  $9:49:17$& $ -56:02:27$&    $5\times5$ &EC&   n&     1st generation, shock,core      \\ 
\astrobj{Camargo 383} &279.50&  -1.52&  $9:50:35$& $ -55:56:17$&    $7\times7$ &EC&   n&                                     \\ 
\astrobj{Camargo 384} &279.52&  -1.73&  $9:49:44$& $ -56:06:48$&    $6\times6$ &EC&   n&         core                        \\ 
\astrobj{Camargo 385} &279.57&  -1.42&  $9:51:27$& $ -55:54:12$&    $6\times6$ &EC&   y&                                     \\ 
\astrobj{Camargo 386} &279.67&  -1.02&  $9:53:47$& $ -55:39:02$&    $5\times5$ &EGr&  n&            prominent                \\ 
\astrobj{Camargo 387} &279.93&  -0.40&  $9:57:56$& $ -55:19:49$&   $11\times11$ &EGr& n&                                     \\  
\astrobj{Camargo 388} &280.00&  -1.20&  $9:54:47$& $ -55:59:49$&    $6\times6$ &EC&   n&   winds, impressive                  \\ 
\astrobj{Camargo 389} &280.02&  -1.42&  $9:53:57$& $ -56:10:56$&    $7\times7$ &EC&   n&    impressive, core                  \\ 
\astrobj{Camargo 390} &280.37&  -1.68&  $9:54:46$& $ -56:36:20$&    $8\times8$ &EGr&  n&      winds                          \\ 
\astrobj{Camargo 391} &280.43&  -1.23&  $9:57:09$& $ -56:17:09$&    $7\times7$ &EC&   n&       poor core                     \\   
\astrobj{Camargo 392} &280.43&  -1.74&  $9:54:48$& $ -56:41:15$&    $8\times8$ &EC&   n&       small core, winds,impressive   \\ 
\astrobj{Camargo 393} &280.44&  -1.83&  $9:54:30$& $ -56:45:46$&    $7\times7$ &EGr&  n&     winds                           \\ 
\astrobj{Camargo 394} &280.46&  -1.02&  $9:58:12$& $ -56:08:24$&    $6\times6$ &EC&   n&                                     \\ 
\astrobj{Camargo 395} &280.51&  -0.99&  $9:58:38$& $ -56:08:48$&    $6\times6$ &EC&   n&          dusty core                 \\ 
\astrobj{Camargo 396} &280.57&  -0.47& $10:01:11$& $ -55:46:01$&    $6\times6$ &EC&   n&           loose core                \\ 
\astrobj{Camargo 397} &280.58&  -1.44&  $9:57:02$& $ -56:32:23$&    $8\times8$ &EGr&  n&          cavities                   \\ 
\astrobj{Camargo 398} &280.62&  -1.19&  $9:58:22$& $ -56:22:13$&    $6\times6$ &EC&   y&      prominent, dusty core           \\ 
\astrobj{Camargo 399} &280.65&  -1.27&  $9:58:12$& $ -56:27:03$&    $7\times7$ &EC&   y&    impressive, central concentration \\ 
\astrobj{Camargo 400} &280.66&  -0.88&  $9:59:56$& $-56:09:03$&    $6\times6$ &EC&    n&    impressive                       \\
\astrobj{Camargo 401} &280.69&  -1.15&  $9:58:58$& $-56:22:58$&    $6\times6$ &EGr&   n&                                     \\
\astrobj{Camargo 402} &280.76&  -1.33&  $9:58:34$& $-56:34:13$&    $10\times10$ &EC&  y&    dusty core                       \\  
\astrobj{Camargo 403} &280.78&  -1.06&  $9:59:50$& $-56:21:58$&    $7\times5$ &EC&    n&     dusty core                      \\
\astrobj{Camargo 404} &280.89&  -0.55& $10:02:41$& $-56:01:30$&    $7\times7$ &EGr&   n&        bubble                       \\
\astrobj{Camargo 405} &280.99&  -1.53&  $9:59:00$& $-56:51:42$&    $6\times6$ &EC&    y&    wind,impressive                  \\
\astrobj{Camargo 406} &281.05&  -1.55&  $9:59:15$& $-56:54:46$&    $6\times6$ &EC&    y&    dusty core                       \\
\astrobj{Camargo 407} &281.06&  -1.73&  $9:58:29$& $-57:04:15$&    $5\times5$ &EC&    n&   core                              \\
\astrobj{Camargo 408} &281.08&  -2.55&  $9:54:53$& $-57:43:36$&    $9\times9$ &EGr&   n&   inc. promin 9´ sharp edge dust neb \\
\astrobj{Camargo 409} &281.09&  -1.59&  $9:59:17$& $-56:58:39$&    $4\times4$ &EC&    y&   protocluster? impressive, rel DBS 125 \\
\astrobj{Camargo 410} &281.12&  -1.15& $10:01:26$& $-56:38:49$&    $7\times7$ &EGr&   n&     cavity                          \\ 
\astrobj{Camargo 411} &281.22&  -1.26& $10:01:33$& $-56:47:15$&    $7\times7$ &EC&    y&    compact core, stellar and dusty   \\
\astrobj{Camargo 412} &281.26&  -1.69&  $9:59:52$& $-57:09:29$&    $6\times6$ &EC&    y&      core                           \\
\astrobj{Camargo 413} &281.30&  -2.25&  $9:57:32$& $-57:37:38$&    $6\times6$ &EC&    n&   loose poor core                   \\
\astrobj{Camargo 414} &281.42&  -1.80& $10:00:18$& $-57:20:34$&    $12\times12$ &EC&  n&    inc. subclusters, 2nd gener, impres\\ \hline \tablebreak
\astrobj{Camargo 415} &281.50&  -2.55&  $9:57:23$& $-57:59:20$&    $6\times6$ &EC&    y&    core, impressive                 \\
\astrobj{Camargo 416} &281.53&  -1.82& $10:00:54$& $-57:25:49$&    $7\times7$ &EC &   n&    winds, impressive core            \\
\astrobj{Camargo 417} &281.56&  -2.48&  $9:58:02$& $-57:50:00$&    $6\times6$ &EC &   n&    dusty core, impressive           \\
\astrobj{Camargo 418} &281.63&  -1.29& $10:03:49$& $-57:03:57$&    $7\times7$ &EC &   n&    poor core                        \\
\astrobj{Camargo 419} &281.64&  -2.54&  $9:58:14$& $-58:04:06$&    $8\times8$ &EC &   n&   core, prominent                   \\
\astrobj{Camargo 420} &281.65&  -2.07& $10:00:28$& $-57:41:57$&    $4\times4$ &ECC&   n&    protocl? bubble, core, dust \& 3 sts \\
\astrobj{Camargo 421} &281.69&  -2.06& $10:00:44$& $-57:42:48$&    $6\times6$ &EC &   y&    core, impressive                 \\
\astrobj{Camargo 422} &281.70&  -1.10& $10:05:05$& $-56:57:02$&    $6\times6$ &EC &   y&    small dusty core                 \\
\astrobj{Camargo 423} &281.81&  -1.09& $10:05:46$& $-57:00:16$&    $6\times6$ &EC &   n&     poor core, impressive           \\
\astrobj{Camargo 424} &281.86&  -1.65& $10:03:37$& $-57:29:24$&    $2\times2$ &EC &   y&     dusty core                      \\
\astrobj{Camargo 425} &281.87&  -2.12& $10:01:33$& $-57:51:59$&    $7\times7$ &EGr&   y&         wind very poor core?        \\
\astrobj{Camargo 426} &281.88&  -1.85& $10:02:49$& $-57:39:50$&    $7\times7$ &EC &   n&    impressive, core                 \\
\astrobj{Camargo 427} &281.94&  -1.52& $10:04:36$& $-57:25:52$&    $7\times7$ &EC &   n&    small dusty core                 \\
\astrobj{Camargo 428} &281.98&  -1.37& $10:05:32$& $-57:19:48$&    $6\times6$ &EC &   y&    small dusty core, prominent      \\
\astrobj{Camargo 429} &282.00&  -1.72& $10:04:07$& $-57:37:50$&    $8\times8$ &EGr&   n&   winds, impressive                 \\
\astrobj{Camargo 430} &282.05&  -1.41& $10:05:48$& $-57:24:21$&    $5\times5$ &EC &   n&   small impressive core             \\
\astrobj{Camargo 431} &282.06&  -1.85& $10:03:55$& $-57:46:09$&    $6\times6$ &EC &   y&  small dusty core, winds            \\
\astrobj{Camargo 432} &282.06&  -2.12& $10:02:44$& $-57:58:51$&    $7\times7$ &EC &   n&  poor core, prominent, bubble         \\   
\astrobj{Camargo 433} &282.16&  -1.57& $10:05:47$& $-57:35:51$&    $6\times6$ &EC &   n&        loose dusty core             \\
\astrobj{Camargo 434} &282.22&  -2.13& $10:03:36$& $-58:05:04$&    $6\times6$ &EC &   n&       small dusty core              \\
\astrobj{Camargo 435} &282.35&  -1.85& $10:05:42$& $-57:56:11$&    $6\times6$ &EC &   n&   unique long core, impressive      \\
\astrobj{Camargo 436} &282.52&  -1.98& $10:06:07$& $-58:08:11$&    $6\times6$ &EC &   n&    loose core, impressive           \\
\astrobj{Camargo 437} &289.29&  -3.93& $10:44:38$&  $-63:23:22$&    $5\times5$ &EC&   n&    complex core struct in W3 and W4 \\

\enddata
\vspace{-0.5cm}
\tablecomments{Cols. $2-5$: Central coordinates. Col. $6$: cluster size. Col. $7$: object type - OC means open cluster, OCC open cluster candidate, EC embedded cluster, ECC embedded cluster candidate, EGr stellar group, and SFN star-forming nebulae. Col. $8$: cross-identifications with \citet{Avedisova02} SFRs or candidates within 5' of the central coordinates, y and n means yes and not.}.

\end{deluxetable}

\end{document}